\documentclass[twocolumn]{aastex631}
\submitjournal{ApJS on February 16, 2025}

\usepackage{courier}
\usepackage{multirow}
\usepackage{threeparttable}
\usepackage{booktabs}
\usepackage{color}
\usepackage[section]{placeins}
\usepackage{graphicx}
\usepackage{subfigure}
\usepackage{parskip}
\usepackage{longtable}
\usepackage{supertabular}
\usepackage{bm}

\begin{document}

\title{The Impact of $^{12}$C($\alpha, \gamma$)$^{16}$O Reaction on the
  Presupernova Evolution and Supernova Explodability of Massive Stars}

\correspondingauthor{Wenyu Xin, Ken'ichi Nomoto}
\email{xinwenyu16@mails.ucas.ac.cn, nomoto@astron.s.u-tokyo.ac.jp}

\author[0000-0003-3646-9356]{Wenyu Xin}
\affiliation{National Astronomical Observatories,
Chinese Academy of Sciences, Beijing 100101, China}
\affiliation{School of Astronomy and Space Science, University of Chinese Academy of 
Sciences, Beijing 100101, China}
\affiliation{Institute for Frontiers in Astronomy and Astrophysics,
Beijing Normal University, Beijing, 102206, China}
\affiliation{School of Physics and Astronomy, Beijing Normal University, Beijing, 100875, China}

\author[0000-0001-9553-0685]{Ken'ichi Nomoto}
\affiliation{Kavli Institute for the Physics and Mathematics of the Universe (WPI), The University of Tokyo Institutes for Advanced Study, \\
The University of Tokyo, Kashiwa, Chiba 277-8583, Japan}


\author[0000-0002-8980-945X]{Gang Zhao}
\affiliation{National Astronomical Observatories,
Chinese Academy of Sciences, Beijing 100101, China}
\affiliation{School of Astronomy and Space Science, University of Chinese Academy of
Sciences, Beijing 100101, China}



\begin{abstract}

Among the uncertainties of stellar evolution theory, we investigate
how the $^{12}$C($\alpha, \gamma$)$^{16}$O reaction rate affects the
evolution of massive stars for the initial masses of $M ({\rm ZAMS})=$
13 - 40 M$_\odot$ and the solar metallicity.  We show that the {\sl
  explodability} of these stars, i.e., which of a neutron star (NS) or
a black hole (BH) is formed, is sensitive to the strength of
convective shell burning of C and O, and thus the mass fractions of C
($X$(C)) and O in the shell.  For the small $^{12}$C($\alpha,
\gamma$)$^{16}$O reaction rate that yields larger $X$(C), $X$(C) is
further enhanced by mixing of C from the overlying layer and then C
shell burning is strengthened.  The extra heating by C shell burning
tends to prevent the contraction of outer layers and decrease the {\sl
  compactness parameter} at $M_r$ = 2.5 M$_\odot$.  This effect leads
to the formation of smaller mass cores of Si and Fe and steeper
density and pressure gradients at the O burning shell in the
presupernova models. If the pressure gradient there is steeper, the
model is more likely to explode to form a NS rather than a BH.  We
describe the pressure gradient against $M_r$ with $V/U$ and the
density drop with $1/U$, where $U$ and $V$ are non-dimensional
variables to describe the stellar structure.  We estimate the critical
values of $V/U$ and $1/U$ at the O-burning shell above which the model
is more likely to explode.  We conclude that the smaller
$^{12}$C($\alpha, \gamma$)$^{16}$O reaction rate makes the mass range
of $M ({\rm ZAMS})$ that forms a NS larger.

\end{abstract}



\section{Introduction} \label{sec:intro}

In the Universe, stars are formed with various masses, evolve, and
die.  Understanding the nature and evolution of stars is
essentially crucial in understanding the evolution of the Universe
\citep[e.g.,][]{1996snih.book.....A}.
Extensive studies on spherical stars have shown that the evolution of
stars are basically determined by their Zero-Age Main-Sequence masses
$M ({\rm ZAMS})$ \citep[e.g.,][]{1962PThPS..22....1H}.
In particular, stars with
$M ({\rm ZAMS}) \gtrsim$ 8 M$_\odot$ evolve through hydrostatic
nuclear burning and end their lives as core-collapse supernovae
\citep[e.g.,][]{2013ARA&A..51..457N}.

The exact explosion mechanism of core-collapse supernovae has
not been well understood and the important properties of the
explosion, e.g., the kinetic energy, has not been well-reproduced
despite the extensive hydrodynamical simulations
\citep[e.g.,][]{1980SSRv...25..155S}. Possible sources of difficulties in the
core-collapse supernova modeling include the complicated physical
processes, such as neutrino transport during core-collapse and
the difficulty in numerical treatment of those processes.

Another source of difficulty would be inaccuracies of the progenitor's
evolution and presupernova structure \citep[e.g.,][]{2017suex.book.....B}.
Such inaccuracies may stem from uncertainties
in the complicated evolution processes, which include mass loss
\citep[e.g.,][]{Renzo_2017}, convection
\citep[e.g.,][]{Woosley_2002}, rotation
\citep[e.g.,][]{2008Presupernova, 2016CONVECTIVE}, magnetic fields
\citep[e.g.,][]{2009PhT....62i..52M},
and nuclear reactions \citep[e.g.,][]{1993PhR...227...65W, 2013ApJS..207...18S,
2018ApJS..234...19F}.

A possible example of such inaccuracies of progenitor's evolution is
the ``mass-gap'' problem of black holes (BHs). The current stellar
evolution theory predicts that stars of 140 M$_\odot$ $\lesssim$
$M ({\rm ZAMS})$  $\lesssim$ 300 M$_\odot$ undergo thermonuclear
explosions triggered by electron-positron pair-creation instability
and are disrupted completely without leaving the compact object
\citep[pair-instability supernova: PISN]{1967PhRvL..18..379B}
\footnote{For discussion on PISNe in this section, we
assume that stars have low enough metallicities for a wind-type
mass loss to be negligible \citep[see, e.g.,][for the wind effect
on the core mass]{2015cdem.confE..15L}.}.

In close binary systems, stars with
$140 M_\odot \lesssim M ({\rm ZAMS}) \lesssim
300 M_\odot$ form He stars with $M$(He) = 65 - 140 M$_\odot$,
which undergo PISNe.  The upper mass limit of BHs formed from $M$(He)
$\lesssim$ 65 M$_\odot$ has been estimated to be $\sim$ 50 M$_\odot$, while
the lower mass limit of BHs from $M$(He) $\gtrsim$ 140 M$_\odot$ is $\sim$ 140
M$_\odot$ \citep{Woosley_2017}.  Therefore, the ``mass-gap'' of $\sim 50 - 140$
M$_\odot$ due to PISNe has been predicted to exist for BHs in close
binary systems.

However, the gravitational-wave (GW) signal GW190521 detected by
LIGO/Virgo seems to originate from a merger of binary BHs
($85_{-14}^{+21}$ M$_{\odot}$ and $66_{-18}^{+17}$ M$_{\odot}$;
\citealt{Abbott_2020}).  The probability that at least one of them
falls in the BH mass-gap due to PISNe is 99.0\%.

This is a challenge to the stellar evolution theory.  It has motivated
many attempts to examine the effects of uncertainties involved in the
stellar evolution, such as the reaction rate \citep{Farmer_2019, Farmer_2020,
2021MNRAS.501.4514C}, rotation \citep{2020A&A...640L..18M, Mapelli_2020},
Super-Eddington accretion\citep{2020ApJ...897..100V},
magnetic fields\citep{10.1093/mnras/staa237}
convection and overshooting\citep{2020ApJ...904L..13R, 2021ApJ...912L..31W,
2020MNRAS.496.1967K}.

Among those attempts is the study of the effects of the uncertainty of
the $^{12}$C$(\alpha, \gamma)^{16}$O reaction rate \citep{Farmer_2019,
Farmer_2020, 2021MNRAS.501.4514C}.
Those works have found that if the $^{12}$C$(\alpha, \gamma)^{16}$O
rate is smaller, the mass range of PISNe and thus the BH mass gap move
to higher masses.  The upper limit of the BH mass is $\sim$ 59, 71,
and 94 M$_\odot$, if the rate is smaller by 1$\sigma$, 2$\sigma$, and 3$\sigma$,
respectively \citep{Farmer_2020}.

To explain the observed BH masses of GW190521, the $^{12}$C$(\alpha,
\gamma)^{16}$O rate might be significantly smaller than the currently
adopted in the stellar evolution calculations, although some
combination with other effects might be necessary.

Motivated by these works, we study the effect of the $^{12}$C$(\alpha,
\gamma)^{16}$O rate on the evolution of massive stars.
The effect of the $^{12}$C$(\alpha, \gamma)^{16}$O rate has been
studied on nucleosynthesis yields \citep{1993PhR...227...65W,
2007ApJ...671..821T, 2013ApJ...769....2W},
carbon burning \citep{2016ApJ...821...38S, 2020MNRAS.492.2578S},
and compactness \citep{2014Sukhbold}.
In the present work, we study the effects of this rate on the stellar
structure and evolution through Fe core-collapse more systematically.
In particular, we will focus on 
how the rate uncertainly affects the presupernova structure
and the explodability, i.e., whether the progenitors explode or collapse  
\citep{2011APS..DNP.CG005O, 2012ApJ...757...69U, 2016ApJ...818..124E,
 2016MNRAS.460..742M}.

More specifically, to clarify the mass and the rate dependencies, we
evolve stars with $M ({\rm ZAMS}) = 13 - 40$ M$_\odot$ from the ZAMS
to Fe core-collapse with MESA by changing the
$^{12}$C$(\alpha, \gamma)^{16}$O rate in the range of $\pm3\sigma$.
Effects on nucleosynthesis yields, including metallicity effects, will
be reported in the forthcoming paper.

The uncertainties in the stellar structure and evolution also stem
from the difference in computer codes and numerical methods.  Thus our
aim of the present study includes the comparison of our MESA results
with other numerical results obtained with different codes such as
Kepler \citep{1993PhR...227...65W, 2014Sukhbold},
FRANEC \citep{2020ApJ...890...43C}, and previous
results by MESA \citep{2016ApJS..227...22F}.

In Section~\ref{sec:model}, we describe the models and some basic
definitions used in this work including non-dimensional variables $U$
and $V$.
In Section~\ref{sec:h_burn}, we present the evolution of H and He
burning and discuss the effect of $^{12}$C$(\alpha, \gamma)^{16}$O
rate on the mass fraction of $^{12}$C during He burning and the C
ignition stages.
In Section~\ref{sec:carbon}, we study how this reaction rate affects
the evolution from C burning to the Fe core-collapse and the
presupernova core structure for $M ({\rm ZAMS})=$ 28, 25 and 30
M$_{\odot}$.
In Section \ref{sec:xi_rate}, we study the explodability of the presupernova models
for all ranges of $M ({\rm ZAMS})$ and $\sigma$ studied here, showing
the criterion of the explodability including the critical values of
$V/U$ and $1/U$.  We then show how the mass range $M ({\rm ZAMS})$ of
stars to form a neutron star (NS) (or a BH) depends on the $^{12}$C$(\alpha,
\gamma)^{16}$O rate.
In Section~\ref{sec:summary}, summaries and discussion are given.

\section{Models and Parameters} \label{sec:model}

We employ the Modules for Experiments in Stellar Astrophysics (MESA, version 12778;
\citealt{2011ApJS..192....3P, 2013ApJS..208....4P, 2015ApJS..220...15P, 
2018ApJS..234...34P, 2019ApJS..243...10P, 2023ApJS..265...15J}) to evolve
massive stars from ZAMS to Fe core-collapse.
The packages ``25M\_pre\_ms\_to\_core\_collapse" (examples in the test suite directory
of MESA) are modified to build our models.

\subsection{Reactions and Nuclear Network} \label{sec:nuclear}

\begin{table}[htb]
\centering
\caption{Nuclides included in the nuclear reaction network of mesa\_128.net.}
\label{tab:network}
\begin{tabular}{cccccc}
\toprule
Element & $A_{\rm min}$ & $A_{\rm max}$ & Element & $A_{\rm min}$ & $A_{\rm max}$ \\
\midrule
n       & 1    & 1    & S       & 31   & 34   \\
H       & 1    & 2    & Cl      & 35   & 37   \\
He      & 3    & 4    & Ar      & 35   & 38   \\
Li      & 7    & 7    & K       & 39   & 43   \\
Be\tnote{1}      & 7    & 10   & Ca      & 39   & 44   \\
B       & 8    & 8    & Sc      & 43   & 46   \\
C       & 12   & 13   & Ti      & 44   & 48   \\
N       & 13   & 15   & V       & 47   & 51   \\
O       & 14   & 18   & Cr      & 48   & 57   \\
F       & 17   & 19   & Mn      & 51   & 56   \\
Ne      & 18   & 22   & Fe      & 52   & 58   \\
Na      & 21   & 24   & Co      & 55   & 60   \\
Mg      & 23   & 26   & Ni      & 55   & 61   \\
Al      & 25   & 28   & Cu      & 59   & 62   \\
Si      & 27   & 30   & Zn      & 60   & 64   \\
P       & 30   & 32   &         &      &      \\
\bottomrule
\end{tabular}
\begin{tablenotes}
\footnotesize
\item[1] $^{8}$Be is not included.
\end{tablenotes}
\end{table}

To study nucleosynthesis, we adopt a large nuclear
reaction network consisting of 128 isotopes (\texttt{mesa\_128.net},
\citealt{2011ApJS..192....3P, 1999ApJS..124..241T,
2000ApJS..129..377T}). The isotopes included in this network are
listed in Table~\ref{tab:network}.
\citet{2016ApJS..227...22F} studied the evolution with several sizes
of network and concluded that $\sim$ 127 isotopes are needed to get
convergence of various quantities of stellar models at the $\sim$ 10\%
level.

For weak interactions, the tabulation of
\citet{2000NuPhA.673..481L, ODA1994231, 1985ApJ...293....1F} are
adopted. In this work, the reaction rates of 3$\alpha$ from
\citet{ANGULO19993}, $^{12}$C$(\alpha, \gamma)^{16}$O from
\citet{2002ApJ...567..643K}, and the $^{12}$C+$^{12}$C rate from
\citet{2018Natur.557..687T} are adopted. 
Other reaction rates are
taken from \texttt{REACLIB} \citep{Cyburt_2010}.

Regarding the uncertainty of the $^{12}$C$(\alpha, \gamma)^{16}$O rate,
all direct measurements of this reaction rate have been
performed at the energies higher than E$_{\rm c.m.}=$ 891 keV
\citep{Hammer_2005}, which is much higher than the energy (E$_{\rm
  c.m.}=$ 300 keV) corresponding to the temperature during helium
burning \citep{PhysRevC.92.045802}.

The cross-section decreases exponentially at low energies, resulting
from the small Coulomb penetrability. Therefore, it is necessary to
extrapolate from the high energy to obtain the reaction rate in
the astrophysics environment. However, most estimates are still far
from the uncertainty of better than 10\% required by stellar models
\citep[e.g.,][]{Woosley_2002}.

\begin{table}
\centering
\caption{The corresponding relationship between multiplier factors,
f$_{C12\alpha}$, and the $\sigma$ of $^{12}$C$(\alpha,\gamma)^{16}$O
reaction rates for the temperature during the He burning stage.}
\begin{tabular}{cccccccc}
\toprule
$\sigma$      & -3   & -2   & -1  & 0 & 1 & 2  &   3    \\
\midrule
f$_{C12\alpha}$   & 0.37 & 0.52 & 0.72 & 1 & 1.28 & 1.93  & 2.69  \\
\bottomrule
\end{tabular}
\label{tab:multiply}
\end{table}

Since the uncertainty of the reaction rate follows the
log-normal distribution, we can calculate the lower and upper rates given
by $\pm 3 \sigma$ for $^{12}$C$(\alpha,\gamma)^{16}$O to cover
99\% of the probability distribution of the rate
\citep{2013ApJS..207...18S}. The $\sigma$ here is defined by
the following formula:
\begin{equation}
e^{\sigma}=\sqrt{\frac{x_{\rm high}}{x_{\rm low}}}
\end{equation}
where $x_{\rm high}$ and $x_{\rm low}$ represent the high and low rates of the
reactions, respectively. In the present work, we use $\sigma_{C12\alpha}$
to represent the uncertainty of the $^{12}$C$(\alpha, \gamma)^{16}$O rate.
We also show the value of the multipliers corresponding to each $\sigma$ in
Table \ref{tab:multiply}.

\begin{figure*}[htb!]
\centering
\begin{minipage}[c]{0.75\textwidth}
\centerline{$M({\rm ZAMS})=28 $ M$_\odot$}
\includegraphics [width=132mm]{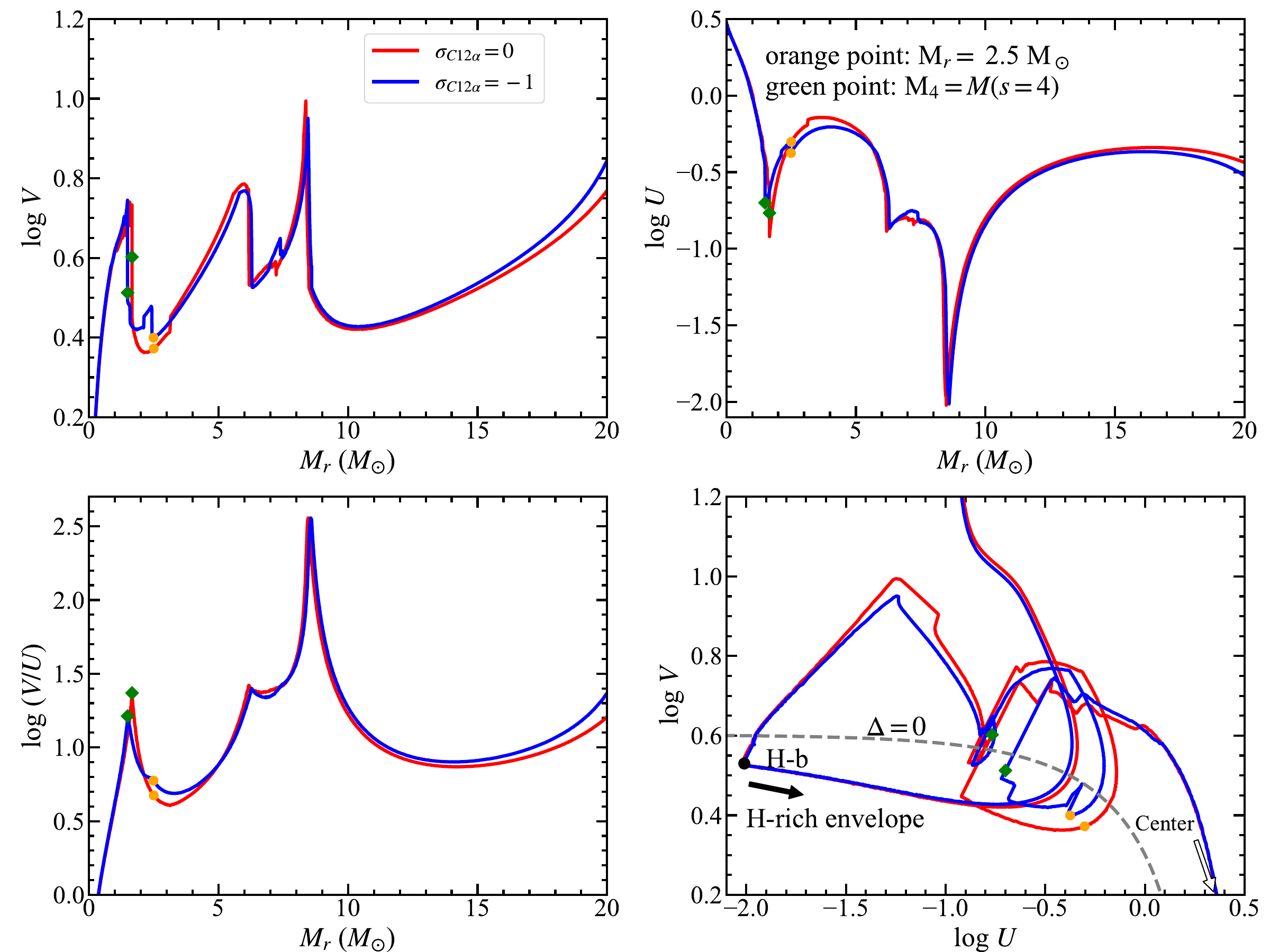}
\end{minipage}%
\caption{The $U-V$ curves at $t=t_{\rm f}$ for $M ({\rm ZAMS}) =$ 28 M$_\odot$.
The black points indicate the place where $M_r = 2.5$ M$_{\odot}$, while the
green squares show the place where $M_r = M(s=4)$, which is defined in
Equation (\ref{equ:m4mu4}).
In the log $U$ - log $V$ plot, we show the star's center with the
white arrow. The location of the hydrogen burning shell and the direction
of the H-rich envelope are indicated with the black point and arrow.
The gray dashed line indicates the $\Delta \equiv 2U+V-4=0$ curve.
\label{fig:28M_uv}}
\end{figure*}

\subsection{Input Physics} \label{sec:input}
We calculate the evolution of 82 models in this work. For
$\sigma_{C12\alpha}=$ 0, we evolve stars with the initial masses of $M ({\rm ZAMS}) = 13 - 40$
M$_{\odot}$ for a mass interval of 1 M$_{\odot}$. We also evolve
stars with $M ({\rm ZAMS}) =$ 20, 23, 25, 28, 30, 32, 35, 38, and 40 M$_{\odot}$ for
$\sigma_{C12\alpha}$=$\pm$1, $\pm$2, and $\pm$3.

For the convection zone, we use the mixing-length theory with the ratio
between the mixing-length and the pressure scale height of
  $\alpha_{\rm mlt}=$ 3.0 when the mass fraction
of hydrogen is larger than 0.5 and $\alpha_{\rm mlt}=$ 1.5 for other
stage. We adopt the exponential scheme with $f_{0}=$ 0.004 and $f=$ 0.01
for the overshooting. The Dutch scheme is adopted for the mass loss with
\texttt{Dutch\_scaling\_factor=0.5}. Three main mass loss
prescriptions included in Dutch wind are from
\citet{1988A&AS...72..259D} for cool stars,
\citet{2001A&A...369..574V} for hot hydrogen-rich stars,
and \citet{2000A&A...360..227N} for Wolf-Rayet stars.

For the initial elemental abundances, we assume the He mass fraction of $Y=2Z+0.24$
and the H mass fraction of $X=1-Y-Z$.  Here, $Z$ denotes the sum of the initial mass
fractions of heavy elements, i.e., C and heavier elements.
We assume the solar abundance ratios from \citet{1989GeCoA..53..197A} 
for isotopic abundance ratios among heavy elements.

\subsection{Basics of Stellar Evolution} \label{sec:basic}

In this section, we describe some relations and definitions,
which will be used later.

\subsubsection{U-V Curves} \label{subsec:uv}

The hydrostatic equilibrium of a spherical star gives two non-dimensional quantities,
which has been defined in, e.g., \citet{1958ses..book.....S}, \citet{1962PThPS..22....1H},
\citet{1980SSRv...25..155S}, \citet{1984ApJ...277..791N}, \citet{2000ApJ...538..837S}
and \citet{2013sse..book.....K}:

\begin{equation} \label{equ:u}
U\equiv \frac{{\rm dln} M_r}{{\rm dln} r}=\frac{4\pi r^3\rho}{M_r}
\end{equation}

\begin{equation} \label{equ:v}
V\equiv -\frac{{\rm dln} P}{{\rm dln} r}= \frac{GM_r\rho}{rP}
\end{equation}
From Equations~(\ref{equ:u}) and (\ref{equ:v}), we obtain $V/U$ as
\begin{equation} \label{equ:vu}
\frac{V}{U} = - \frac{{\rm dln} P}{{\rm dln} M_r} = \frac{G M_r^2}{4\pi r^4P}
\end{equation}
Differentiating Equations~(\ref{equ:u}) and (\ref{equ:v}), we also obtain
\begin{equation} \label{equ:dlnr}
{\rm d ln} r = - \frac{1}{\Delta} ({\rm dln} U - {\rm dln V})
\end{equation}

\begin{equation} \label{equ:delta}
\Delta \equiv 2U + 4V - 4   
\end{equation}

Here $V/U$ and $1/U$, respectively, show the steepness of the {\sl gradients}
of log $P$ and log $r$ against log $M_r$, which are
crucially important for the {\sl explodability} of stars as will be
discussed later (Section \ref{sec:m4mu4_core}).
Note also that $U$ is the ratio between the local density at the enclosed mass
$M_r$ and the mean density within the sphere of radius $r$.
If $U$ is very small (i.e., $1/U$ is very large), it indicates the existence
of almost a {\sl density jump}, which is also critical for explodability.
$V$ is the ratio between gravitational energy and thermal energy.

Figure \ref{fig:28M_uv} shows the $U-V$ curves of the $M ({\rm ZAMS}) =$ 
28 M$_\odot$ star at the final stage when the central temperature
reaches log $T_{\rm c}$ (K) = 10.0.  Here, we show the curves for
$\sigma_{C12\alpha}= 0$ and $-1$.  Although the two curves look
similar, important differences exist between the two cases, as
will be discussed in Section~\ref{sec:uv_curve}.

In the log $U$ - log $V$ plot, we show the star's center with the
white arrow.  The location of the hydrogen burning shell at
$M_r=$ 8.5 M$_\odot$ and the direction of the H-rich envelope are
indicated by the black point and arrow.  
The dashed line indicates the curve of $\Delta \equiv
2U+V-4=0$ \citep{1980SSRv...25..155S}. The loop of the log $U$ - log
$V$ curve shows a relatively large jump across $\Delta = 0$.
The location of the above jump is typically a nuclear burning shell
at the core edge.  This jump from large $V$ to small $U$ in the
deep core is critical for {\sl explodability}.

The black points indicate the place where $M_r = 2.5$ M$_{\odot}$,
while the green squares show the place where $M_r = M(s=4)$, which is
defined in Equation (\ref{equ:m4mu4}).  These points will be discussed
in later sections.

\subsubsection{Negative Specific Heat}

Another important quantity related to hydrostatic equilibrium is the
effective mass ($M_{\rm eff}$) giving the relation between the central
quantities and the mass \citep{1988PhR...163...13N}.

\begin{equation} \label{equ:meff}
\frac{P_c^3}{\rho_c^4}=4\pi {\rm G}^3(\frac{M_{\rm eff}}{\phi})^2
\end{equation}
where the $P_{c}$ and $\rho_{c}$ are the central pressure and density,
respectively; $\phi$ is a dimensionless mass that depends on the
polytropic index $N$ in the central region as $\phi=$ 10.73 and 16.14
for $N=$ 1.5 and 3, respectively \citep{1980SSRv...25..155S}.

Here the density ($\rho$) and pressure ($P$) are related by the
polytropic index as
\begin{equation}
\frac{{\rm dln} \rho}{{\rm dln} P}=\frac{N}{N+1}
\end{equation}

As will be discussed in later sections, $M_{\rm eff}$ corresponds to the core mass
enclosed within the most active nuclear burning shell.

For an ideal gas of $P = ({\rm k}/\mu {\rm H}) \rho T$,
where $\mu$ is the mean
molecular weight, H the unit of atomic mass, and k the Boltzmann
constant, Equation (\ref{equ:meff}) gives
\begin{equation} 
\frac{T_c^3}{\rho_c} \propto {\rm G}^3 M_{\rm eff}^2
\label{equ:trho}
\end{equation}
The specific entropy $s$ 
\begin{equation} 
s = \frac{\rm k}{\mu \rm H} {\rm ln} (\frac{T^{3/2}}{\rho}) + {\rm C}_1 
\label{equ:s1}
\end{equation}
is then given at the center as
\begin{equation} 
\frac{\mu \rm H}{\rm k} s_c = {\rm ln}(\frac{M_{\rm eff}^2}{T_c^{3/2}}) + {\rm C}_2 
= {\rm ln}(\frac{M_{\rm eff}}{\rho_c^{1/2}}) + {\rm C}_3
\label{equ:s2}
\end{equation}
where C$_1$, C$_2$, and C$_3$ are additional constants.  Equation
\ref{equ:s2} shows that for the same $T_c$ (i.e., for roughly the same
nuclear burning stage), $s_c$ is higher (and $\rho_c$ is lower) for
larger $M_{\rm eff}$, i.e., higher $s_c$ is necessary to sustain
larger mass against self-gravity.  When the nuclear reaction is not
active, the stellar core loses entropy by radiation and neutrinos and
thus $s_c$ decreases.  Equation (\ref{equ:s2}) shows that both $T_c$ and
$\rho_c$ increase as $s_c$ decreases for given $M_{\rm eff}$ as shown
by 
\begin{equation} 
C_g = \frac{{\rm d} s_c}{{\rm d ln} T_c} = - 1.5 \frac{\rm k}{\mu \rm H}
\label{equ:cg}
\end{equation}
Equation (\ref{equ:cg}) \citep{1988PhR...163...13N} implies that the stellar core
has the {\sl negative specific heat}.  The increase in $\rho_c$ means
the stellar core contracts and releases gravitational energy, part of
which goes into internal energy and the rest goes into radiation and
neutrino losses.

For discussion on the explodability of the presupernova models,
several criteria have been proposed. One is the compactness parameter
($\xi_{M_r/{\rm M}_{\odot}}$)
from \citet{2011APS..DNP.CG005O}:
\begin{equation} \label{equ:compact}
\xi_{M_r/{\rm M}_{\odot}} = \frac{M_r/{\rm M}_{\odot}}{r(M_r)/1000\,{\rm km}}
\end{equation}
where $M_r$ is the enclosed mass at the radius of $r(M_r)$.
\citet{2011APS..DNP.CG005O} proposed $M_r/{\rm M}_{\odot} =$ 2.5, i.e.,
$\xi_{2.5}$ for the actual criterion.

The others are $M_4$ and $\mu_4$ introduced by
\citet{2016ApJ...818..124E, 2020ApJ...890...51E}:
\begin{equation} \label{equ:m4mu4}
\mu_4 = \frac{{\rm d} m/{\rm M}_\odot}{{\rm d}r/1000\, {\rm km}} \Big|_{s=4}
\end{equation}
where $M_4$ and $r_4$ are the mass and radial coordinates where the 
specific entropy is $s=4$ (hereafter $s$ is given in units of erg g$^{-1}$K$^{-1}$). 
$\mu_4$ and $\mu_4M_4$ are related to the accretion
rate and accretion luminosity during the Fe core-collapse.

\section{Hydrogen Burning and Helium burning} \label{sec:h_burn}

\subsection{Hydrogen Burning}

\begin{table}[htb!]
\centering
\caption{Lost masses until the end of H burning, $\Delta M_{\rm H}$, He burning,
$\Delta M_{\rm He}$, and Fe core-collapse, $\Delta M_{\rm CC}$, 
as well as the final mass, $M(\rm final)$.}
\label{tab:reduced}
\begin{tabular}{ccccc}
\toprule
$M$({\rm Initial}) &$\Delta M_{\rm H}$ & $\Delta M_{\rm He}$ & $\Delta M_{\rm CC}$ & $M(\rm final)$\\
(M$_{\odot}$) &(M$_{\odot}$)  & (M$_{\odot}$)  & (M$_{\odot}$)  & (M$_{\odot}$) \\
\midrule
13           & 0.17  & 0.60  & 0.64  & 12.36 \\
14           & 0.19  & 0.79  & 0.83  & 13.17 \\
15           & 0.22  & 0.98  & 1.02  & 13.98 \\
16           & 0.25  & 1.43  & 1.48  & 14.52 \\
17           & 0.29  & 1.81  & 1.86  & 15.14 \\
18           & 0.33  & 1.66  & 1.71  & 16.29 \\
19           & 0.37  & 2.49  & 2.55  & 16.45 \\
20           & 0.39  & 2.51  & 2.58  & 17.42 \\
21           & 0.42  & 2.75  & 2.82  & 18.18 \\
22           & 0.46  & 2.87  & 2.94  & 19.06 \\
23           & 0.51  & 2.93  & 3.01  & 19.99 \\
24           & 0.55  & 4.07  & 4.15  & 19.85 \\
25           & 0.61  & 3.88  & 3.97  & 21.03 \\
26           & 0.62  & 4.61  & 4.70  & 21.30 \\
27           & 0.69  & 4.96  & 5.06  & 21.94 \\
28           & 0.78  & 5.76  & 5.85  & 22.15 \\
29           & 0.89  & 5.76  & 5.86  & 23.14 \\
30           & 0.99  & 6.07  & 6.17  & 23.83 \\
31           & 1.11  & 7.60  & 7.70  & 23.93 \\
32           & 1.23  & 7.85  & 7.97  & 24.03 \\
33           & 1.35  & 8.37  & 8.51  & 24.49 \\
34           & 1.48  & 8.50  & 8.63  & 25.37 \\
35           & 1.62  & 9.69  & 9.81  & 25.19 \\
36           & 1.77  & 10.62 & 10.74 & 25.26 \\
37           & 1.91  & 11.25 & 11.38 & 25.62 \\
38           & 2.06  & 12.06 & 12.12 & 25.88  \\
39           & 2.22  & 12.12 & 12.24 & 26.76 \\
40           & 2.38  & 12.85 & 12.97 & 27.03 \\
\bottomrule
\end{tabular}
\end{table}

\begin{figure}
\centering
\begin{minipage}[c]{0.42\textwidth}
\centerline{$M({\rm ZAMS})=28 $ M$_\odot$}
\includegraphics [width=75mm]{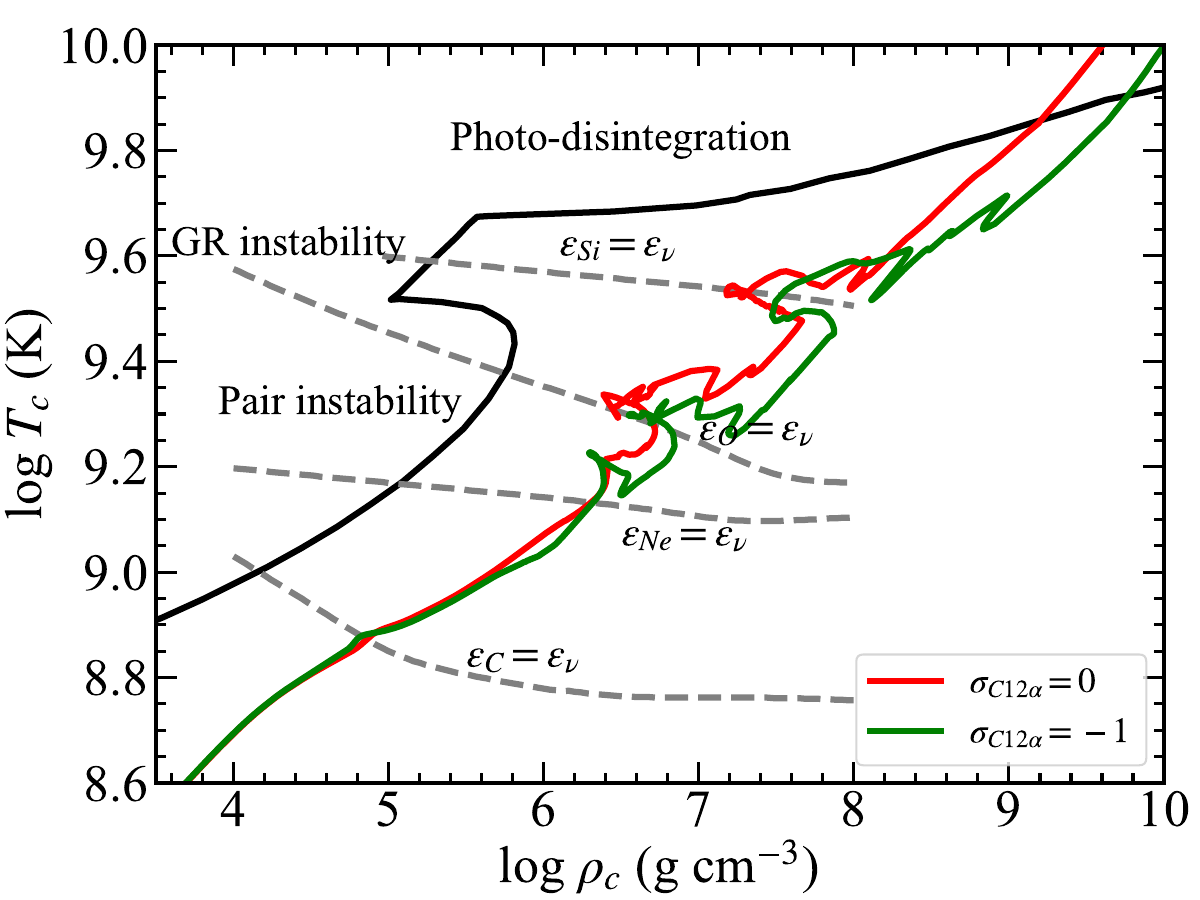}
\end{minipage}%
\caption{The central temperature against the central density for the evolution of the star with $M ({\rm ZAMS}) =$ 28 M$_{\odot}$, $Z =$ 0.02,
$\sigma_{C12\alpha}=$ 0 and $-$1. The blue dashed lines show the ignition 
lines of C burning, Ne burning, O burning and Si burning,
where the energy generation rate
by nuclear burning equals the energy loss rate by neutrino emissions.
In the region on the left of the black line, stars are dynamically unstable
due to the electron-positron pair creation (indicated as ``pair instability'') (\citealt{2009ApJ...706.1184O}), general relativistic effects (``GR instability'') 
(see, e.g., \citealt{1966PASJ...18..384O}),
and the photo-disintegration of matter in nuclear
statistical equilibrium (NSE) at $Y_{\rm e}=$ 0.5 (``photo-disintegration'') (\citealt{2009ApJ...706.1184O}).
\label{fig:trho}}
\end{figure}

We evolve the stars from H burning with $X=$ 0.70, $Y=$ 0.28, and $Z=$ 0.02
through the Fe core-collapse for $\sigma_{C12\alpha}=$ 0 and $-$1.
Figure~\ref{fig:trho} shows the evolution of stars of $M ({\rm ZAMS}) =$
28 M$_{\odot}$ in the central density and temperature diagram.
 The blue dashed lines show the ignition 
lines of C burning (for $X$(C) $=0.5$), Ne burning, O burning and Si burning,
where the energy generation rate
by nuclear burning equals the energy loss rate by neutrino emissions.
In the region on the left of the black line, stars are dynamically unstable
due to the electron-positron pair creation (indicated as ``pair instability'') (\citealt{2009ApJ...706.1184O}), general relativistic effects (``GR instability'') 
(see, e.g., \citealt{1966PASJ...18..384O}),
and the photo-disintegration of matter in nuclear
statistical equilibrium (NSE) at $Y_{\rm e}=$ 0.5 (``photo-disintegration'') (\citealt{2009ApJ...706.1184O}).

H burning forms a He core.  When the mass fractions of H at the center
becomes $X < 10^{-4}$, which is the end of H burning, we define the He
core mass, $M$(He), as an enclosed mass, $M_r$,
at the He core boundary where $X$ where $X$
changes inwardly from $X \ge 10^{-4}$ to $X < 10^{-4}$.

\begin{figure}[htbp]
\centering
\begin{minipage}[c]{0.42\textwidth}
\includegraphics [width=75mm]{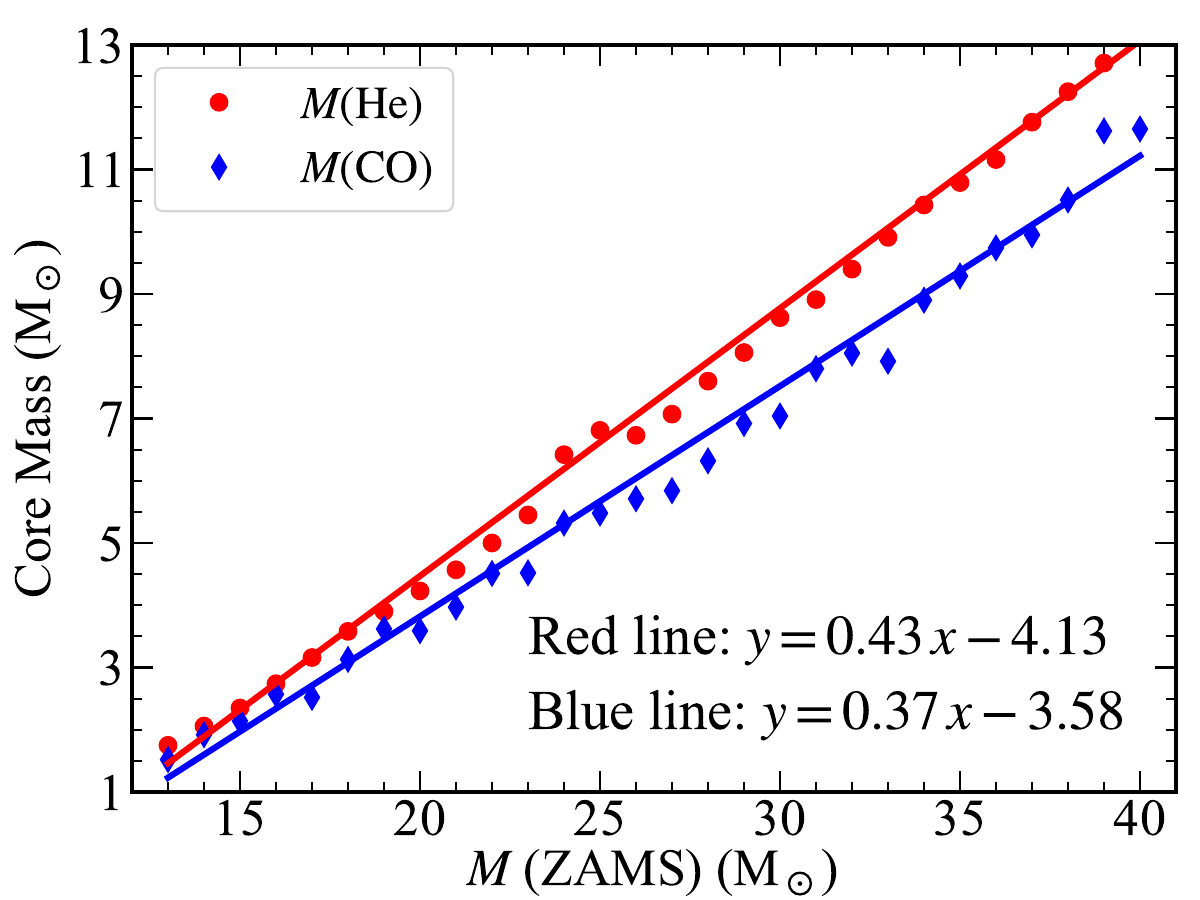}
\end{minipage}%
\caption{The core mass as a function of $M ({\rm ZAMS})$ for $\sigma_{C12\alpha}=$ 0.
The red points and blue diamonds represent $M$(He) and $M$(CO), respectively. The red
and blue straight lines are obtained by fitting the data.
\label{fig:he_zams}}
\end{figure}

\begin{figure}[htbp]
\centering
\begin{minipage}[c]{0.42\textwidth}
\includegraphics [width=75mm]{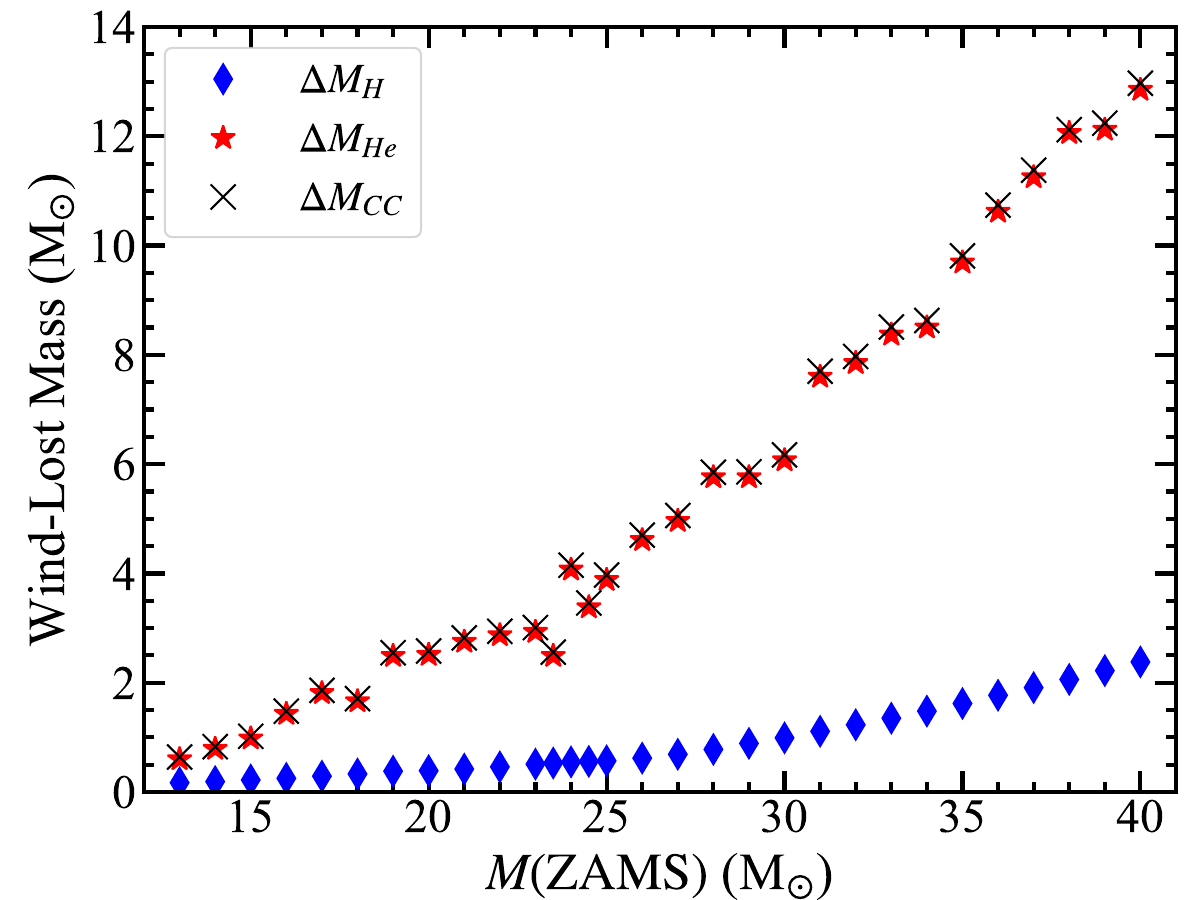}
\end{minipage}%

\caption{Lost masses by the stellar wind until the end of H burning (blue diamonds),
He burning (red stars), and Fe core-collapse (black crosses), respectively,
as a function of $M ({\rm ZAMS})$.
\label{fig:reduced}}
\end{figure}

In Figure~\ref{fig:he_zams}, we show $M$(He) (M$_\odot$) as a function
of $M ({\rm ZAMS})$ by the red points. We obtain the gray straight line of

\begin{equation}
M({\rm He}) = 0.43 M ({\rm ZAMS}) - 4.13\,{\rm M}_\odot  \label{he_zams}
\end{equation}

by fitting the red points.
(Of course, the $M$(He) - $M ({\rm ZAMS})$ relation does not depend on
$\sigma_{C12\alpha}$.)

Some deviations from the line may be due to different spatial
and temporal resolutions in some models, which would affect the
convective overshooting and thus the convective core mass.
Our $M$(He) - $M ({\rm ZAMS})$ relation is in good agreement with other works
\citep[e.g.,][]{2014Sukhbold}.

For $Z=$ 0.02, the wind mass loss is important during H and He burning.
The accumulated lost masses at the end of H burning $\Delta M_{\rm H}$, He burning
$\Delta M_{\rm He}$, and Fe core-collapse $\Delta M_{\rm CC}$, are summarized
in Table~\ref{tab:reduced} and Figure~\ref{fig:reduced}.

\begin{figure}
\centering
\begin{minipage}[c]{0.42\textwidth}
\includegraphics [width=75mm]{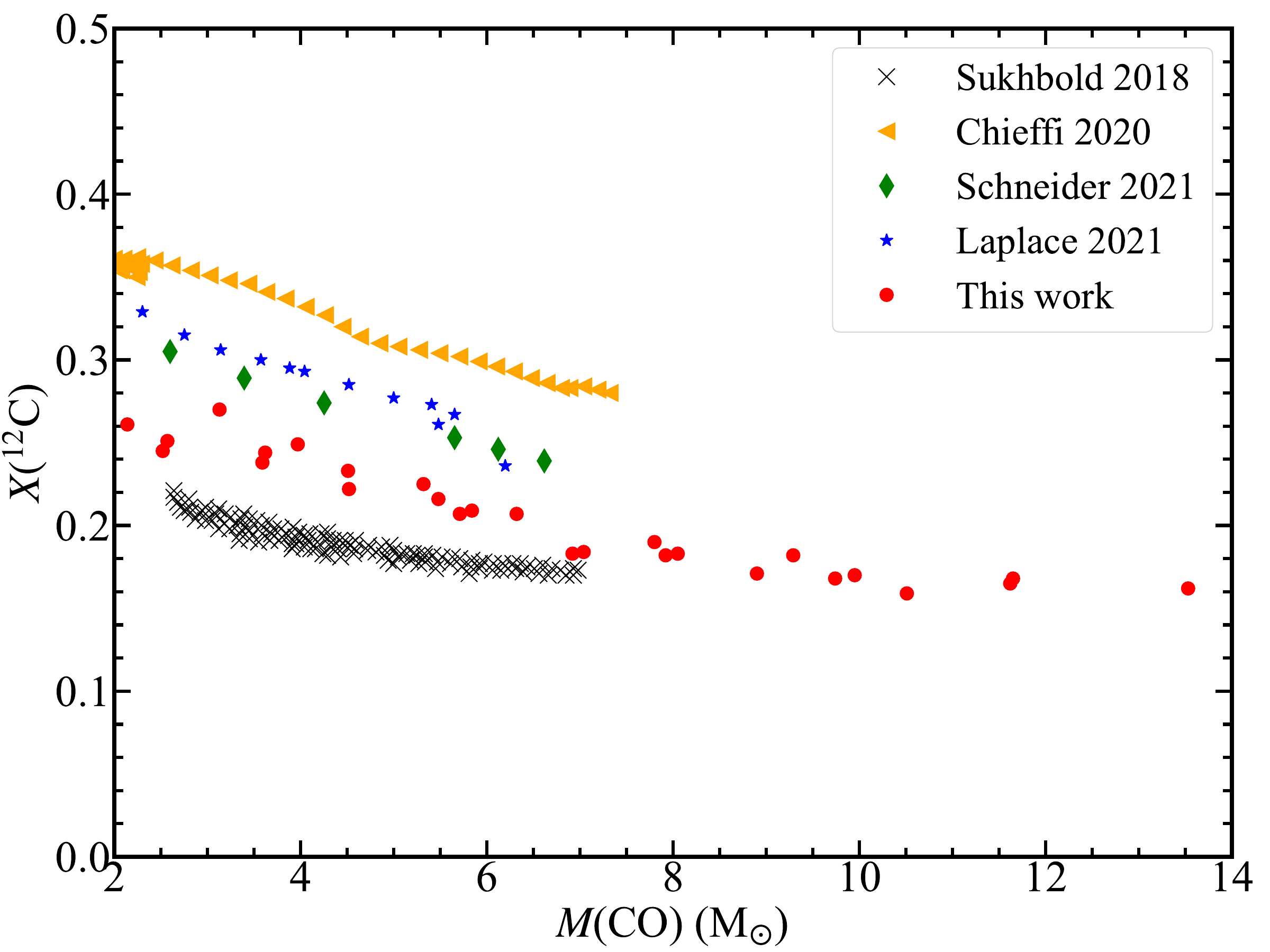}
\end{minipage}%
\caption{The mass fraction of $^{12}$C, $X$($^{12}$C), at the center as a
function of the CO core mass, $M$(CO), for $\sigma_{C12\alpha}=$ 0.
The red points indicate our results. The grey cross, green squares
blue stars and orange triangle are those from \citet{2018ApJ...860...93S},
\citet{Schneider_2020}, \citet{Laplace_2021} and \citet{2020ApJ...890...43C}.
\label{fig:xc}}
\end{figure}

\begin{figure} 
\centering
\begin{minipage}[c]{0.42\textwidth}
\includegraphics [width=75mm]{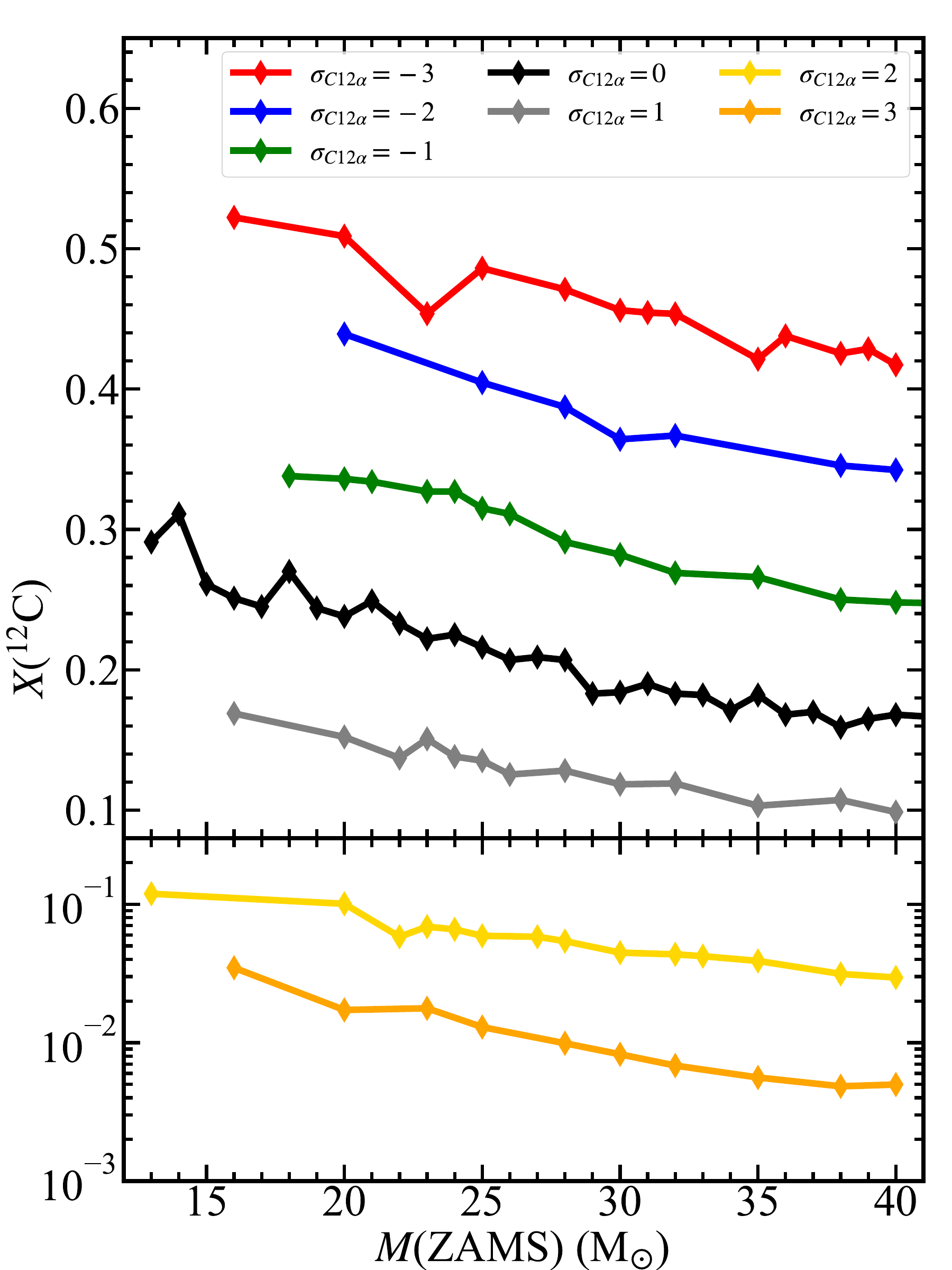}
\end{minipage}%
\caption{$X$($^{12}$C) at the center when the mass
fraction of He in the center becomes lower than
10$^{-4}$, as functions of $M ({\rm ZAMS})$ and $\sigma_{C12\alpha}$.
\label{fig:xc_2d}}
\end{figure}

\subsection{Helium Burning} \label{sec:hec_burn}

He burning synthesizes mainly $^{12}$C and $^{16}$O. As $X(^{4}$He) decreases,
the central temperature increases to be high enough
for the $^{12}$C$(\alpha, \gamma)^{16}$O reaction to convert $^{12}$C to $^{16}$O.

We define the CO core mass, $M$(CO), as $M_r$ at the CO core boundary where
$X(^{4}$He) inwardly decreases below $10^{-4}$.
We show $M$(CO) as a function of $M ({\rm ZAMS})$ by the blue points in
Figure~\ref{fig:he_zams}.
We obtain the green straight line of
\begin{equation}
M({\rm CO}) = 0.37 M ({\rm ZAMS}) - 3.58\,{\rm M}_\odot \label{co_zams}
\end{equation}
by fitting the blue points.

\begin{table*}[htbp]
\centering
\caption{Important evolutionary stages for $M({\rm ZAMS})=$ 28 M$_{\odot}$ and $\sigma_{C12\alpha} =$ 0 and $-$1.}
\label{tab:stages}
\begin{tabular}{cccccc}
\toprule
                               & \multicolumn{2}{c}{$\sigma_{C12\alpha} =$ 0} & \multicolumn{2}{c}{$\sigma_{C12\alpha} =-$ 1} &                      \\
Stages                         & log $\tau$ (yr)     & Convective Region ($M_r$)    &  log $\tau$ (yr)   & Convective Region ($M_r$)            & Comment              \\
\midrule
He EX                          & 4                      & $X$($^{12}$C)=0.176      & 4                    & $X$($^{12}$C)=0.264                 &                      \\
C CB                           & 3 - 1                      & radiative                & 3 - 1                    & radiative                           &                      \\
1st C SB                       & 1 - 0                    & radiative                & 2 - 1                & radiative                           & CO core contraction  \\
C EX                           & 0                        &                          & 0.3                    &                                     & NeO core rapid contraction  \\
2nd C SB                       & 0 - $-0.8$               & 1.7 - 4.0                & 1 - 0.1                & 1.2 - 3.2                           &                      \\
Ne CB                          & $-0.3$ - $-0.6$          & 0.0 - 0.1                & 0.2 - $-0.2$           & 0.0 - 0.1                           &                      \\
Ne SB                          & $-0.6$ - $-0.9$          & 0.1 - 1.0                & $-0.2$                 & 0.1 - 0.6                           &                      \\
O CB                           & $-0.9$ - $-2.6$          & 0.0 - 0.6                & $-0.2$ - $-1.8$        & 0.0 - 0.9                           &                      \\
3rd C SB                       & $-0.9$ - $-2.6$          & 2.2 - 6.5                & 0.1 - $-5$             &  2.0 - 7.0                          & Most important difference   \\
1st O SB                       & $-2.6$ - $-3.4$          & 0.9 - 1.5                & $-1.8$ - $-2.8$        &  0.9 - 1.5                          &                       \\
2nd O SB                       & $-3.6$ - $-4.5$          & 1.5 - 2.2                & $-2.8$ - $-3.0$        &  1.2 - 1.8                          &                      \\ 
3rd O SB                       & $-4.5$ - $-5$            & 1.7 - 2.2                & $-3.5$ - $-5.0$        &  1.2 - 2.0                          & Bottom of O SB gives $M$(Si) \\
Si CB                          & $-3.4$ - $-3.6$          & 0.0 - 0.4                & $-2.9$ - $-3.5$        &  0.0 - 1.0                          &                      \\
1st Si SB                      & $-3.6$ - $-3.8$          & 0.4 - 0.8                & $-3.7$ - $-4.0$        &  0.8 - 1.2                          &                      \\     
2nd Si SB                      & $-4.1$ - $-5.0$          & 0.9 - 1.5                & $-4.6$ - $-5.0$        &  1.2 - 1.5                          & Bottom of Si SB gives $M$(Fe) \\  
\bottomrule
\end{tabular}
\begin{tablenotes}
\footnotesize
\item CB, SB, and EX denote central burning, shell burning, and exhaustion stages, respectively.
\end{tablenotes}
\end{table*}

\begin{figure*}[htbp]
\centering
\begin{minipage}[c]{0.48\textwidth}
\centerline{$M({\rm ZAMS})=28 $ M$_\odot$, $\sigma_{C12\alpha}=0$}
\includegraphics [width=85mm]{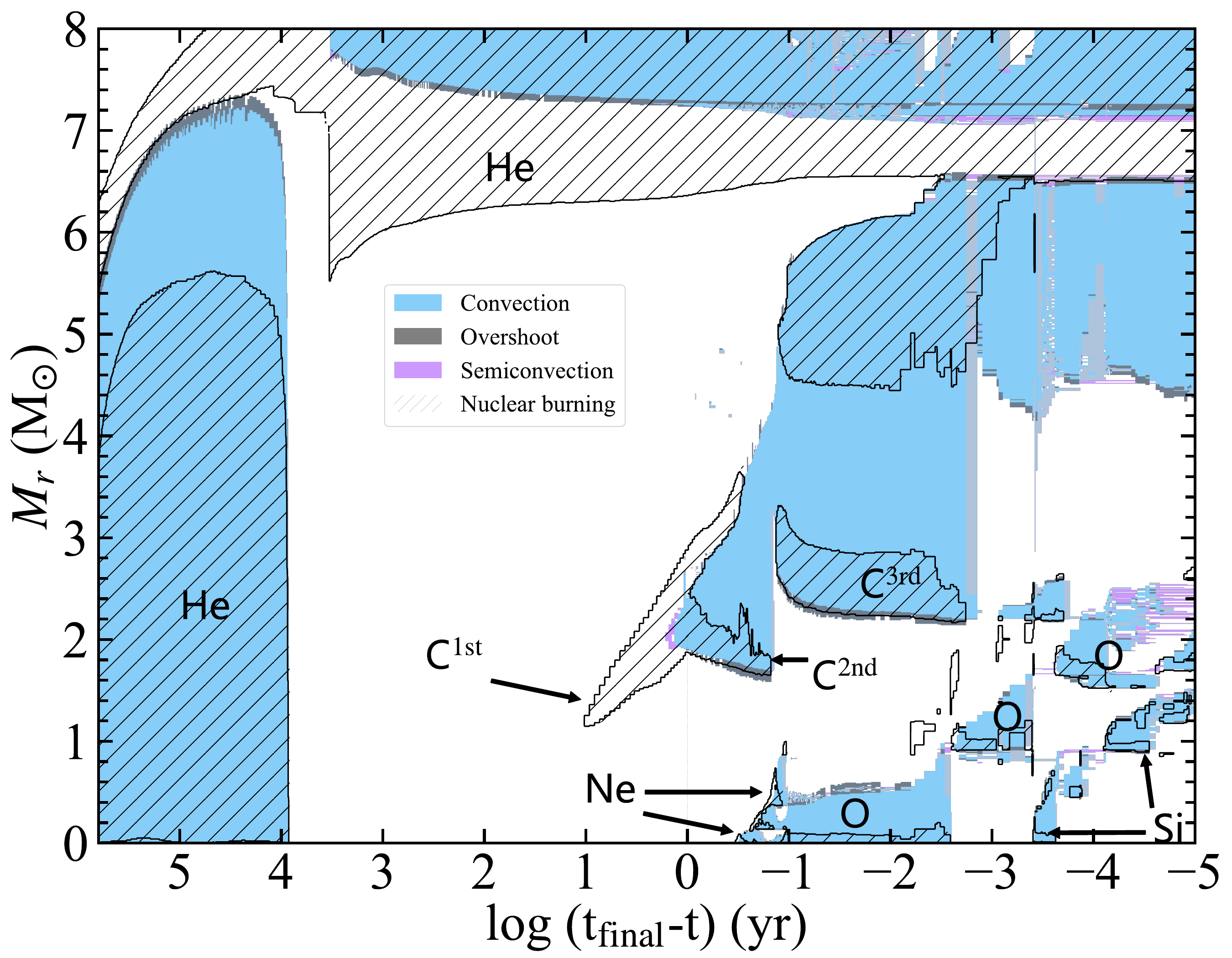}
\end{minipage}%
\begin{minipage}[c]{0.48\textwidth}
\centerline{$M({\rm ZAMS})=28 $ M$_\odot$, $\sigma_{C12\alpha}=-1$}
\includegraphics [width=85mm]{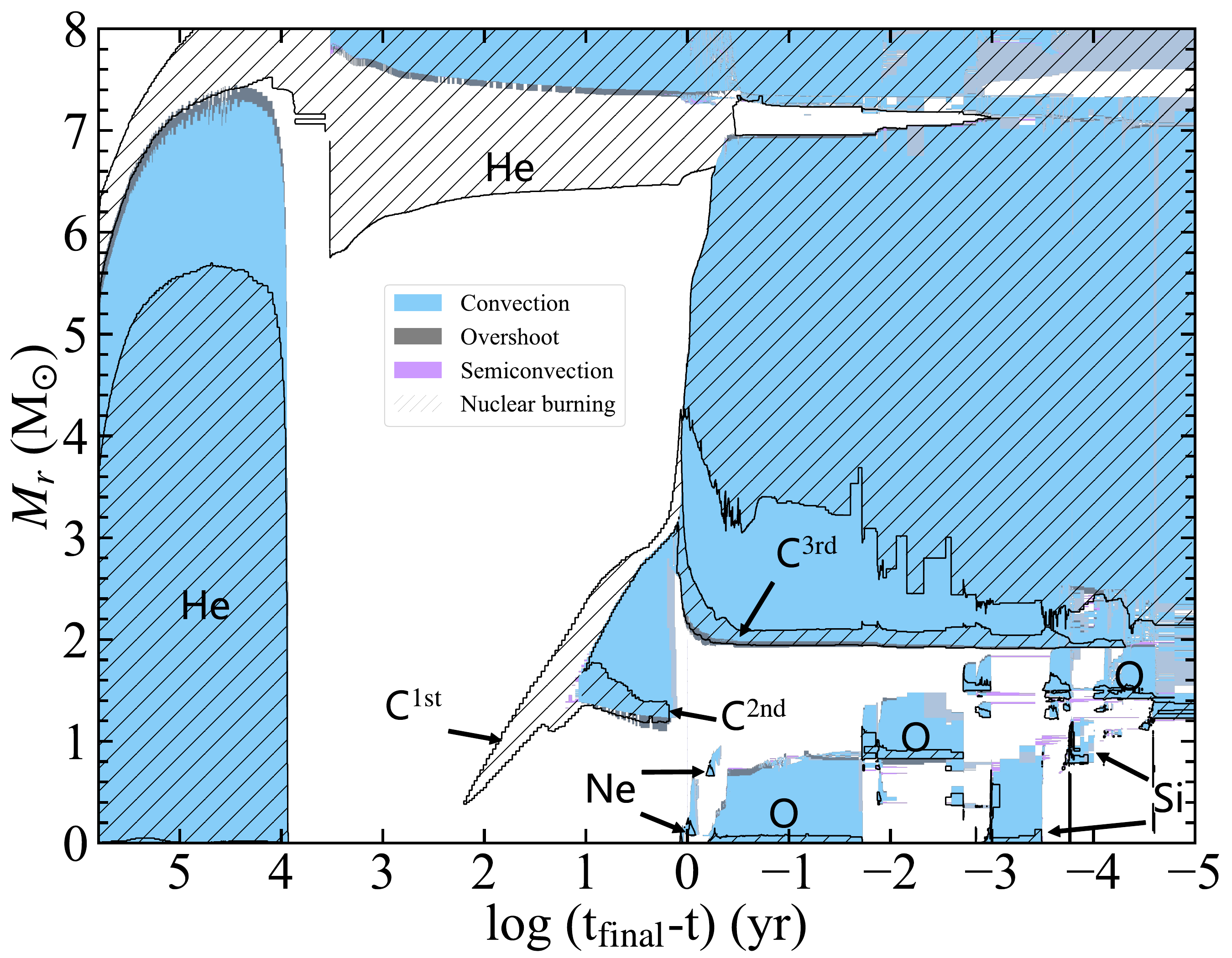}
\end{minipage}%
\caption{Kippenhahn diagrams of stars with $M {\rm (ZAMS)}$ = 28 M$_{\odot}$ for
$\sigma_{C12\alpha}=$ 0 (left) and $-1$ (right).
The inner part of $M_r = 0 - 8$ M$_{\odot}$ is shown.
The blue, grey and pink regions show where the convection, overshoot and
semiconvection mixing occurs. The black-hatched region indicates where
the nuclear burning occurs.
\label{fig:28M_core}}
\end{figure*}

$X$($^{12}$C) after He burning mainly
depends on $\sigma_{C12\alpha}$ and is an important quantity for later
stellar evolution, as will be discussed in Section~\ref{sec:carbon}.
We thus compare $X$($^{12}$C) in our work with others.
As shown in Figure~\ref{fig:xc}, $X$($^{12}$C) in our models behaves similarly to
other MESA works from \citet{Laplace_2021} and \citet{Schneider_2020}.
On the other hand, our result is located between those calculated by the KEPLER
code \citep{2018ApJ...860...93S} and the FRANEC code \citep{2020ApJ...890...43C}.
These differences stem not only from different sources of the 
$^{12}$C$(\alpha,\gamma)^{16}$O rate but also depend on other effects
such as convective overshooting during the late phase of core He burning,
3D convective mixing in shell burning, as well as the binary stripping effect
which is discussed in \citep{Schneider_2020}.

Figure \ref{fig:xc_2d} shows $X$($^{12}$C) at the center when $X$($^4$He) becomes lower than
10$^{-4}$ as functions of $M ({\rm ZAMS})$ and $\sigma_{C12\alpha}$.
We note that the effect of $\sigma_{C12\alpha}$ on $X$($^{12}$C) is much larger than that of $M ({\rm ZAMS})$.
For example, $X$($^{12}$C) decreases from 0.51 to 0.02 as
$\sigma_{C12\alpha}$ increases from $-3$ to 3 for $M ({\rm ZAMS}) =$ 20 M$_{\odot}$.

\subsection{Carbon Ignition} \label{sec:c_ign}

After the core He burning, the CO core contracts.
Figure~\ref{fig:trho} shows that the evolution of the central
density and temperature reaches the dashed line of the ignition of C
burning ($^{12}$C+$^{12}$C), where the nuclear energy generation rate
($\epsilon_{\rm C}$) is equal to the energy loss rate by neutrino
emissions ($\epsilon_\nu$).  
However, that the C ignition line is obtained for $X$($^{12}$C)
= 0.5.  With smaller $X$($^{12}$C), C burning is less energetic because
$\epsilon_{\rm C}$ depends on $X$($^{12}$C)$^2$.  If $X$($^{12}$C) is
smaller than a certain value, $\epsilon_{\rm C} < \epsilon_\nu$.  Then, C
burning does not form a convective core and proceeds radiatively.

\begin{table}[htb]
\caption{The critical masses for different $\sigma_{C12\alpha}$s.}
\label{tab:mcrit}
\begin{tabular}{cccccc}
\toprule
$\sigma_{C12\alpha}$ & -3 & -2 & -1 & 0  & 1  \\
\midrule
$M_{\rm crit}$  & 26 & 24 & 22 & 20 & 16 \\
\bottomrule
\end{tabular}
\end{table}

Figure \ref{fig:xc_2d} shows that $X$($^{12}$C) is smaller
for larger $M ({\rm ZAMS})$ and larger $^{12}$C$(\alpha,\gamma)^{16}$O rate.
Thus C burning proceeds radiatively for stars with $M({\rm ZAMS}) > M_{\rm crit} \sim$ 
20 M$_\odot$ in our models \citep[also][]{2020ApJ...890...43C, 2021ApJ...916...79C}
as seen in Table~\ref{tab:mcrit} and will be discussed in detail in our forthcoming paper.
For the 28 M$_\odot$ stars, no convective core of C burning is formed
for both $\sigma_{C12\alpha} = 0$ and $-1$ as seen in Figure \ref{fig:28M_core}.

\section{Carbon Burning to Fe core-collapse} \label{sec:carbon}

In the following subsections, we study how the
$^{12}$C$(\alpha,\gamma)^{16}$O rate affects the presupernova core structure
and the explodability
(e.g., \citealp{1993PhR...227...65W,1996ApJ...457..834T})
by comparing the evolutionary features for several cases.
We particularly focus on the stars with $M ({\rm ZAMS})$ = 28 
M$_\odot$, 25 M$_\odot$, and 35 M$_\odot$
because of the following reasons.  Hereafter, we call these stars as
the ``28 M$_\odot$ star'', ``25 M$_\odot$ star'',
and ``35 M$_\odot$ star'', respectively.

\noindent
(1) The evolution of the 28 M$_\odot$ star is a typical case in which
shows a large difference between $\sigma_{C12\alpha}=$ 0 and $-1$.

\noindent
(2) The evolution of the 25 M$_\odot$ star shows a rather small
difference between $\sigma_{C12\alpha}=$ 0 and $-1$, but there appears
an important difference near the final stages of evolution.
A significant difference occurs at earlier stages for $\sigma_{C12\alpha}=-2$.

\noindent
(3) The evolution of the 35 M$_{\odot}$ star is a typical case where a
significant difference appears for $\sigma_{C12\alpha}=$ 1.

\noindent
We discuss the 28 M$_\odot$ models in detail, and the 25 M$_\odot$ and
35 M$_\odot$ models in short for comparisons.

\begin{figure*}[htbp]
\centering
\begin{minipage}[c]{0.85\textwidth}
\centerline{$M({\rm ZAMS})=28 $ M$_\odot$}
\includegraphics [width=150mm]{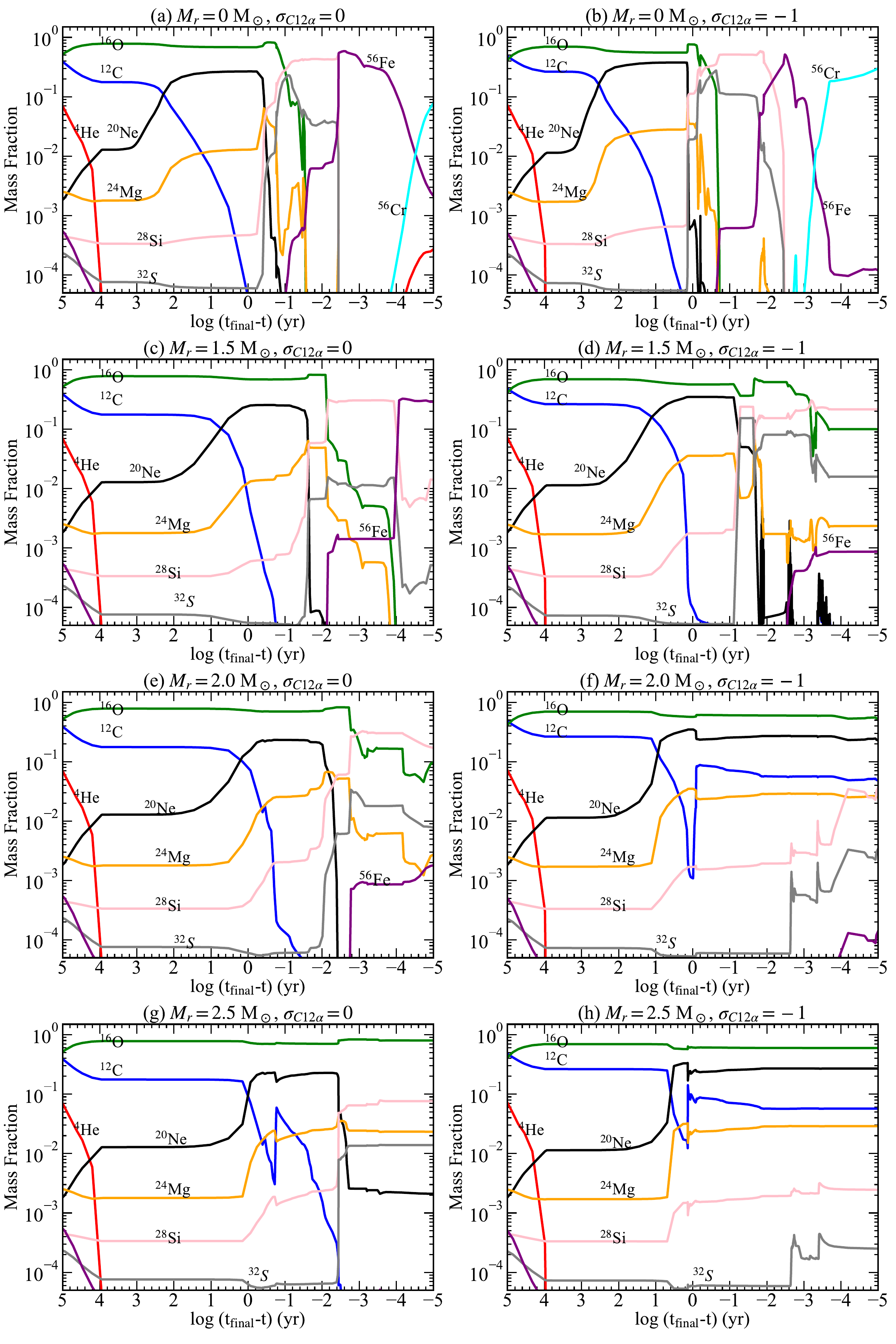}
\end{minipage}%
\caption{The time evolution of the mass fractions of some isotopes
at the center, $M_r=$1.5 M$_\odot$, $M_r=$2.0 M$_\odot$ and $M_r=$2.5
M$_\odot$ for the $M ({\rm ZAMS})$ = 28 M$_{\odot}$ star with
$\sigma_{C12\alpha}=$ 0 (left) and -1 (right).
\label{fig:28M_central_Xcore}}
\end{figure*}

\subsection{Chemical Evolution of $28$ M$_\odot$ Star} \label{sec:28M}

After He exhaustion, the star evolves through the gravitational
contraction of the core and nuclear burning at the center and outer
shells until the formation of the Fe core, as shown in Figure
\ref{fig:trho}.

\begin{figure*}[htbp]
\centering
\begin{minipage}[c]{0.9\textwidth}
\centerline{$M({\rm ZAMS})=28 $ M$_\odot$}
\includegraphics [width=160mm]{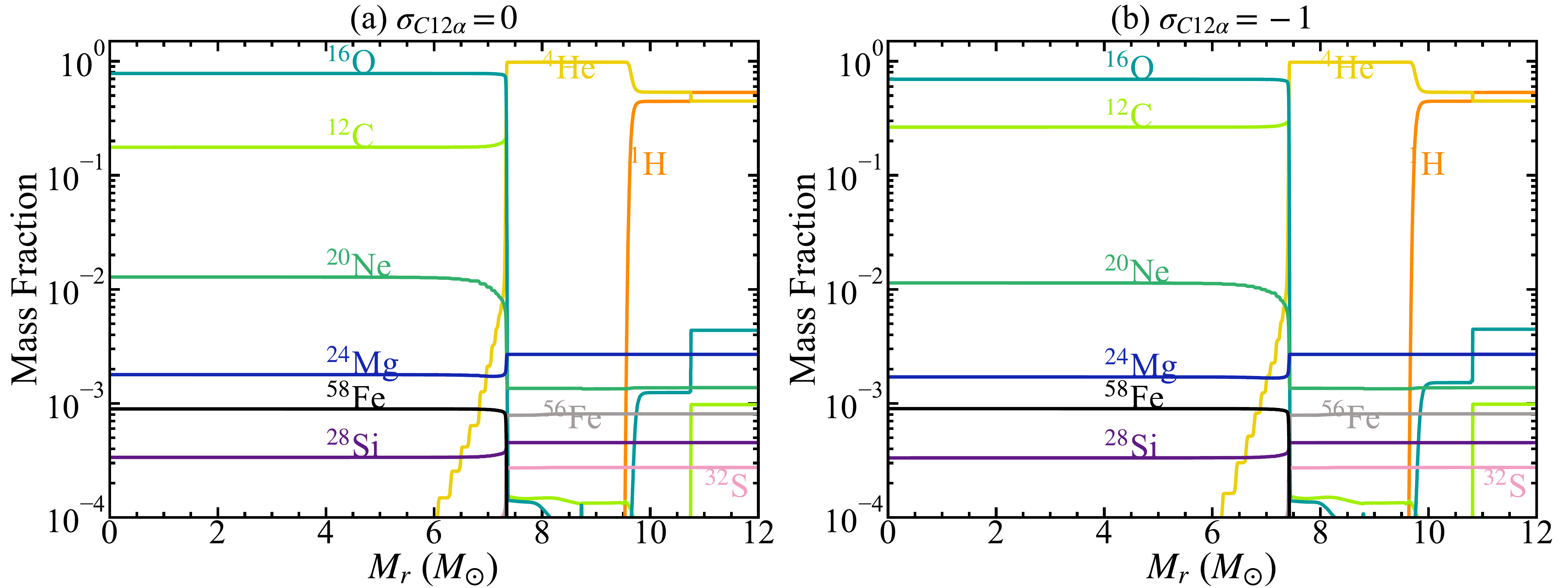}
\end{minipage}%
\caption{Abundance distributions of stars with $M({\rm ZAMS}) = $
28 M$_{\odot}$ for $\sigma_{C12\alpha}=$ 0 (left) and $-$1 (right) at
the end of He burning ($\tau \sim 10^4$ yr).
\label{fig:28M_abund1}}
\end{figure*}

\begin{figure*}[htbp]
\centering
\begin{minipage}[c]{0.9\textwidth}
\centerline{$M({\rm ZAMS})=28 $ M$_\odot$}
\includegraphics [width=160mm]{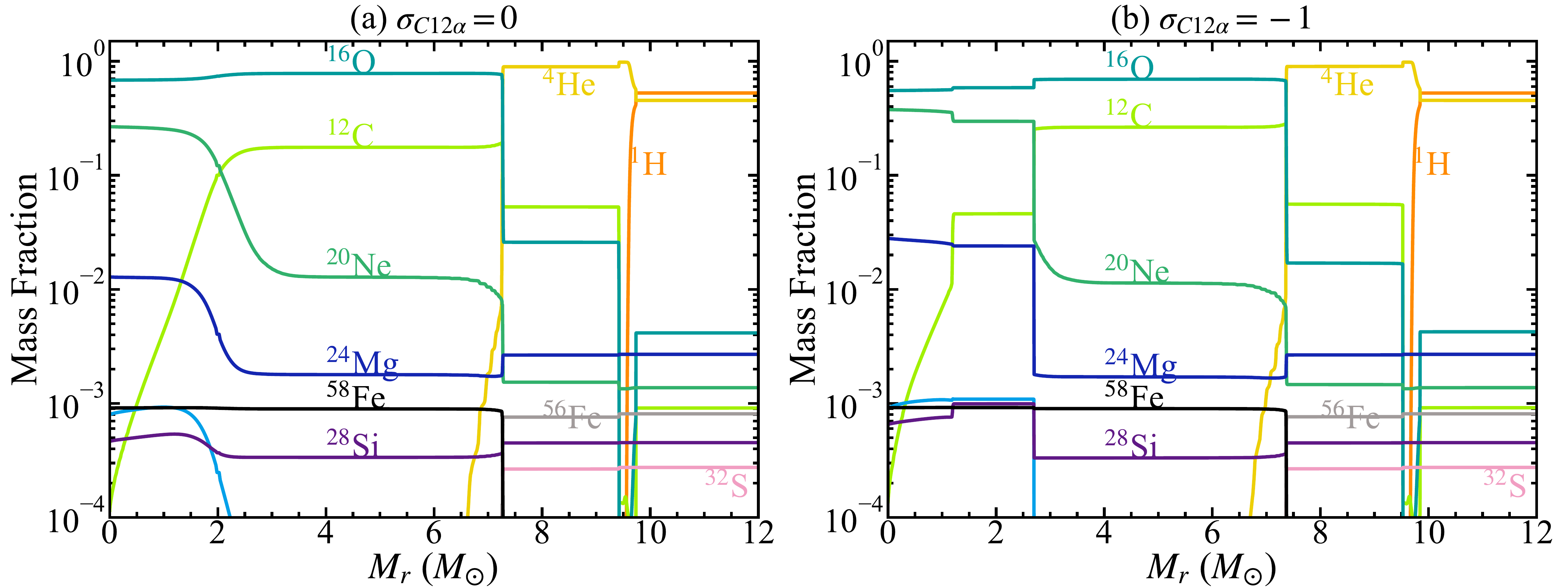}
\end{minipage}%
\caption{Abundance distributions of stars with $M({\rm ZAMS}) = $
28 M$_{\odot}$ for $\sigma_{C12\alpha}=$ 0 (left) and $-$1 (right) at
the end of C burning ($\tau \sim 10^{0.3-0}$ yr).
\label{fig:28M_abund2}}
\end{figure*}

\begin{figure*}[htbp]
\centering
\begin{minipage}[c]{0.9\textwidth}
\centerline{$M({\rm ZAMS})=28 $ M$_\odot$}
\includegraphics [width=160mm]{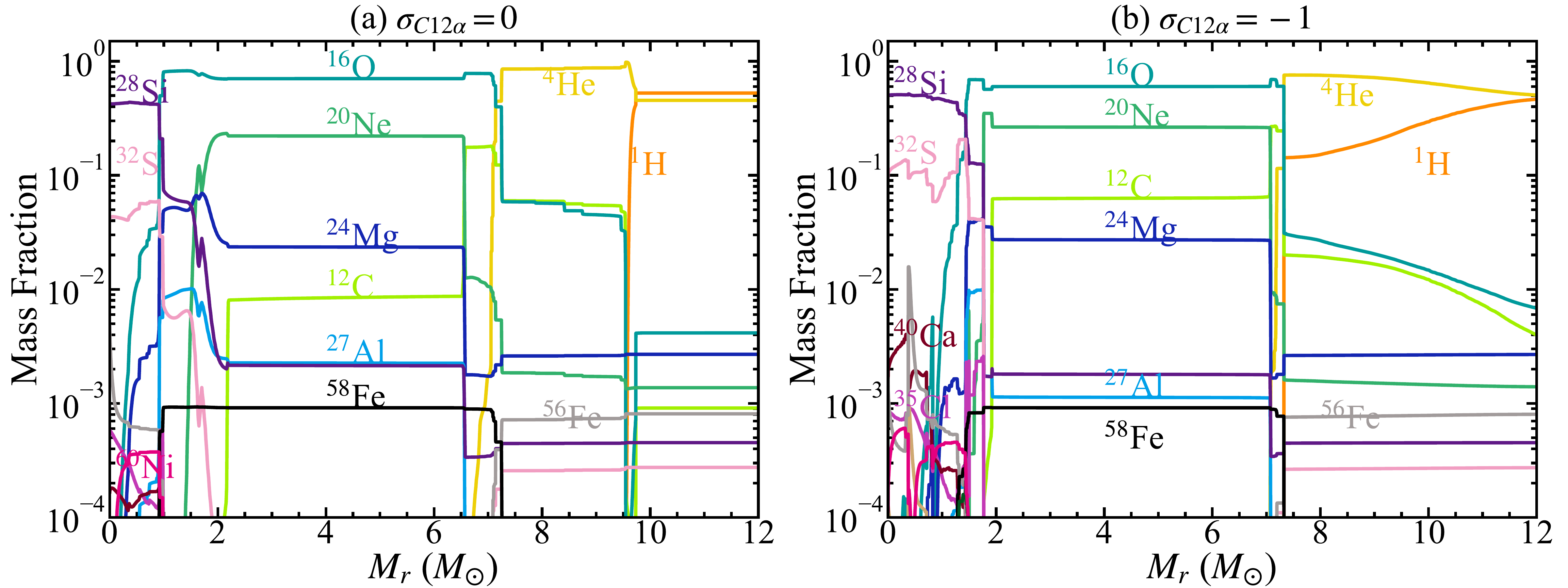}
\end{minipage}%
\caption{Abundance distributions of stars with $M({\rm ZAMS}) =$
28 M$_{\odot}$ at the end of O burning: $\tau \sim 10^{-1.6}$
yr for $\sigma_{C12\alpha}=0$ (left) and $\tau \sim 10^{-0.7}$
yr for $\sigma_{C12\alpha}=-1$ (right).
\label{fig:28M_abund3}}
\end{figure*}

\begin{figure*}[htbp]
\centering
\begin{minipage}[c]{0.9\textwidth}
\centerline{$M({\rm ZAMS})=28 $ M$_\odot$}
\includegraphics [width=160mm]{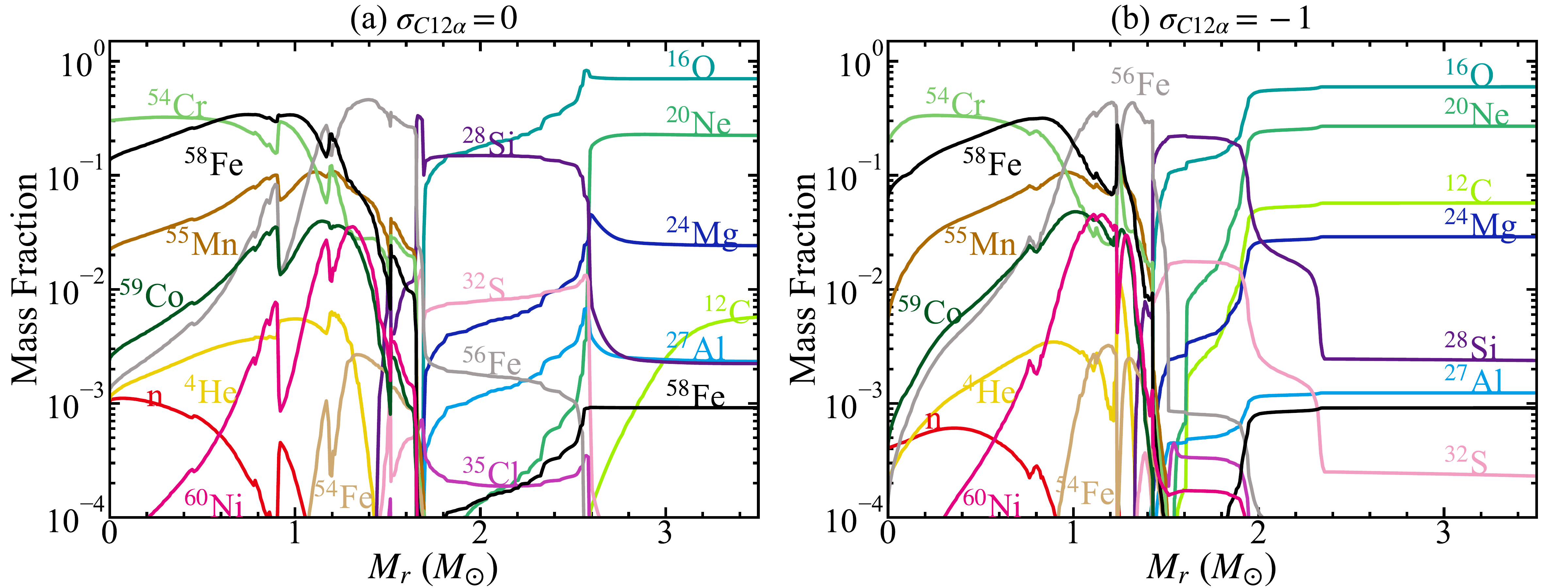}
\end{minipage}%

\caption{Abundance distributions in the inner 3.5 M$_\odot$ of stars
with $M({\rm ZAMS}) = $ 28 M$_{\odot}$ for $\sigma_{C12\alpha}=$ 0
(left) and $-$1 (right) at log $T_{\rm c}$ (K) $=$ 10.0 ($\tau=t_{\rm f}$).
\label{fig:28M_abund4}}
\end{figure*}

\begin{figure*}[htbp]
\centering
\begin{minipage}[c]{0.9\textwidth}
\centerline{$M({\rm ZAMS})=28 $ M$_\odot$}
\includegraphics [width=160mm]{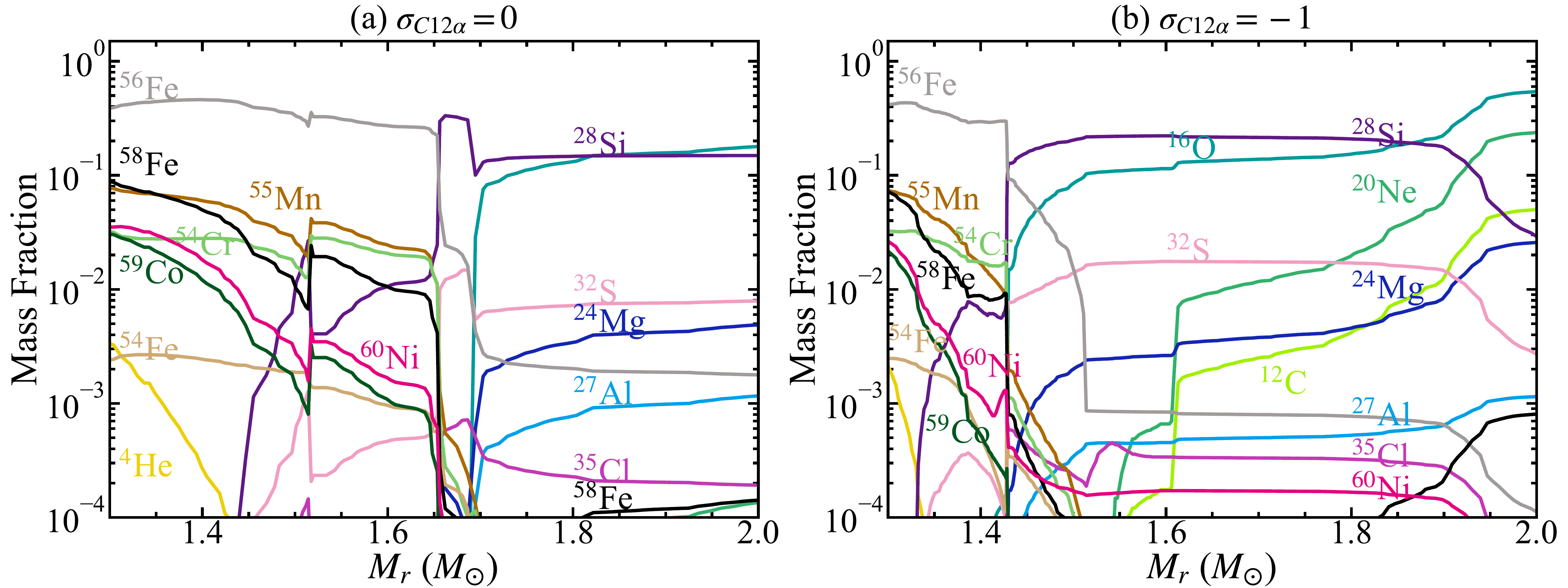}
\end{minipage}%

\caption{Detailed abundance distributions between $M_r$=1.2 - 2.0 M$_\odot$ of
stars with $M({\rm ZAMS}) = $ 28 M$_{\odot}$ for
$\sigma_{C12\alpha}=$ 0 (left) and $-$1 (right) at log $T_{\rm c}$ (K) = 10.0
($\tau=t_{\rm f}$).
\label{fig:28M_abund5}}
\end{figure*}

These evolutionary stages are summarized in Table \ref{tab:stages} for
the 28 M$_{\odot}$ star, where CB, SB, and EX denote central
burning, shell burning, and exhaustion stages, respectively.
Here the time
\begin{equation}
\tau = t_{\rm f} -t  \label{tau}
\end{equation}
is measured from the final stage at
$t_{\rm f}$ where the central density reaches log $T_{\rm c}$ (K) = 10.0.
The evolutionary behavior and the time depend on
the $^{12}$C$(\alpha,\gamma)^{16}$O rate as seen from the comparison
between $\sigma_{C12\alpha} =$ 0 and $-$1 in Table \ref{tab:stages}.

The following figures show the evolutionary changes in various stellar
quantities in the 28 M$_{\odot}$ star from He burning to the beginning
of Fe core-collapse.  The dependence on the
$^{12}$C$(\alpha,\gamma)^{16}$O rate is seen from the comparison
between $\sigma_{C12\alpha} =$ 0 and $-$1 in these figures.

Figures~\ref{fig:28M_core} show Kippenhahn Diagrams.
The inner part of $M_r = 0 - 8$ M$_{\odot}$ is shown.
Nuclear burning (at the center and outer shells) produces a convective
region as indicated by the blue region.
The hatched region shows where the energy generation rate of nuclear
reactions are larger than the neutrino loss rate.  
We should see that several convective shells are formed, which largely 
affect the stellar structure and evolution, 
as will be discussed in later sections.

Figure~\ref{fig:28M_central_Xcore} shows the time evolution of the
mass fractions of some isotopes at the center
(Figure~\ref{fig:28M_central_Xcore} (a, b)), $M_r=1.5$
M$_\odot$ (Figure~\ref{fig:28M_central_Xcore} (c, d)),
$M_r=2.0$ M$_\odot$ (Figure~\ref{fig:28M_central_Xcore} (e, f)), 
and $M_r=2.5$ M$_\odot$ (Figure~\ref{fig:28M_central_Xcore} (g, h)) for the
28 M$_{\odot}$ star from He burning through Fe core-collapse for
$\sigma_{C12\alpha}=$ 0 (Figure~\ref{fig:28M_central_Xcore} (a, c, e, g)),
and $\sigma_{C12\alpha}= -1$ (Figure~\ref{fig:28M_central_Xcore} (b, d, f, h)).
In the following subsections, the chemical evolution of the central core is
described as shown in the abundance evolution at the center
(Figure~\ref{fig:28M_central_Xcore} (a, b)).
Then, the evolution of the outer layers and the burning of carbon and
oxygen shells will be discussed.

\subsubsection {Central Carbon Burning}

The abundance distributions before C burning are shown in Figure
\ref{fig:28M_abund1}.
Note $X$($^{12}$C) in the CO core is smaller for $\sigma_{C12\alpha} = 0$ (left) than
that for $\sigma_{C12\alpha} = -1$ (right).

As shown by decreasing $X$($^{12}$C) at the center in Figure~\ref{fig:28M_central_Xcore},
significant C burning takes place at $\tau \sim 10^{2.6}$ yr.
As seen from Kippenhahan diagrams (Figs.~\ref{fig:28M_core}),
no convective core appears, i.e., C burns radiatively because
the neutrino energy loss rate exceeds the energy generation rate for 
$M({\rm ZAMS}) > M_{\rm crit} \sim$ 20 M$_\odot$
(see Table~\ref{tab:mcrit} and subsection \ref{sec:c_ign}).
As a result, the CO core continues to contract as seen in the smooth increase in
$T_c$ and $\rho_c$ in Figure~\ref{fig:trho}.
There is no large difference in the central evolution between
$\sigma_{C12\alpha} =$ 0 and $-1$ during central C burning.

\subsubsection {ONeMg Core Contraction and Neon Burning}

After C-exhaustion in the central region, an ONeMg core composed of O,
Ne, Mg, and Na are formed.  The abundance distributions in the central
ONeMg core around $\tau \sim 10^{0.3 - 0}$ yr are shown in
Figures~\ref{fig:28M_abund2}.  

In the contracting ONeMg core, Ne is ignited at the center and burns
convectively around $\tau \sim 10^{-0.3}$ yr and $10^{0.2}$ yr for
$\sigma_{C12\alpha} =$ 0 and -1, respectively
(Figs.~\ref{fig:28M_core} and \ref{fig:28M_central_Xcore}(a, b)).

\subsubsection {Oxygen Burning}

After Ne exhaustion, an OSi core composed of O, Mg, Si, and S is formed.
O burning takes place around $\tau \sim 10^{-0.2} - 10^{-1.0}$ yr
(Figs.~\ref{fig:28M_central_Xcore}(a, b)), forming a convective core 
for both $\sigma_{C12\alpha} = 0$ and $-1$.  

Then, a Si-rich core composed of Si, S, Ar, and Ca is formed.
Abundance distributions with the Si-rich core are shown in
Figure~\ref{fig:28M_abund3} at $\tau \sim 10^{-1.6}$ yr for
$\sigma_{C12\alpha}=0$ (left) and $\tau \sim 10^{-0.7}$ yr for
$\sigma_{C12\alpha}=-1$ (right).

\subsubsection {Silicon Burning and Fe Core Contraction} \label{sec:28M_final}

Si burning takes place convectively for both $\sigma_{C12\alpha} = 0$
(Figure \ref{fig:28M_central_Xcore}(a, c)) and $\sigma_{C12\alpha} = -1$
(Figure \ref{fig:28M_central_Xcore}(b)), and a Fe core is formed.  The
Fe core contraction is accelerated by photo-disintegration of Fe-peak
species and becomes dynamically unstable.  We stop our calculations at
log $T_{\rm c}$ (K) = 10.0 ($t=t_{\rm f}$) because the reaction rates
provided by the database (REACLIB) of MESA reach only log $T$(K) = 10.0.

The abundance distributions at the final stage of $t=t_{\rm f}$
are shown in Figure~\ref{fig:28M_abund4}, and the
detailed distributions in the Fe core are
shown in Figure~\ref{fig:28M_abund5} for both $\sigma_{C12\alpha}$.
(The $Y_e$ distribution will be shown later in Figure~\ref{fig:28M_logp_rho}.)

The Fe core masses $M$(Fe) at the boundary with $X$(Si) $= 10^{-3}$ are
1.51 M$_\odot$ and 1.36 M$_\odot$ for $\sigma_{C12\alpha} = 0$
and $\sigma_{C12\alpha} = -1$, respectively.  

\begin{figure*}[htbp]
\centering
\begin{minipage}[c]{0.75\textwidth}
\centerline{$M({\rm ZAMS})=28 $ M$_\odot$}
\includegraphics [width=132mm]{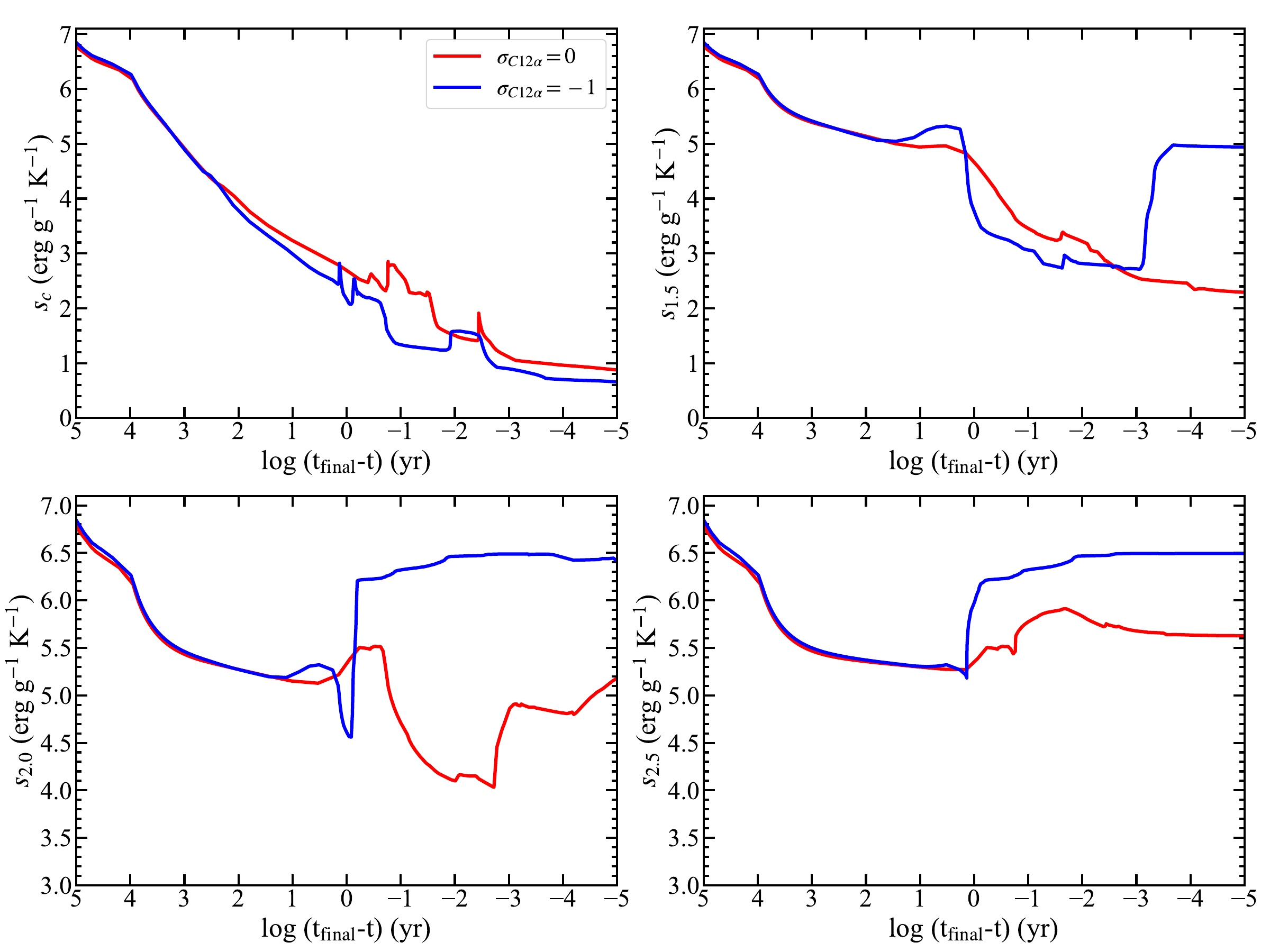}
\end{minipage}%
\caption{The time evolution of specific entropy in units of erg g$^{-1}$K$^{-1}$
 at the center ($s_{\rm c}$), $M_r=$ 1.5
M$_\odot$ ($s_{1.5}$), $M_r=$ 2.0 M$_\odot$ ($s_{2.0}$) and $M_r=$ 2.5
M$_\odot$ ($s_{2.5}$) of $M {\rm (ZAMS)}$ = 28 M$_{\odot}$ star with
$\sigma_{C12\alpha}=$ 0 and $-$1.
\label{fig:28M_entr}}
\end{figure*}

\begin{figure*}[htbp]
\centering
\begin{minipage}[c]{0.75\textwidth}
\centerline{$M({\rm ZAMS})=28 $ M$_\odot$}
\includegraphics [width=132mm]{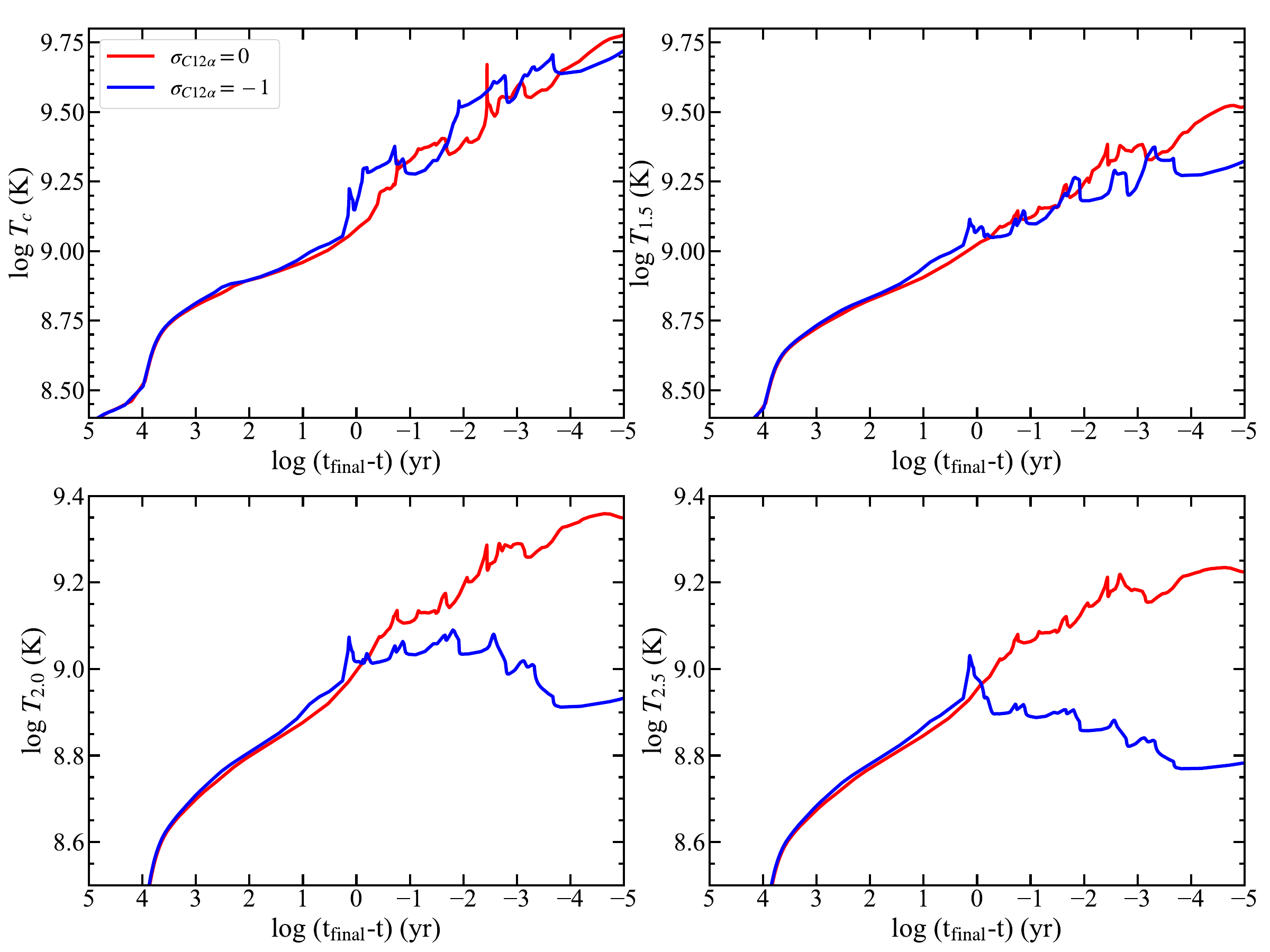}
\end{minipage}%
\caption{The time evolution of temperature at the center ($T_{\rm c}$), $M_r=$ 1.5
M$_\odot$ ($T_{1.5}$), $M_r=$ 2.0 M$_\odot$ ($T_{2.0}$) and $M_r=$ 2.5
M$_\odot$ ($T_{2.5}$) of $M {\rm (ZAMS)}$ = 28 M$_{\odot}$ star with
$\sigma_{C12\alpha}=$ 0 and $-$1.
\label{fig:28M_temp}}
\end{figure*}

\subsubsection {Carbon Shell Burning} \label{sec:c_shell}

The difference in the evolution between $\sigma_{C12\alpha} = 0$ and
$-1$ more clearly appears in C shell burning (C SB in Table
\ref{tab:stages}) compared with central core burning, as seen in
Kippenhahan diagrams (Figs.~\ref{fig:28M_core}) and summarized in
Table \ref{tab:stages}.  In this subsection, we describe the differences
which can be seen mostly in the abundance patterns in
Figures~\ref{fig:28M_central_Xcore}.  Differences in physical
quantities will be described in the next subsection.

(1) First C Shell Burning (1st C SB) in the CO Core: 
As discussed in the earlier subsection, the CO core contracts
even during central C burning.  As a result of the
temperature increase in the outer shell, the energy generation rate of
C burning exceeds the neutrino loss rate ($\epsilon >$ 0) around 
$\tau \sim 10^{2.2}$ yr and $\sim - 10^1$ yr for $\sigma_{C12\alpha} =-1$
and 0, respectively (Table \ref{tab:stages}); Figure \ref{fig:28M_core}).
The first C shell burning takes place radiatively.

(2) Second C Shell Burning (2nd C SB) in the ONeMg Core: 
As the ONeMg core contracts around $\tau \sim 10^{1.5 - 0.0}$ yr,
the second C shell burning is ignited and produces a convective shell for
both $\sigma_{C12\alpha}$.  This is seen from the {\sl decrease} in
$X$($^{12}$C) at $M_r =$ 1.5 and 2.0 M$_\odot$ in
Figures~\ref{fig:28M_central_Xcore}(c)-(f).

(3) The third C Shell Burning (3rd C SB): 
The 3rd C shell burning behavior depends clearly on $\sigma_{C12\alpha}$.

(3-1) For $\sigma_{C12\alpha} = -1$, C shell-burning forms a
convective shell at $\tau \sim 10^0$ yr. It extends from $M_r \sim 2.0
$ M$_\odot$ to 6.9 M$_\odot$ and mixes some C from the outer layer
into the convective shell as seen from the {\sl increase} in
$X$($^{12}$C) at $M_r = 2.0$ M$_\odot$ and 2.5 M$_\odot$ in
Figure~\ref{fig:28M_central_Xcore}(f) and (h), respectively.  Owing to
this enhancement of $X$($^{12}$C), C shell burning is reactivated and
continues to be active through collapse with the convective shell
extended from $M_r =$ 2.0 M$_\odot$ (Figure \ref{fig:28M_core}).

(3-2) For $\sigma_{C12\alpha} = 0$, on the other hand, such convective
mixing does not occur at $\tau \sim 10^0$ yr because of smaller
$X$($^{12}$C) during earlier C shell burning, Thus no increase in
$X$($^{12}$C) is seen at $M_r = 2.0$ M$_\odot$ in
Figure~\ref{fig:28M_central_Xcore}(e).  Then C is exhausted at $\tau
\sim 10^{-1.5}$ yr.

Such a difference in the convective C shell between
$\sigma_{C12\alpha} = 0$ and $-1$ stems from the difference in the C/O
ratio after the He burning, and leads to an important difference in the
later evolution.

As will be discussed in subsection \ref{sec:uv_curve} on $U-V$ curves, 
the existence of active C-shell burning at $M_r \sim 2.0$
M$_\odot$ makes $M_{\rm eff}$ (Equation \ref{equ:meff})
smaller for $\sigma_{C12\alpha} = -1$ than 0.
This makes the mass of the heavier element core larger for
$\sigma_{C12\alpha} =0$ than $-1$ as will be shown in
\S~\ref{sec:uv_curve}.

In fact, for $\sigma_{C12\alpha} = 0$, the layer around $M_r=$ 2.0
M$_\odot$ becomes a part of the Si-rich core after $\tau= 10^{-2.8}$
yr (Figure \ref{fig:28M_central_Xcore}(e)).  For $\sigma_{C12\alpha}=-1$, 
on the other hand, those layers are still O-rich with some C
(Figure \ref{fig:28M_central_Xcore}(f)).

Similar differences are seen around $M_r=$ 1.5 M$_\odot$. For
$\sigma_{C12\alpha} = 0$, O burns out after $\tau= 10^{-2.0}$ yr and
the layer becomes part of a Fe core after $\tau= 10^{-4.0}$ yr
(Figure \ref{fig:28M_central_Xcore}(c)).  For $\sigma_{C12\alpha} = -1$,
on the other hand, O shell burning maintains its activity until the
collapse and the layer is OSi-rich
(Figure \ref{fig:28M_central_Xcore}(d)). This makes the Fe core mass for
$\sigma_{C12\alpha} = 0$ (1.51 M$_\odot$) larger than for
$\sigma_{C12\alpha} = -1$ (1.36 M$_\odot$) as seen in 
Figure \ref{fig:28M_abund5}.

\subsection{Thermal and Dynamical Evolution of $28$ M$_\odot$ Star}
\label{sec:24Mt}

In association with the chemical evolution of the $28$ M$_\odot$ star
as described in the earlier subsection, the thermal and dynamical
structures of the star evolve as follows with the significant
dependence on $\sigma_{C12\alpha}$.

\subsubsection {Evolution of Entropy and Temperature}

Figure \ref{fig:28M_entr} shows the evolution of specific entropy at
the center ($s_{\rm c}$), $M_r=$ 1.5 M$_{\odot}$ ($s_{1.5}$), $M_r=$ 2.0
M$_{\odot}$ ($s_{2.0}$), and $M_r=$ 2.5 M$_\odot$ ($s_{2.5}$) for
$\sigma_{C12\alpha}=$ 0 and $-$1.

\begin{figure*}[htbp]
\centering
\begin{minipage}[c]{0.75\textwidth}
\centerline{$M({\rm ZAMS})=28 $ M$_\odot$}
\includegraphics [width=132mm]{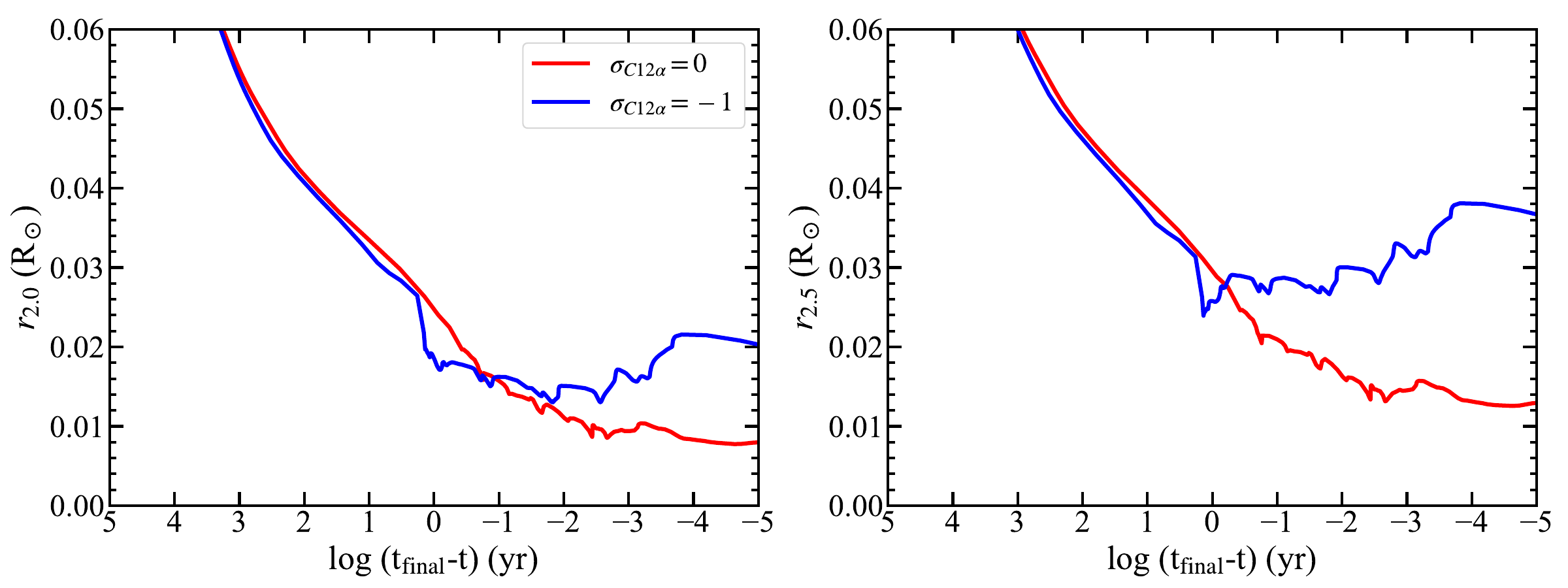}
\end{minipage}%
\caption{The time evolution of radius at $M_r=$ 2.0 M$_{\odot}$
($r_{2.0}$) and $M_r=$ 2.5 M$_{\odot}$ ($r_{2.5}$) of
$M {\rm (ZAMS)}$ = 28 M$_{\odot}$ star with $\sigma_{C12\alpha}=$ 0 and $-$1.
\label{fig:28M_radius}}
\end{figure*}

Figure \ref{fig:28M_temp} shows the evolution of temperatures at the
center ($T_{\rm c}$), $M_r=$ 1.5 M$_\odot$ ($T_{1.5}$), $M_r=$ 2.0 M$_\odot$
($T_{2.0}$) and $M_r=$ 2.5 M$_\odot$ ($T_{2.5}$) for both cases of
$\sigma_{C12\alpha}$.

Even during central C-burning ($\tau \sim 10^{3} - 10^{1}$ yr), $s_{\rm c}$
decreases because the neutrino energy loss rate exceeds the nuclear
energy generation rate.  Then the CO core contracts to increase $T_{\rm c}$
because of the gravothermal effect (Equation \ref{equ:s2} - \ref{equ:cg}).

\begin{figure*}[htbp]
\centering
\begin{minipage}[c]{0.75\textwidth}
\centerline{$M({\rm ZAMS})=28 $ M$_\odot$}
\includegraphics [width=132mm]{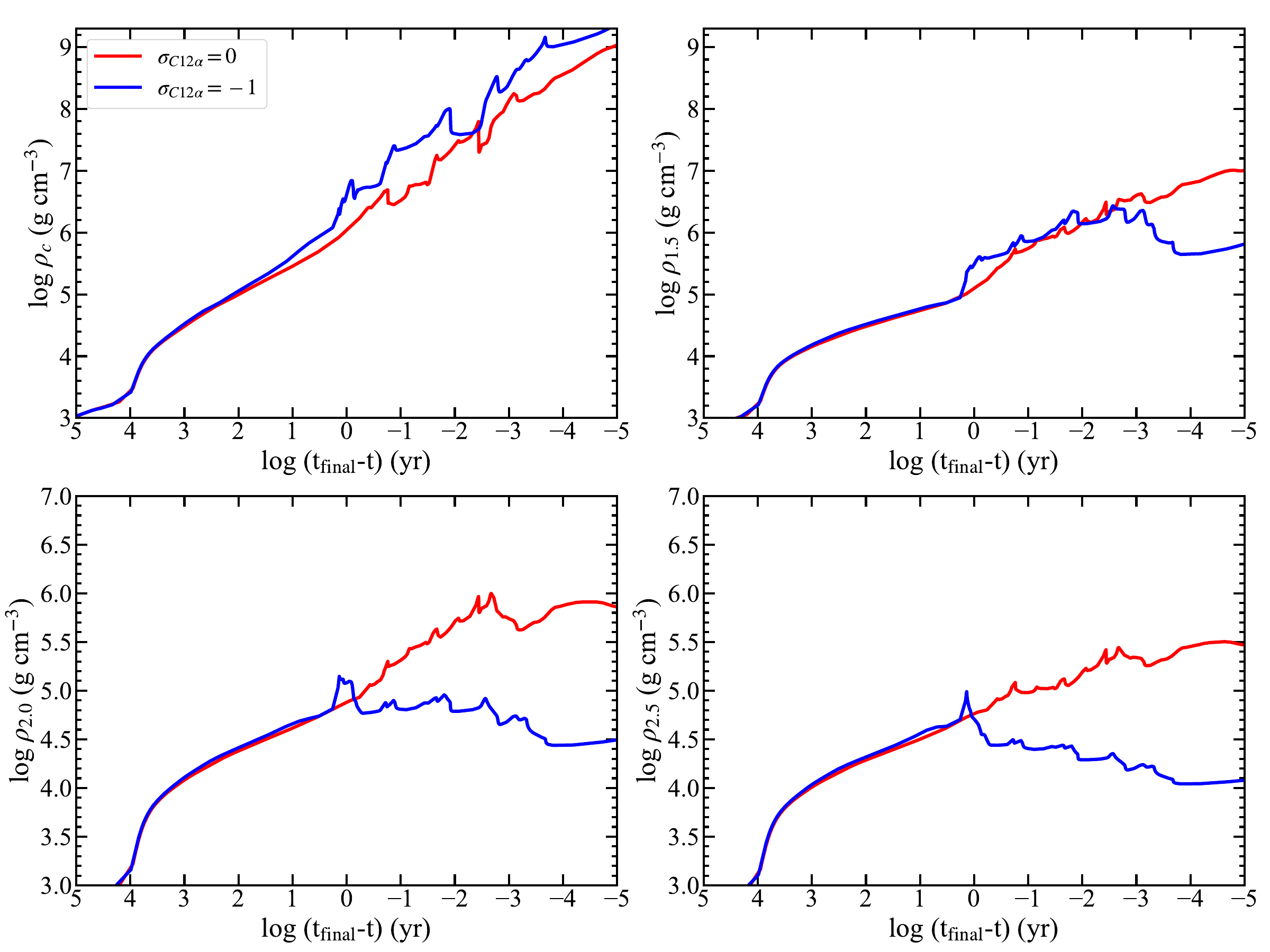}
\end{minipage}%
\caption{The time evolution of density at the center ($\rho_{\rm c}$), $M_r=$ 1.5
M$_\odot$ ($\rho_{1.5}$), $M_r=$ 2.0 M$_\odot$ ($\rho_{2.0}$) and $M_r=$ 2.5
M$_\odot$ ($\rho_{2.5}$) of $M {\rm (ZAMS)}$ = 28 M$_{\odot}$ star with
$\sigma_{C12\alpha}=$ 0 and $-$1.
\label{fig:28M_rho}}
\end{figure*}

\begin{figure*}[htbp]
\centering
\begin{minipage}[c]{0.75\textwidth}
\centerline{$M({\rm ZAMS})=28 $ M$_\odot$}
\includegraphics [width=132mm]{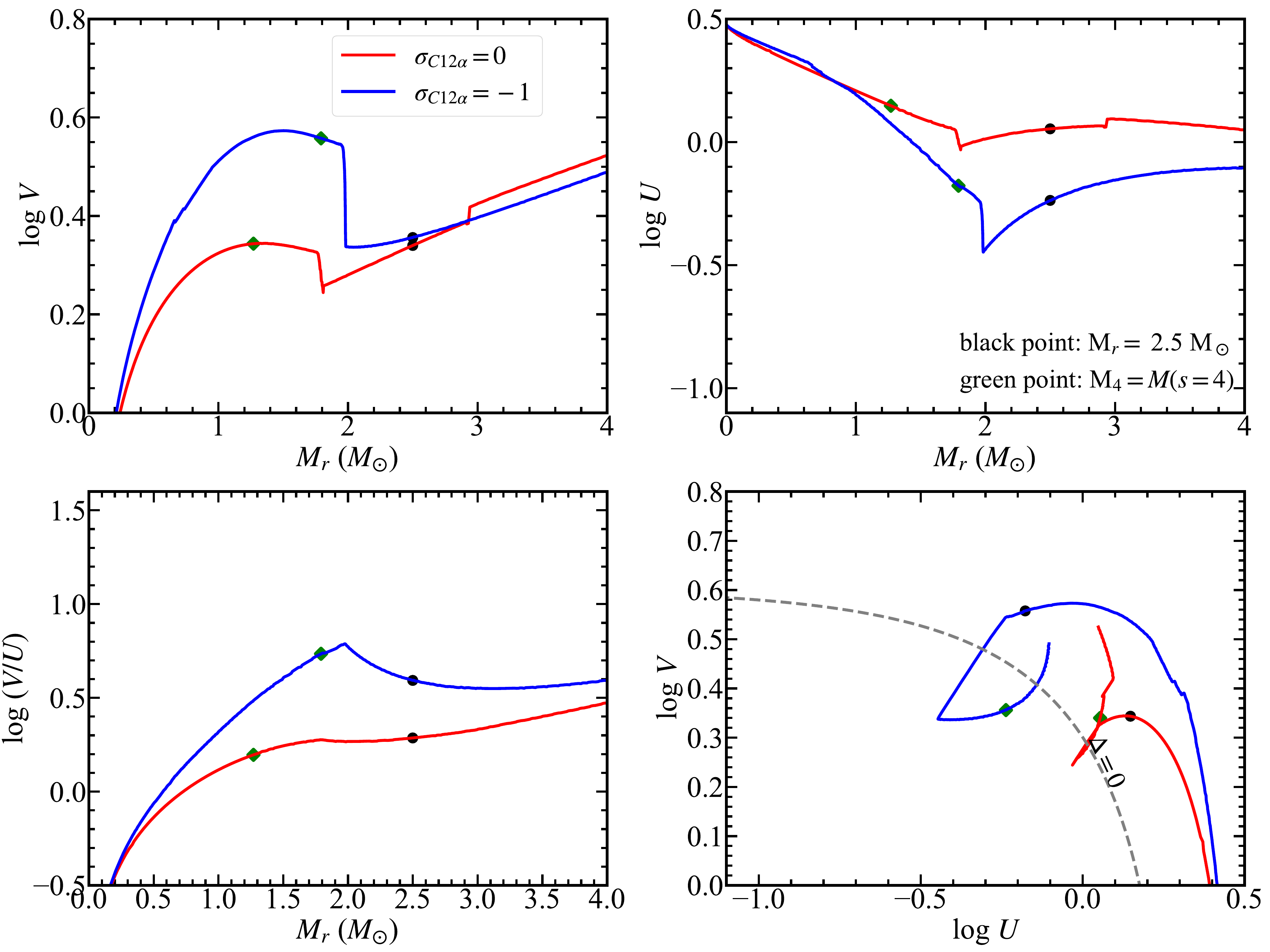}
\end{minipage}%
\caption{The $U-V$ curves of 28 M$_{\odot}$ star at $\tau \sim 10^{-0.3}$ yr.
The orange points represent the place where $M_r = 2.5$ M$_{\odot}$,
while the green diamonds show the place where $M_r = M(s=4)$.
\label{fig:28M_uv1}}
\end{figure*}

Around $\tau \sim 10^{1.2}$ yr, carbon in the central region is almost
exhausted (Figure~\ref{fig:28M_central_Xcore}).  Subsequent 
contraction of the ONeMg core leads to the increase in the core
temperatures ($T_{\rm c}$, $T_{1.5}$, $T_{2.0}$, and $T_{2.5}$)
(Figure \ref{fig:28M_temp}).  Then, the first C-shell burning is ignited
radiatively for both cases of $\sigma_{C12\alpha}$
(Figure \ref{fig:28M_core}).  

The second C-shell burning forms a convective shell for both cases
(Table\ref{tab:stages}).  The resulting heating is higher for
$\sigma_{C12\alpha}=-1$ than 0 as seen in the slightly higher $s_{1.5}$ and
$s_{2.0}$ around $\tau \sim 10^{1.2} - 10^{0}$ yr.  This is because of
larger $X$($^{12}$C) in $\sigma_{C12\alpha}=-1$ although the
difference is not so large.

A large difference between the two cases of $\sigma_{C12\alpha}$
appears around $\tau \sim 10^{0}$ yr near the end of central
C-burning (Figure~\ref{fig:28M_central_Xcore}(a)).

For $\sigma_{C12\alpha}=-1$ (the blue lines), C-shell burning keeps a
large convection zone above $M_r =$ 2.0 M$_\odot$
as seen from the increase in $s_{2.0}$ and $s_{2.5}$ (the blue lines) 
near $\tau = 10^{0}$ yr. Such enhancement of
$s_{2.0}$ occurs because $X$($^{12}$C) is enhanced by mixing of outer
C into the convective shell around $\tau \sim 10^{0}$ yr
(Figure~\ref{fig:28M_central_Xcore}(f)).
This convective C-shell burning continues to exist through Fe core-collapse.
Because of this heating, the increase in $T_{2.0}$ is slow
and $T_{2.5}$ is almost constant,
in contrast to the increase in $T_{\rm c}$ and $T_{1.5}$
due to core contraction through Fe core-collapse.

For $\sigma_{C12\alpha}=0$ (the red lines), on the other hand, $X$($^{12}$C)
at $M_r =$ 2.0 M$_\odot$ is not enhanced
(Figure~\ref{fig:28M_central_Xcore}(e)), so that the increase in
$s_{2.0}$ is small from $\tau \sim 10^{0}$ to $10^{-0.6}$ yr.
It even decreases after $\tau = 10^{-0.6}$ yr due
to the neutrino energy loss. Then $T_{2.0}$ and $T_{2.5}$ continue to
increase as in $T_{\rm c}$ and $T_{1.5}$.

\subsubsection {Evolution of Radius and Density} \label{sec:evo_r_rho}

The difference in the heating effect of C-shell burning appears in
the difference in the evolution of core structure between the two cases.

Figure~\ref{fig:28M_radius} shows the evolution of radius at $M_r=$
2.0 M$_{\odot}$ ($r_{2.0}$) and $M_r=$ 2.5 M$_{\odot}$ ($r_{2.5}$) for
$\sigma_{C12\alpha}=$ 0 (the red line) and $-1$ (the blue lines).

The obvious differences appear at $\tau \sim 10^{0}$ yr, where
$r_{2.5}$ stays almost constant for $\sigma_{C12\alpha}= -1$ (blue),
while $r_{2.5}$ decreases along $r_{2.0}$ (red).
This is due to the heating effect of C-shell burning above $M_r>$ 2.0 M$_{\odot}$.

Figures \ref{fig:28M_rho} show the evolution of the density at the
center ($\rho_{\rm c}$), $M_r=$ 1.5 M$_{\odot}$ ($\rho_{1.5}$), $M_r=$ 2.0
M$_{\odot}$ ($\rho_{2.0}$) and $M_r=$ 2.5 M$_{\odot}$ ($\rho_{2.5}$).

For both cases of $\sigma_{C12\alpha}$, $\rho_{\rm c}$ and $\rho_{1.5}$
continue to increase. For $\sigma_{C12\alpha}= 0$ (red), 
$\rho_{2.0}$ and $\rho_{2.5}$ continue to increase through the
collapse. For $\sigma_{C12\alpha}=-1$ (blue), on the other hand,
$\rho_{2.0}$ keeps almost constant and $\rho_{2.5}$ even decreases.

These evolutions result in a rather large difference in the Fe core
structure as will be discussed in subsection \ref{sec:pre-struc}.
We emphasize that enhancement of $X$($^{12}$C) in the C shell burning causes
such large differences in the thermal and dynamic structure of the
presupernova core.

\begin{figure*}[ht!]
\centering
\begin{minipage}[c]{0.75\textwidth}
\centerline{$M({\rm ZAMS})=28 $ M$_\odot$}
\includegraphics [width=132mm]{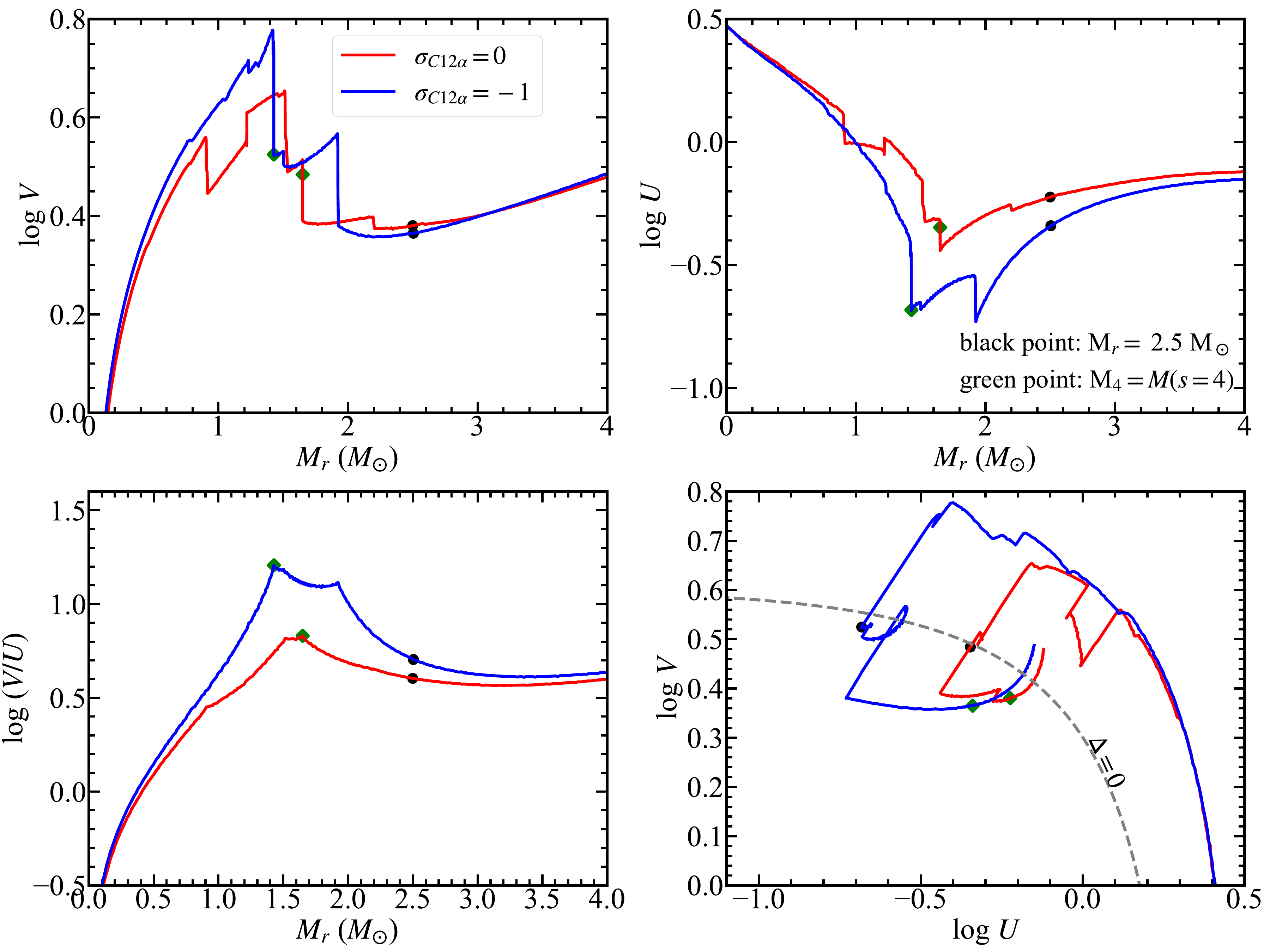}
\end{minipage}%
\caption{The $U-V$ curves of 28 M$_{\odot}$ star at $\tau \sim 10^{-3.3}$ yr.
The orange points represent the place where $M_r = 2.5$ M$_{\odot}$,
while the green diamonds show the place where $M_r = M(s=4)$.
\label{fig:28M_uv2}}
\end{figure*}

\begin{figure*}[ht!]
\centering
\begin{minipage}[c]{0.75\textwidth}
\centerline{$M({\rm ZAMS})=28 $ M$_\odot$}
\includegraphics [width=132mm]{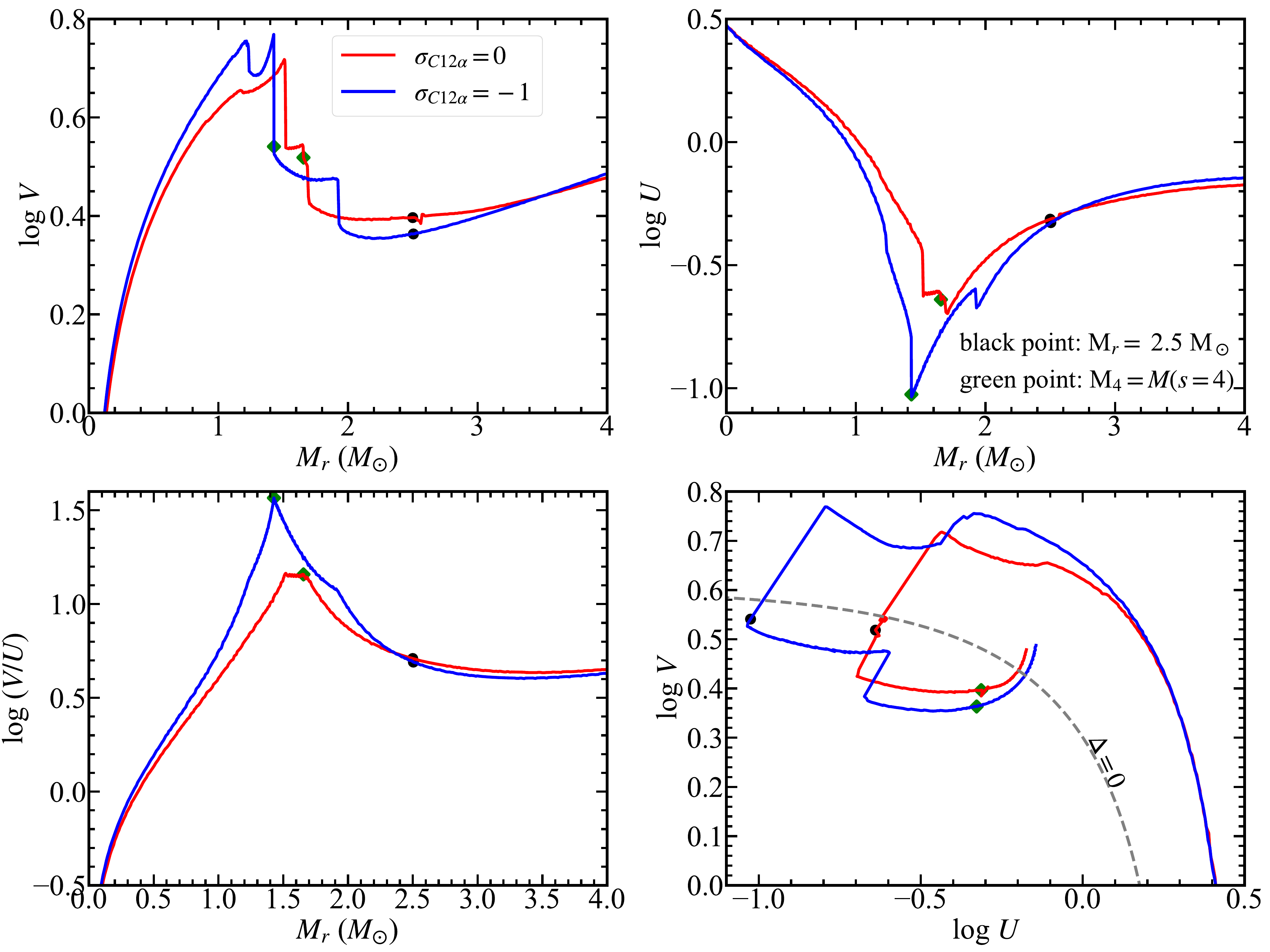}
\end{minipage}%
\caption{The $U-V$ curves of 28 M$_{\odot}$ star at $\tau=t_{\rm f}$.
The orange points represent the place where $M_r = 2.5$ M$_{\odot}$,
while the green diamonds show the place where $M_r = M(s=4)$.
\label{fig:28M_uv_tfinal}}
\end{figure*}

\begin{figure*}[htb]
\centering
\centerline{$M({\rm ZAMS})=28 $ M$_\odot$}
\begin{minipage}[c]{0.43\textwidth}
\includegraphics [width=75mm]{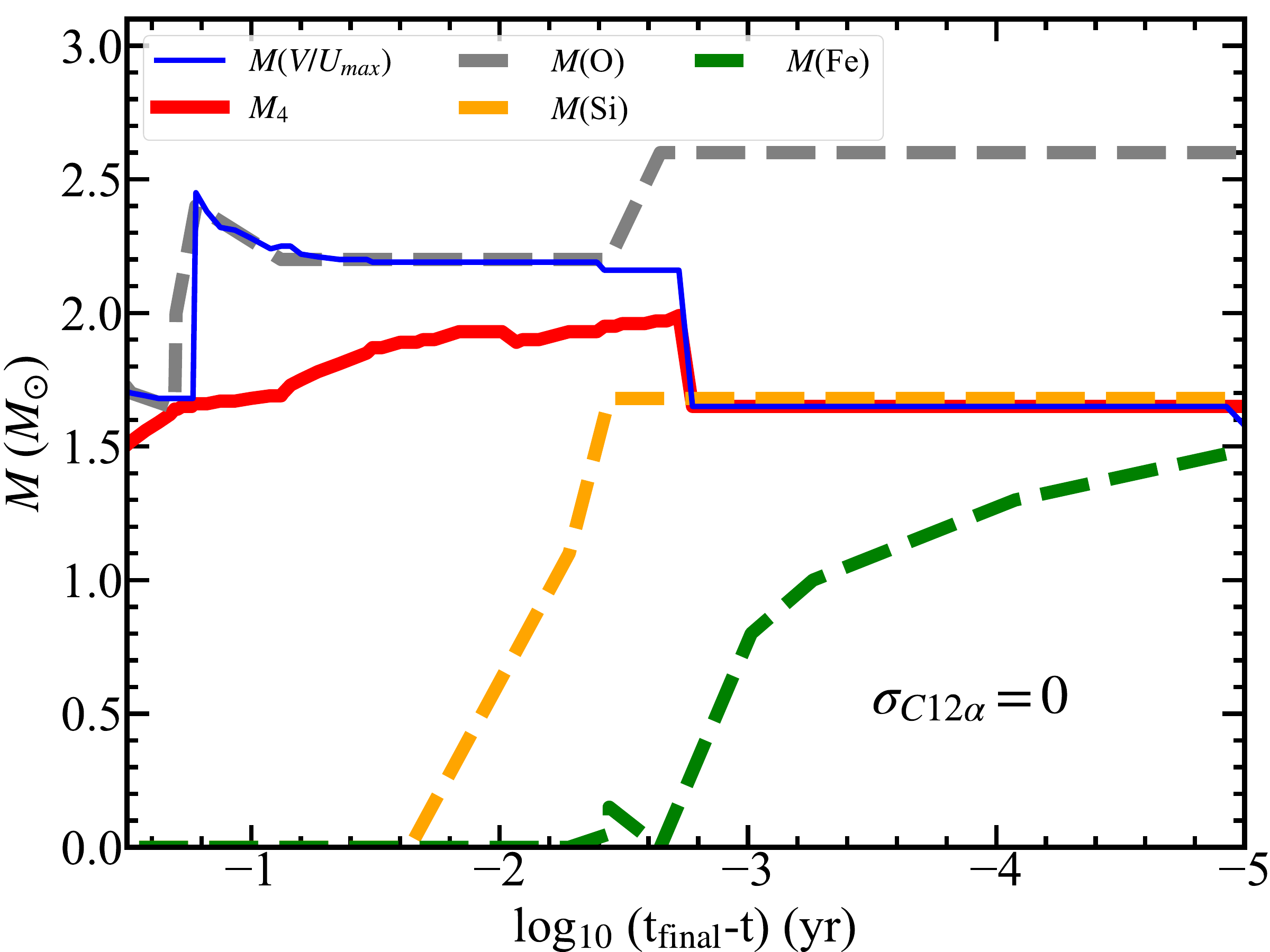}
\end{minipage}%
\begin{minipage}[c]{0.43\textwidth}
\includegraphics [width=75mm]{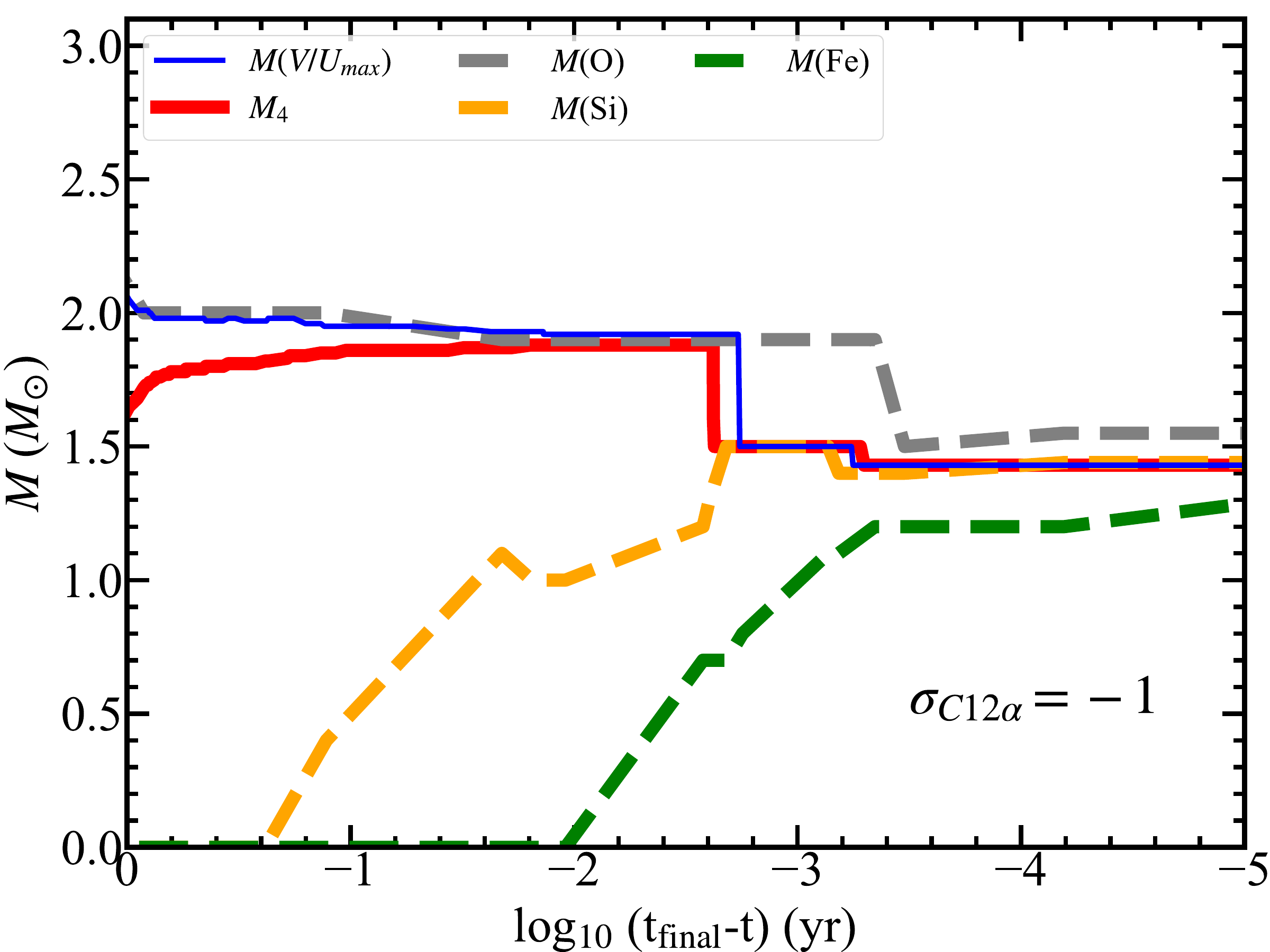}
\end{minipage}%
\caption{The time evolution of core masses of Fe, $M$(Fe), Si,
  $M$(Si), and O, $M$(O), $M_r = M(s=4)$, $M_4$, and $M(V/U_{\rm
    max})$ for $M ({\rm ZAMS})$ = 28 M$_\odot$ with
  $\sigma_{C12\alpha}$ = 0 (left) and $-1$ (right).  The core masses
  are shown with the thick dashed lines, and $M_4$ is the thick solid
  line.  $M(V/U_{\rm max})$ is indicated by the thin solid line. Both
  $M$(Si) and $M_4$ are overlapped with $M(V/U_{\rm max})$ near the
  final stage.
  \label{fig:core_muv}}
\end{figure*}

\subsubsection {Effects of Oxygen Shell Burning}

In the earlier sections, we have discussed the effect of O-shell burning
on the core structure in detail. In Figure \ref{fig:28M_entr}, \ref{fig:28M_temp}
and \ref{fig:28M_rho} show a obvious increase in $s_{1.5}$ and decrease in
$T_{1.5}$ and $\rho_{1.5}$ after log $\tau=$ -3.8 yr.
These changes show that the expansion of the shell near $M_{r}=$ 1.5
M$_{\odot}$ originate from the earlier mixes of O in $M_{r}=$ 1.5
M$_{\odot}$, which has been mentioned in section \ref{sec:28M_final}.

\begin{figure*}[ht!]
\centering
\centerline{$M({\rm ZAMS})=28$ M$_\odot$}
\begin{minipage}[c]{0.42\textwidth}
\centerline{$M({\rm ZAMS})=28 $ M$_\odot$}
\includegraphics [width=75mm]{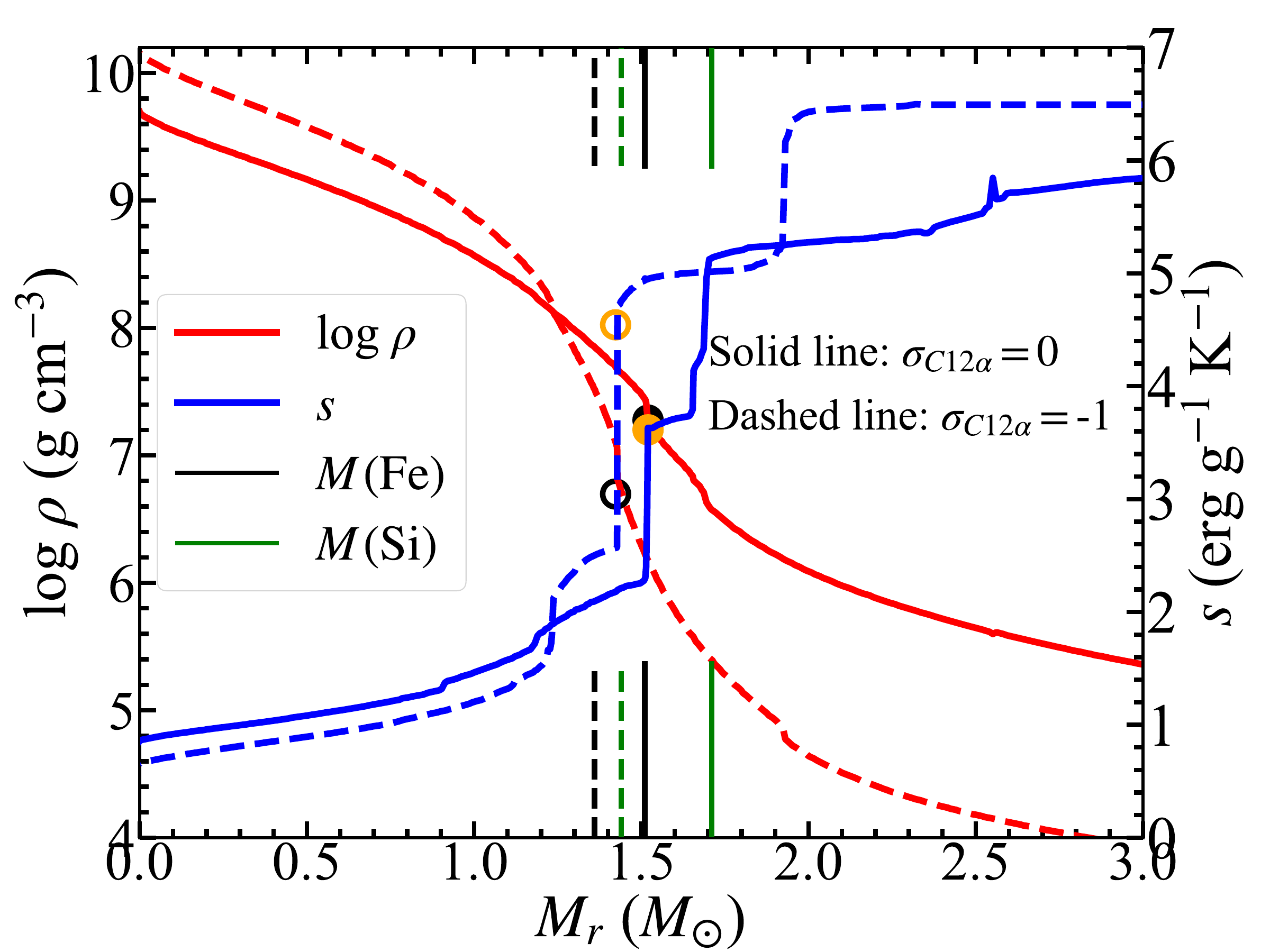}
\end{minipage}%
\begin{minipage}[c]{0.42\textwidth}
\centerline{$M({\rm ZAMS})=28 $ M$_\odot$}
\includegraphics [width=75mm]{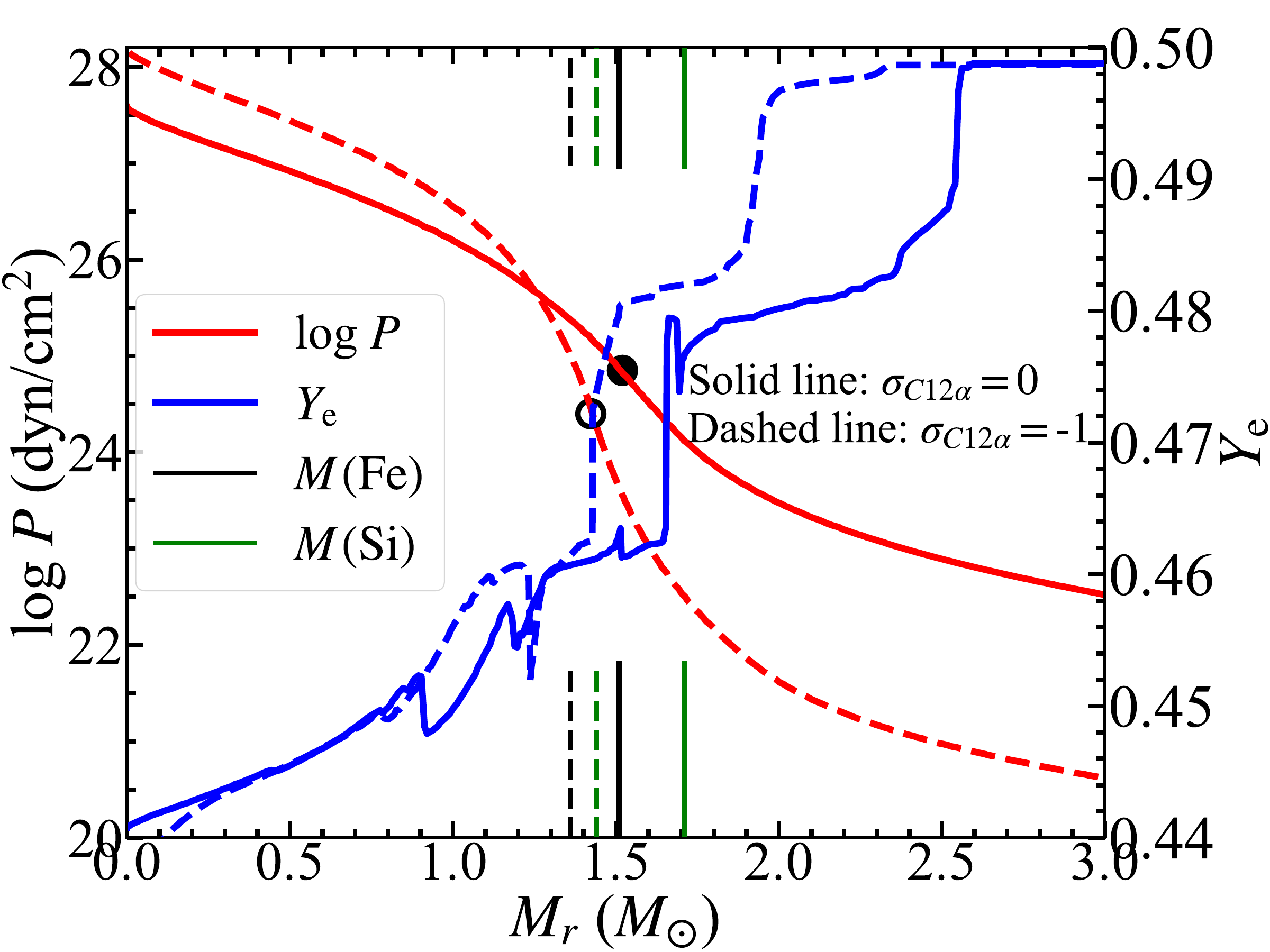}
\end{minipage}%
\caption{(Left) The density and entropy distribution of a star with
$M ({\rm ZAMS})$ = 28 M$_{\odot}$ for $\sigma_{C12\alpha}=$ 0 (solid line)
and -1 (dashed line). The green and black lines show the edges of Si and Fe cores,
respectively. The steepest gradients of log $\rho$ and $s$,
which corresponds to the peak of log ($V/U$), 
are marked respectively by the filled black and orange circles 
(for $\sigma_{C12\alpha}=$ 0) and the open black and orange circles
(for $\sigma_{C12\alpha}=-1$).
(Right) Same as Left but for the pressure and $Y_{\rm e}$.
\label{fig:28M_logp_rho}}
\end{figure*}

\subsubsection {U-V Curves}  \label{sec:uv_curve}

{\bf Nuclear Shell Burning and U-V Curves}:

Nuclear shell burning has an important effect on the structure and
evolution of stars because shell burning forms a sharp increase
(almost a jump) in entropy and a convective shell.  This is seen in
the enhancement of $s_{2.0}$ and $s_{2.5}$ (Figure~\ref{fig:28M_entr})
due to C burning (log $tau \sim 0$ yr) and also $s_{1.5}$ due to
O-burning (log $\tau \sim -3.3$).  Such an enhancement produces a
sharp gradient of entropy as well as other physical quantities against
$M_r$ and $r$.  This enhancement depends on $\sigma_{C12\alpha}$.

The strength of nuclear shell burning and the shell location 
can easily be seen in the $U-V$ curves: See Figures~\ref{fig:28M_uv}
and \ref{fig:28M_uv_tfinal} for the $U-V$ curves of the whole star and
that of the inner part ($M_r=$ 0 - 4 M$_\odot$) at the collapse 
($t= t_{\rm f}$), respectively.  Hereafter, we discuss the $U-V$ curves
of only the inner part ($M_r=$ 0 - 4 M$_\odot$).

As seen from Equations \ref{equ:u}, \ref{equ:v}, and \ref{equ:vu}, the
peaks of $U$ and $V$ indicate the locations of the steeper gradients
of $M_r$ and $P$ against log $r$ compared with the surrounding
regions, while the peaks of $V/U$ and $1/U$ indicate the location of
the steeper gradients of $P$ and $r$ against log $M_r$.

The loop of the log $U$ - log $V$ curve shows a relatively large jump
across $\delta = 0$ (Equation \ref{equ:delta}), where $U$ is minimum and
thus, physical quantities have steep gradients concerning $M_r$.
This implies that the location of the above jump is the core edge in
$M_r$, which is the location of a nuclear burning shell.  

The location of the bottom of the nuclear burning shell is
indicated by the peaks of $V/U$, because nuclear shell burning
produces a large entropy jump, thus creating a larger gradient of
physical quantities concerning $M_r$.  The effect of the larger
entropy jump in $\sigma_{C12\alpha}= -1$ (blue) than 0 (red) appears
in the larger peak values of $V$ (i.e., more centrally condensed) and
$V/U$, the smaller minimum value of $U$ (i.e., more extended), and the
larger loop of $U-V$ in the blue lines than in the red lines.

Figures \ref{fig:28M_uv1}, \ref{fig:28M_uv2}, and
\ref{fig:28M_uv_tfinal} show the time evolution of the $U-V$ curves
around $\tau = 10^{-0.3}$ yr, $10^{-3.3}$ yr, and $t_{\rm f}$,
respectively.  At $\tau = 10^{-0.3}$ yr, strongest nuclear shell
burning appears at $M_r=$ 2.0 M$_\odot$ for $\sigma_{C12\alpha}= -1$,
which is the bottom of convective C burning shell.  
At $\tau \sim 10^{-3.3}$ yr, the location of strongest nuclear burning
shell moves to $M_r =$ 1.5 M$_\odot$, which is O shell burning.

We note that these figures show that the maximum values of $V$, 1/$U$
and $V/U$ are larger in $\sigma_{C12\alpha}=-1$ than 0.  This is
mainly due to stronger C-shell burning with larger $X(^{12}$C).

$M(V/U_{\rm max})$, $M_{\rm eff}$, {\bf Core Masses, and} $s_{\rm c}$:

Among these many peaks in the $U-V$ curves, $M_r$ at the maximum $V/U$
($M(V/U_{\rm max})$) is the mass contained below the strongest burning
shell.
\footnote{Hereafter, we denote the maximum $V/U$, i.e.,
 $(V/U)_{\rm max}$, as $V/U_{\rm max}$.}
Since $V/U_{\rm max}$ appears at the most active nuclear burning
shell, which forms the steepest pressure gradient, $M(V/U_{\rm max})$
can be regarded as $M_{\rm eff}$ in Equation~\ref{equ:meff}.  Then
$M(V/U_{\rm max})$ is strongly related to the evolution of the core as
described in Equation~\ref{equ:meff}, thus being useful to discuss the
explodability in the subsequent sections.

To see the dependence of $M(V/U_{\rm max})$ on $\sigma_{C12\alpha}$,
Figure \ref{fig:core_muv} shows the time evolution of core masses of
Fe ($M$(Fe)), Si ($M$(Si)), and O ($M$(O)), $M_r = M(s=4)$ ($M_4$),
and $M(V/U_{\rm max})$ for $M({\rm ZAMS})$ = 28 M$_\odot$ with
$\sigma_{C12\alpha} = 0$ (left) and $-1$ (right).  It is seen that
$M(V/U_{\rm max})$ is smaller for $\sigma_{C12\alpha}= -1$ than 0,
mainly because of strong C-shell burning at $M_r =$ 2 M$_\odot$.  This
smaller $M(V/U_{\rm max})$ produces the smaller core masses of O, Si,
and Fe.  We also note that $M$(Si) (i.e., $M_r$ at the O-burning
shell) and $M_4$ are almost identical to $M(V/U_{\rm max})$ near the
final stage so that these masses are smaller for
$\sigma_{C12\alpha}= -1$ than 0.

The effect of smaller $M(V/U_{\rm max})$ (and thus $M_{\rm eff}$) for
$\sigma_{C12\alpha} = -1$ than 0 can be seen in the evolution of log
$T_{\rm c}$ - log $\rho_{\rm c}$ (Figure~\ref{fig:trho}).  Here
$s_{\rm c}$ is smaller (and thus $\rho_{\rm c}$ is higher for the same
$T_{\rm c}$) for $\sigma_{C12\alpha} = -1$ than 0 according to
Equations~\ref{equ:s2} and \ref{equ:trho}.  Such evolutions of smaller
$s_{\rm c}$ and higher $\rho_{\rm c}$ (at the same $T_{\rm c}$) are
seen in Figures~\ref{fig:28M_entr} and \ref{fig:28M_rho}, respectively.

Regarding the explodability, the maximum values of log $V/U$ are 1.6
and 1.1 for $\sigma_{C12\alpha}=-1$ and 0, respectively, which show
the difference in the steepness of log $P$ with respect to $M_r$ as
seen in Figure \ref{fig:28M_logp_rho} below.  This indicates that the
star with $\sigma_{C12\alpha}=-1$ is easier to explode, as will be
discussed later.

Thus, the $U-V$ curves are handy diagrams to indicate the
location of the critical points, such as the strongest nuclear shell
burning and $M_4$ by showing ``quantitatively'' the gradients of
quantities with respect to $M_r$ and $r$.

\subsubsection {Presupernova Structure and Compactness} \label{sec:pre-struc}

\begin{figure}[htbp]
\centering
\begin{minipage}[c]{0.4\textwidth}
\centerline{$M({\rm ZAMS})=28 $ M$_\odot$}
\includegraphics [width=75mm]{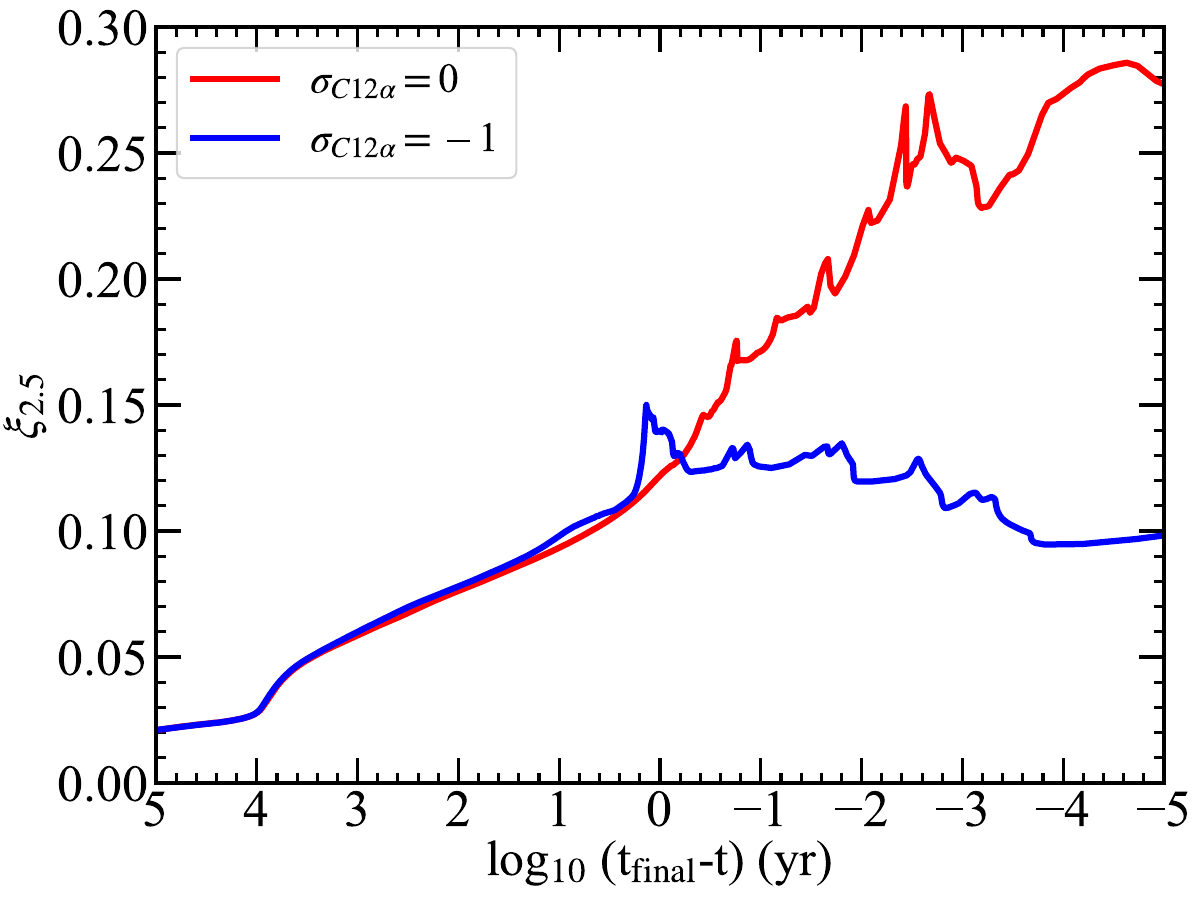}
\end{minipage}%
\caption{The time evolution of $\xi_{2.5}$
for $M ({\rm ZAMS})$ = 28 M$_{\odot}$ and $\sigma_{C12\alpha}=$ 0 and $-$1.
\label{fig:28M_compact}}
\end{figure}

\begin{figure}[htbp]
\centering
\begin{minipage}[c]{0.4\textwidth}
\centerline{$M({\rm ZAMS})=25 $ M$_\odot$}
\includegraphics [width=75mm]{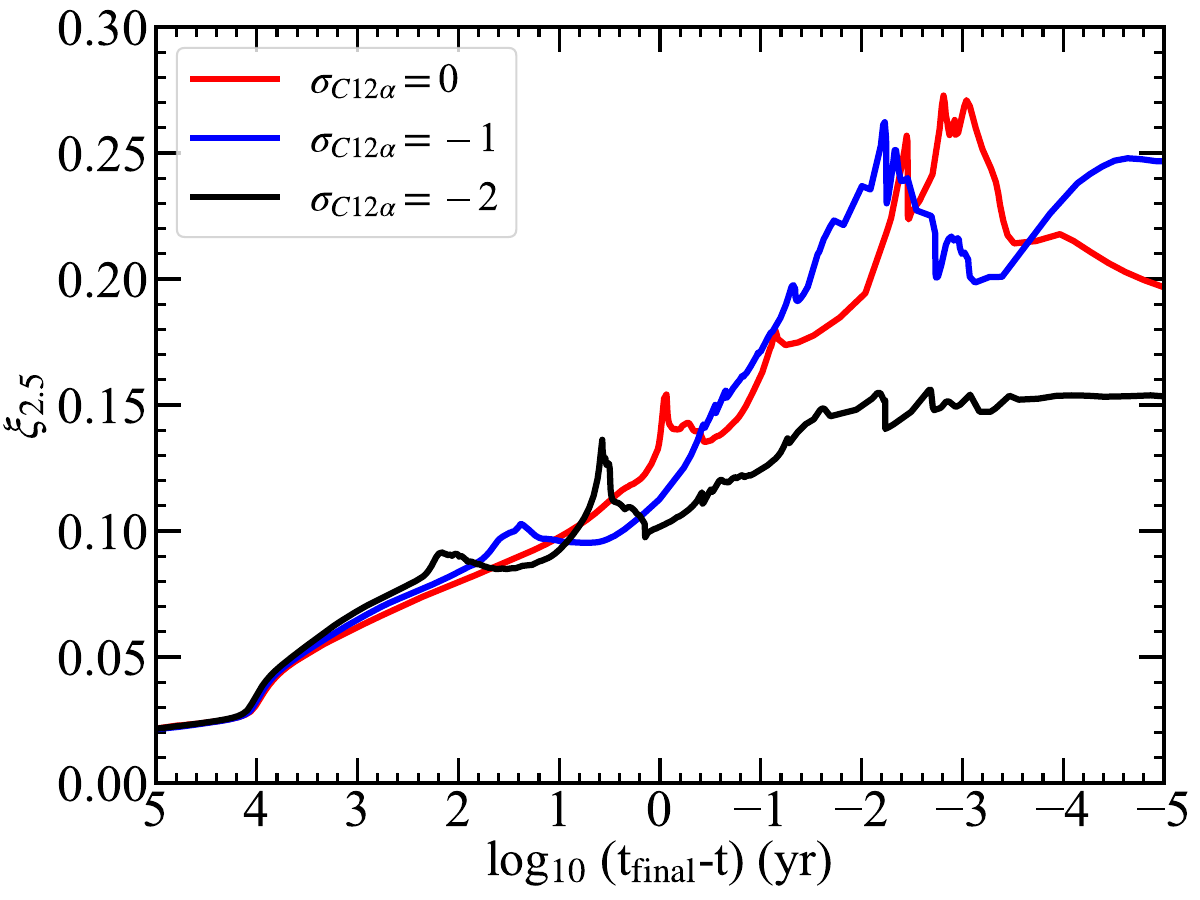}
\end{minipage}%
\caption{The time evolution of $\xi_{2.5}$
for $M ({\rm ZAMS})$ = 25 M$_{\odot}$ and $\sigma_{C12\alpha}=$ 0, $-$1 and $-$2.
respectively.
\label{fig:25M_compact}}
\end{figure}

\begin{figure}[htbp]
\centering
\begin{minipage}[c]{0.4\textwidth}
\centerline{$M({\rm ZAMS})=35 $ M$_\odot$}
\includegraphics [width=75mm]{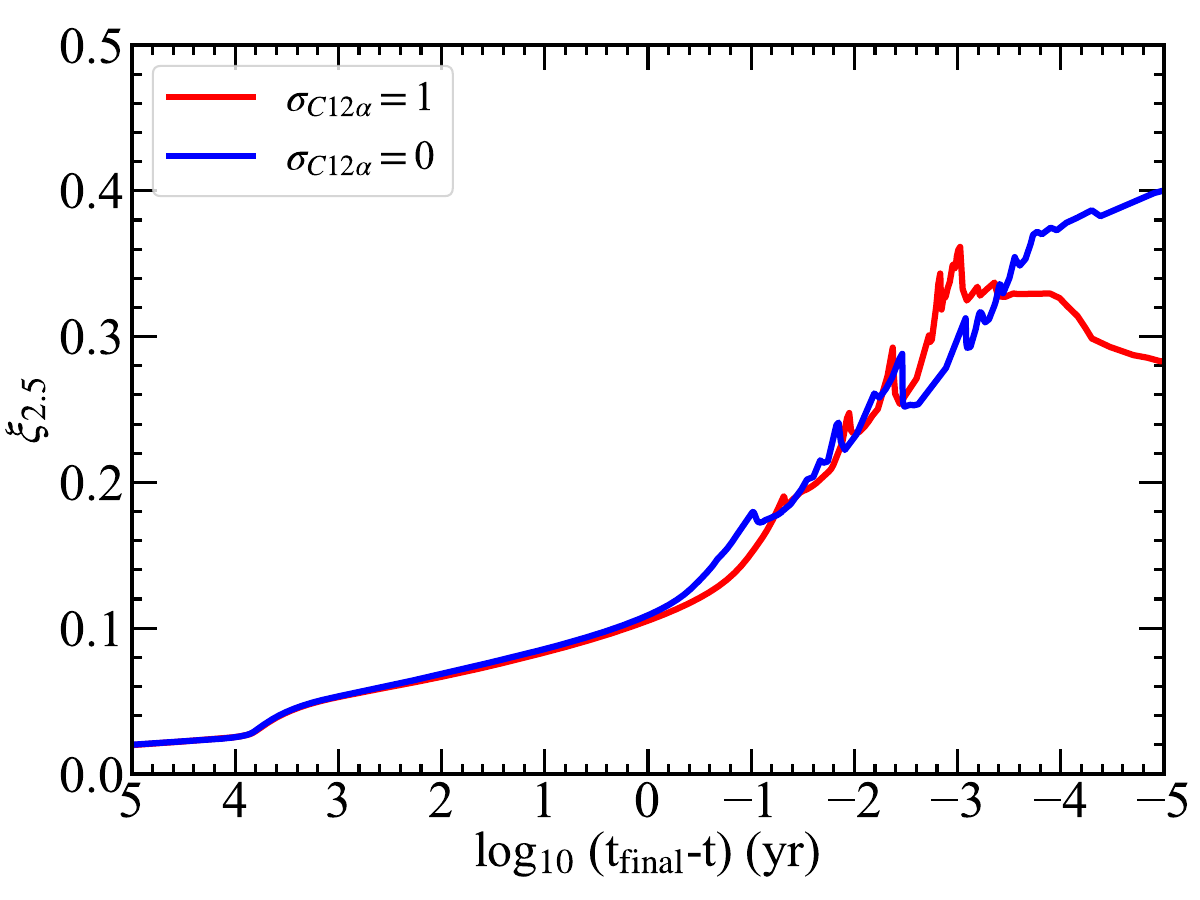}
\end{minipage}%
\caption{The time evolution of $\xi_{2.5}$
for $M ({\rm ZAMS})$ = 35 M$_{\odot}$ with $\sigma_{C12\alpha}=$ 1 and 0.
\label{fig:35M_compact_p10}}
\end{figure}

\begin{figure}
\centering
\begin{minipage}[c]{0.42\textwidth}
\centerline{$M({\rm ZAMS})=25 $ M$_\odot$, $\sigma_{C12\alpha}=0$}
\includegraphics [width=75mm]{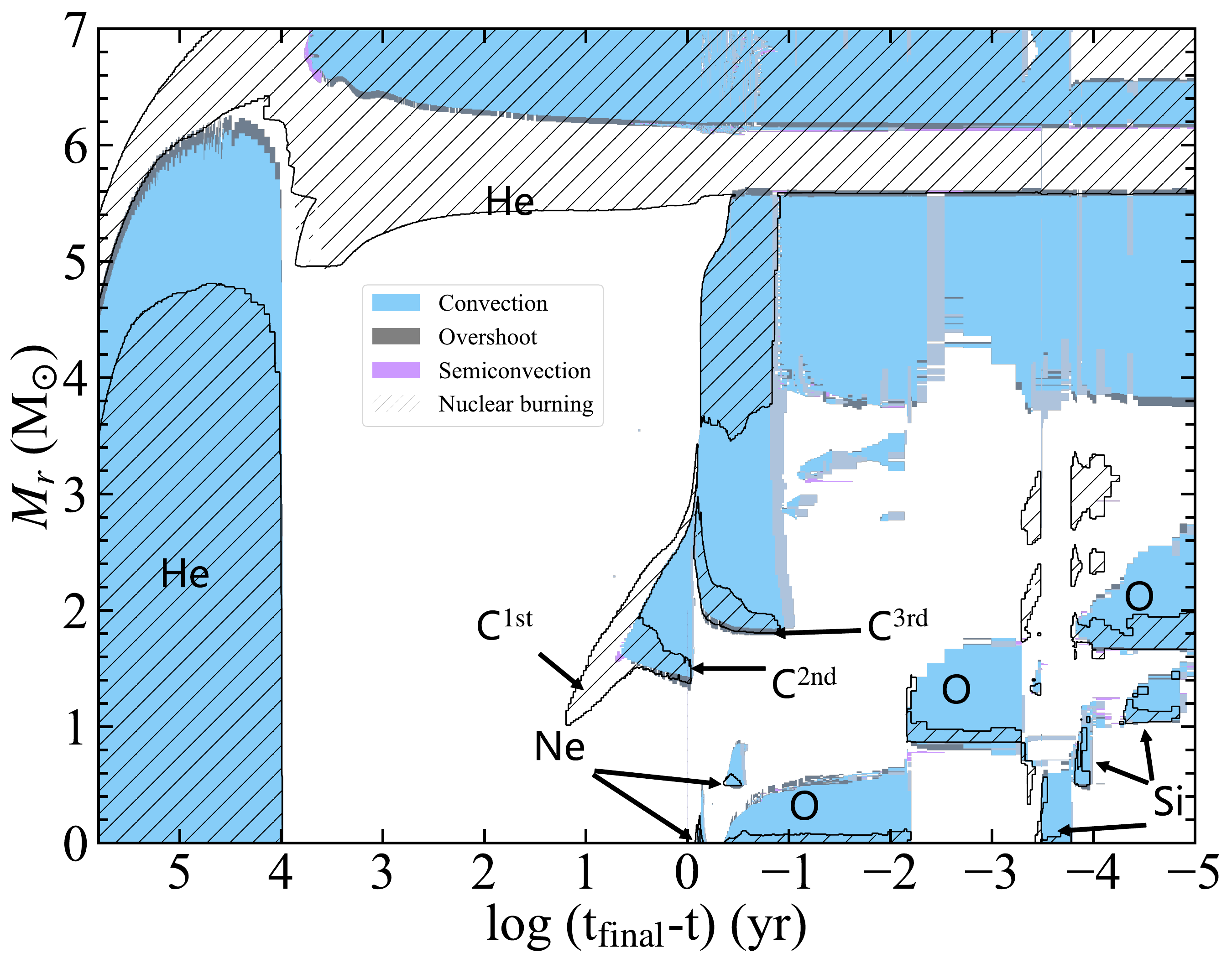}
\centerline{$M({\rm ZAMS})=25 $ M$_\odot$, $\sigma_{C12\alpha}=-1$}
\includegraphics [width=75mm]{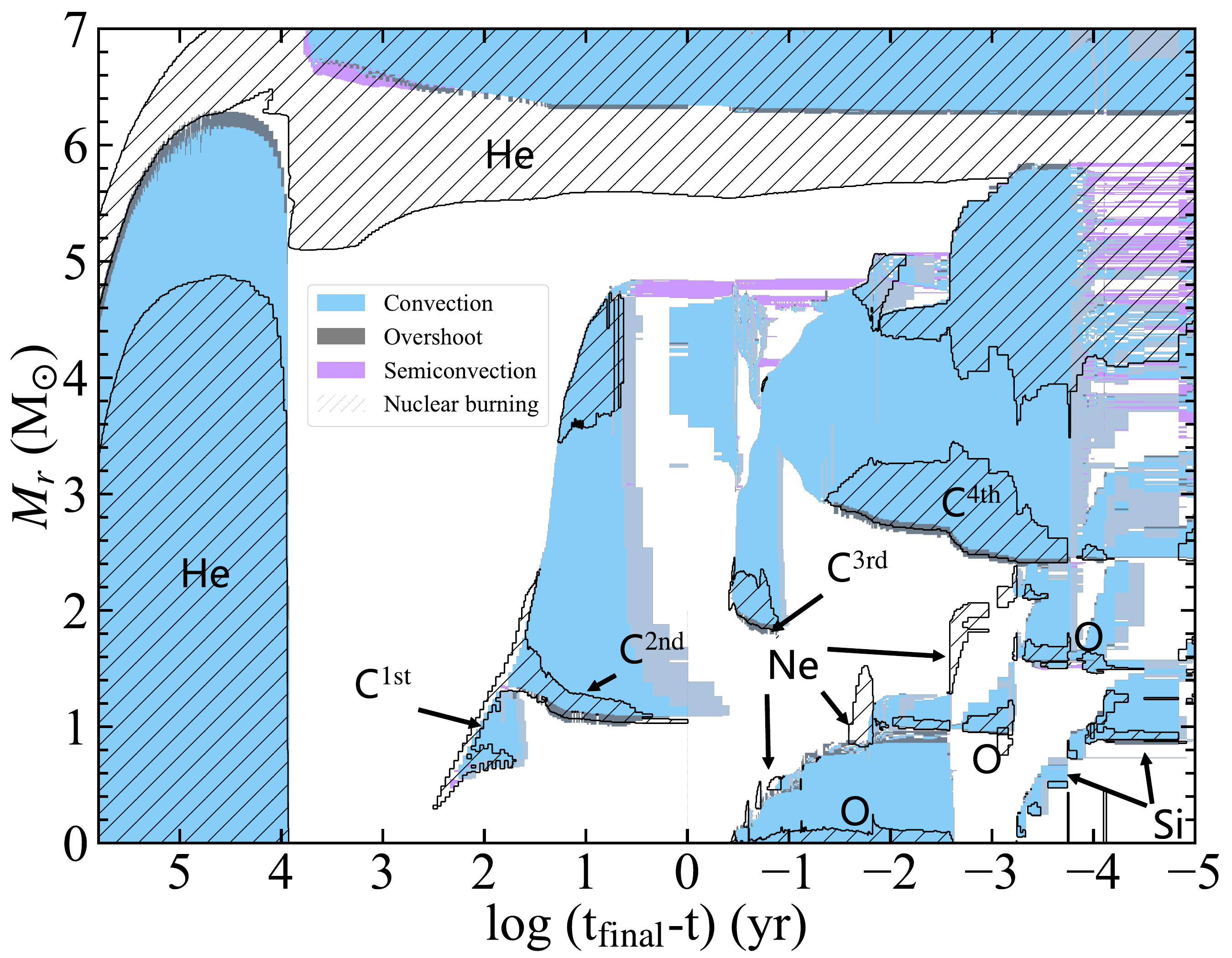}
\centerline{$M({\rm ZAMS})=25 $ M$_\odot$, $\sigma_{C12\alpha}=-2$}
\includegraphics [width=75mm]{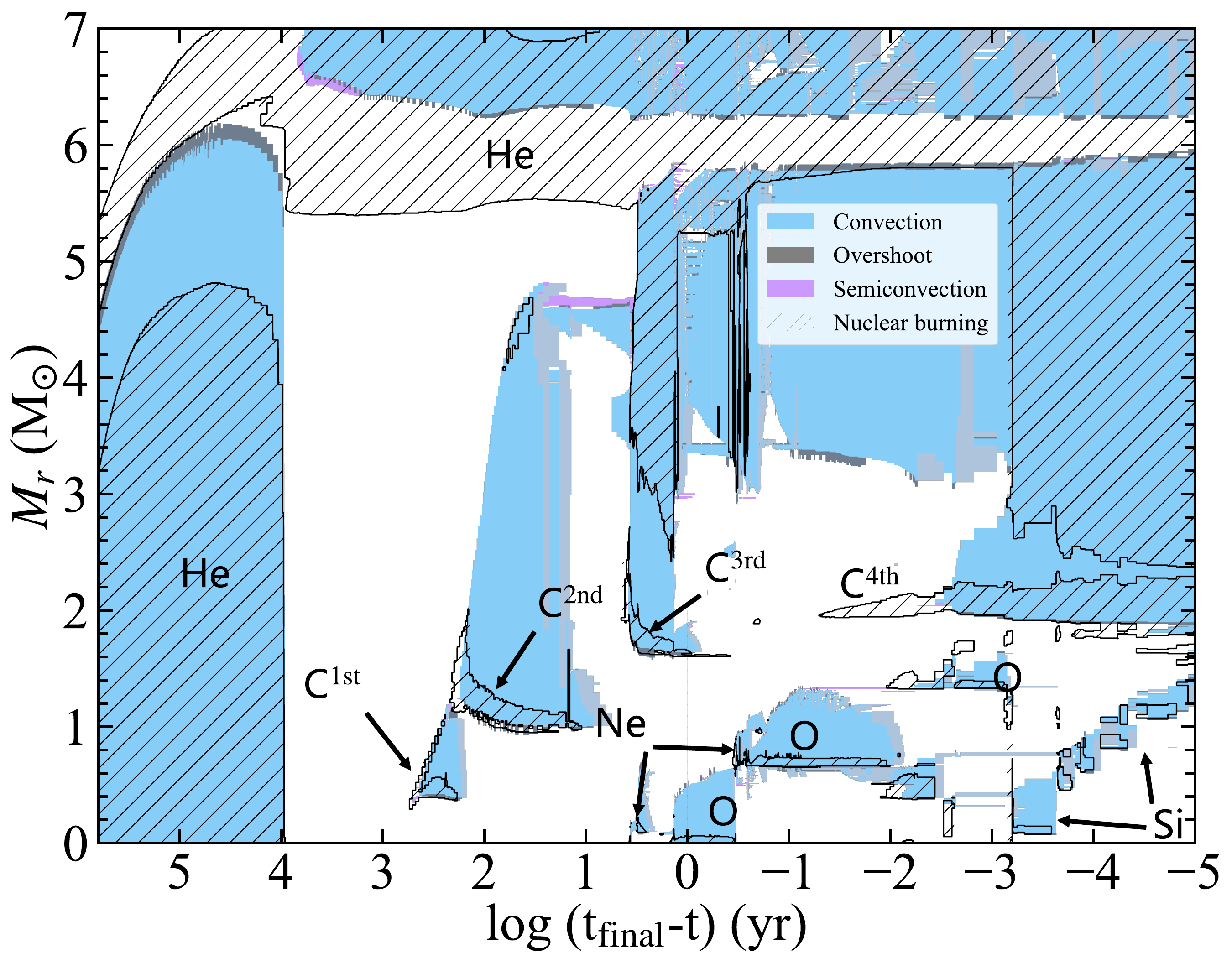}
\end{minipage}%
\caption{Same as Figure~\ref{fig:28M_core}, but for stars with $M {\rm (ZAMS)} =$
25 M$_{\odot}$ for $\sigma_{C12\alpha}=0$ (top), $-1$ (median) and $-2$ (bottom).
\label{fig:25M_core}}
\end{figure}

\begin{figure*}[htbp]
\centering
\begin{minipage}[c]{0.85\textwidth}
\centerline{$M({\rm ZAMS})=25 $ M$_\odot$}
\includegraphics [width=150mm]{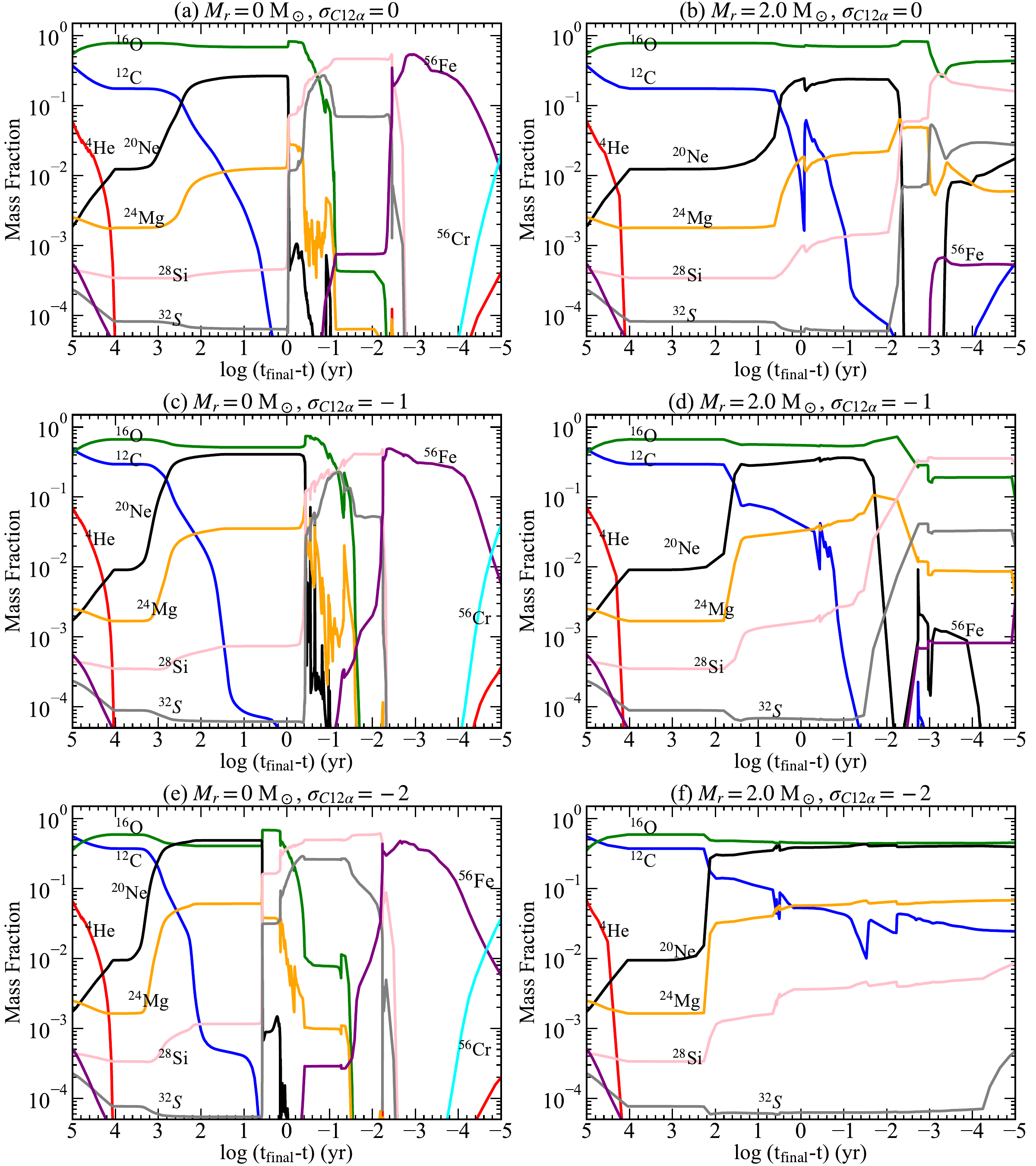}
\end{minipage}%
\caption{The time evolution of the mass fractions of some
isotopes for for 25 M$_{\odot}$ stars at the center (left), and $M_r=$2.0 M$_\odot$ (right)
from He burning through Fe core-collapse for $\sigma_{C12\alpha}=0$
(upper), $-1$ (center), and $-2$ (lower).
\label{fig:25M_central_Xcore}}
\end{figure*}

\begin{figure*}[htbp]
\centering
\begin{minipage}[c]{0.75\textwidth}
\centerline{$M({\rm ZAMS})=25 $ M$_\odot$}
\includegraphics [width=132mm]{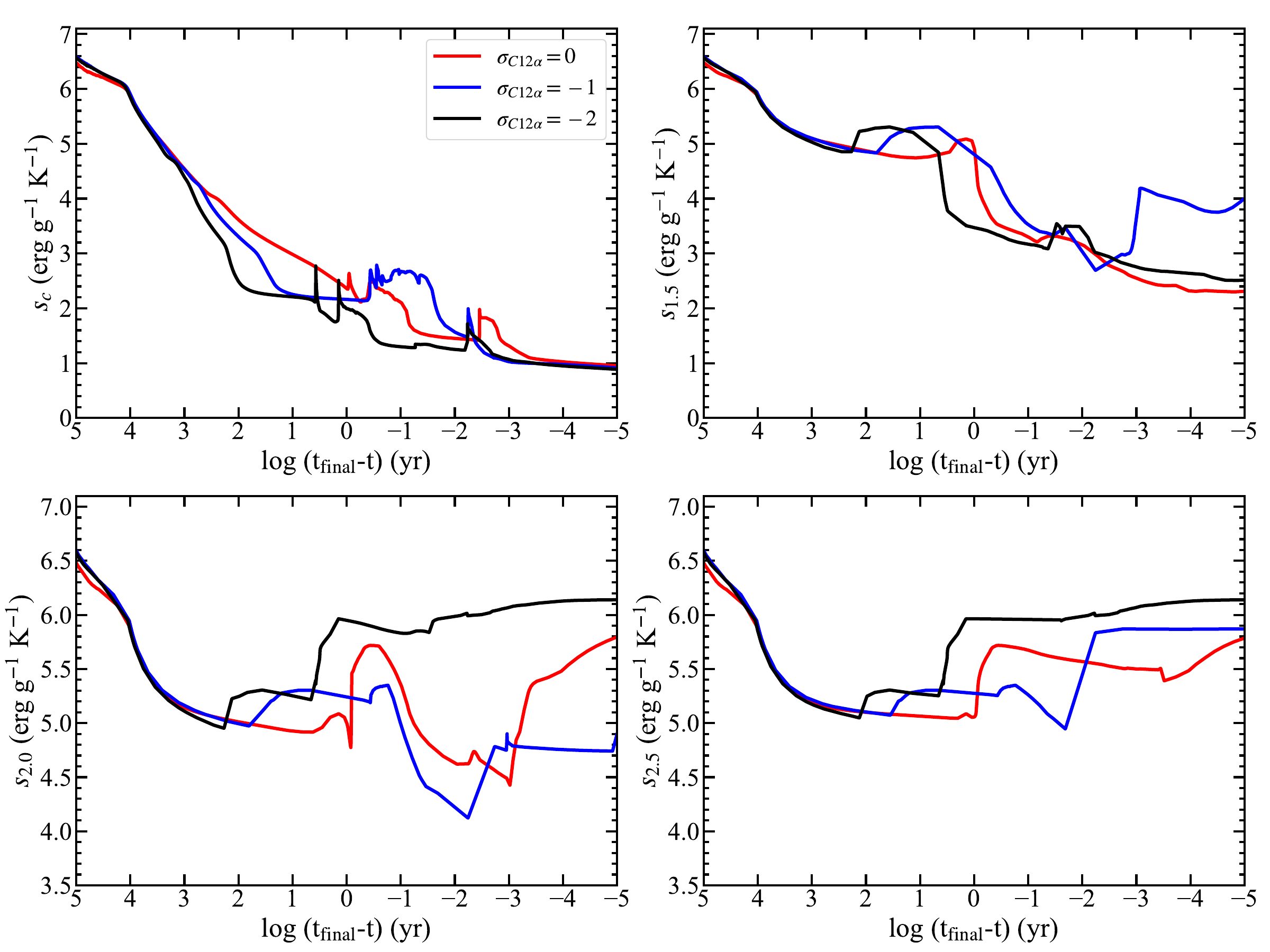}
\end{minipage}%
\caption{The time evolution of specific entropy at the center ($s_{\rm c}$),
$M_r=$ 1.5 M$_\odot$ ($s_{1.5}$), $M_r=$ 2.0 M$_\odot$
($s_{2.0}$) and $M_r=$ 2.5 M$_\odot$ ($s_{2.5}$) of stars
with $M {\rm (ZAMS)}$ = 25 M$_{\odot}$ for $\sigma_{C12\alpha}=$ 0
$-$1 and $-$2.
\label{fig:25M_entr}}
\end{figure*}

\begin{figure*}[htbp]
\centering
\begin{minipage}[c]{0.75\textwidth}
\centerline{$M({\rm ZAMS})=25 $ M$_\odot$}
\includegraphics [width=132mm]{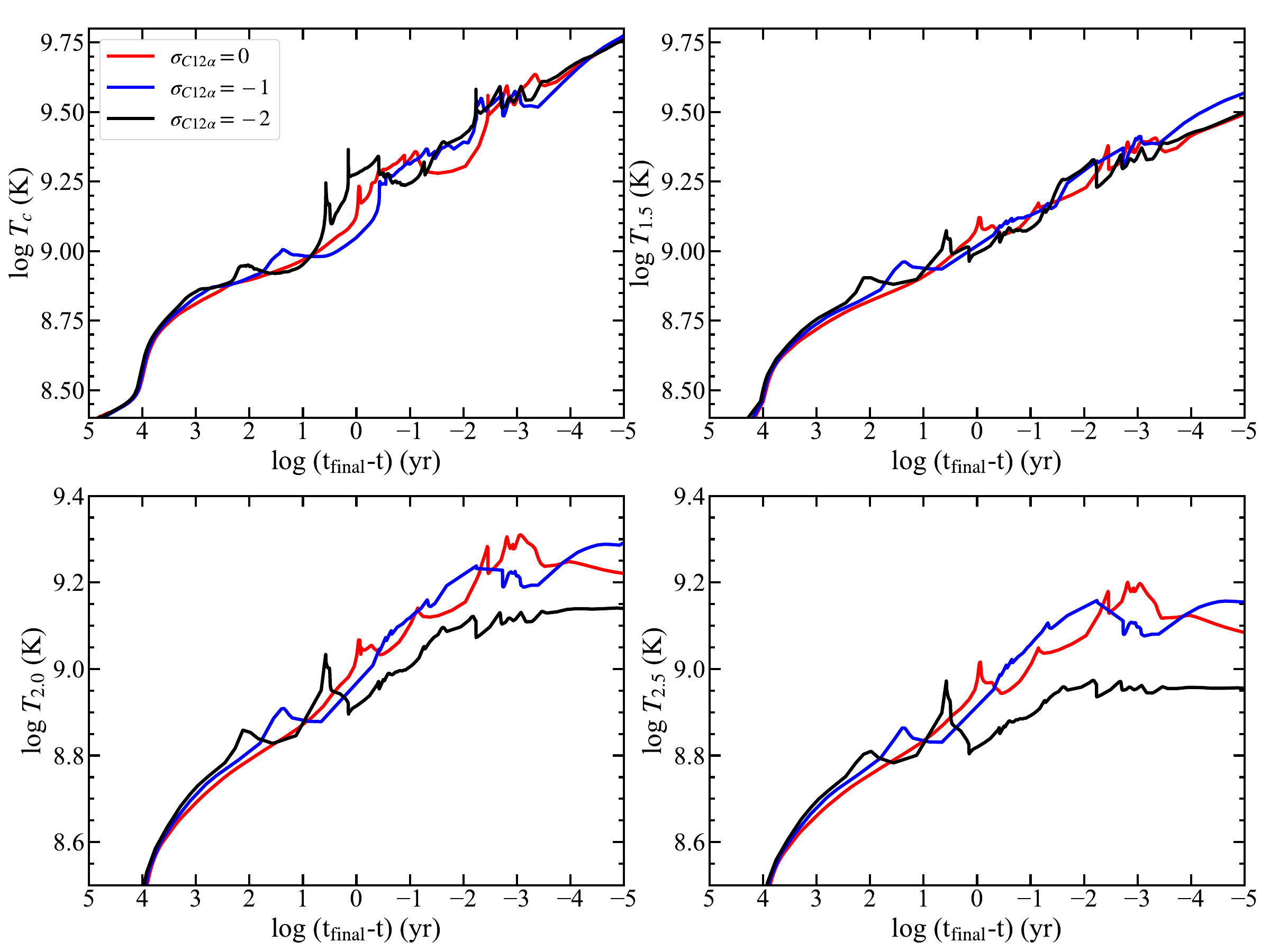}
\end{minipage}%
\caption{The time evolution of temperature at the center ($T_{\rm c}$),
$M_r=$ 1.5 M$_\odot$ ($T_{1.5}$), $M_r=$ 2.0 M$_\odot$
($T_{2.0}$) and $M_r=$ 2.5 M$_\odot$ ($T_{2.5}$) of stars
with $M {\rm (ZAMS)}$ = 25 M$_{\odot}$ for $\sigma_{C12\alpha}=$ 0
$-$1 and $-$2.
\label{fig:25M_temp}}
\end{figure*}

Figure \ref{fig:28M_logp_rho} (left) shows distributions of the log
$\rho$ and $s$ at the beginning of the collapse ($\tau=0$ yr:
$t=t_{\rm f}$), i.e., at log $T_{\rm c}$ (K) = 10.0, where the dashed
lines and solid lines represent $\sigma_{C12\alpha}=0$ and $-1$,
respectively. The green lines and black lines show the
outer edges of the Si core and Fe core, respectively,
which are defined as the locations where the energy generation rates
of shell O burning and shell Si burning is the highest.
Similar distributions of log $P$ and $Y_{\rm e}$ are shown in
Figure \ref{fig:28M_logp_rho} (right).

It is seen that the gradients of log $\rho$ and log $P$ with
respect to $M_r$ are much steeper for $\sigma_{C12\alpha} =-1$ than
$\sigma_{C12\alpha}=0$.  Also, log $\rho_{\rm c}$ is higher and
$s_{\rm c}$ is lower for $\sigma_{C12\alpha} =-1$ than 0 because of
smaller $M(V/U_{\rm max})$ (i.e., smaller $M_{\rm eff}$) for
$\sigma_{C12\alpha} =-1$ as discussed in the subsection
\ref{sec:uv_curve} on the $U-V$ curves based on Equations~\ref{equ:s2}
and \ref{equ:trho}.

The steepest gradient of log $P$ with respect to $M_r$ corresponds to
the maximum log ($V/U$) (Figure~\ref{fig:28M_uv_tfinal}) and is marked
by cyan ($\sigma_{C12\alpha} =0$) and black ($\sigma_{C12\alpha} =-1$)
filled circles. The corresponding steepest points in log $\rho$
(filled circles) and $s$ (open circles) are also indicated.  We note
that $s$ at $V/U_{\rm max}$ is close to 4 but not exactly 4 (see also
Figure \ref{fig:compact_sigma} in the later section). 
However, $M_4$ is almost identical
to $M(V/U_{\rm max})$ because of the extremely sharp gradient of $s$
due to entropy production by O-shell burning.

The difference in the core structure also appears in the evolution of
the compactness parameter $\xi_{2.5}$ after $\tau >$ $10^{0.2}$ yr
(Figure \ref{fig:28M_compact}). For $\sigma_{C12\alpha} =0$ (red),
$\xi_{2.5}$ continues to increase to reach 0.28 at $\tau= -5$ yr.
For $\sigma_{C12\alpha} =-1$ (blue), on the other hand, $\xi_{2.5}$
decreases from 0.15 to 0.10 at $\tau >$ $10^{0.2}$ yr.
It is obvious that the core with $\sigma_{C12\alpha} =0$ (red) is much more
compact with larger $\xi_{2.5}$ than $\sigma_{C12\alpha} =-1$ (blue).
The difference in $\xi_{2.5}$ stems exactly from the difference in the evolution
of radius, where $r_{2.5}$ stays almost constant as shown in Figure
\ref{fig:28M_radius}.

Such a difference in the density and pressure distributions and the
compactness parameter would lead to significant differences in the
core-collapse hydrodynamics and the explodability, as well as associated
explosive nucleosynthesis.

\subsection{Evolutions of 25 and 35 M$_\odot$ Stars}

For the 28 M$_\odot$ star, we have shown how the evolution of
$\xi_{2.5}$ is affected by $\sigma_{C12\alpha}$.  We also have investigated
how the effect of $\sigma_{C12\alpha}$ depends on $M{\rm (ZAMS)}$.
In Figure~\ref{fig:25M_compact}, we show that the
evolution of $\xi_{2.5}$ in the 25 M$_{\odot}$ stars for three cases
of $\sigma_{C12\alpha}=$ 0, $-$1 and $-2$.
We also show in Figure~\ref{fig:35M_compact_p10} the evolution of
$\xi_{2.5}$ in the 35 M$_{\odot}$ stars for the cases of
$\sigma_{C12\alpha}=$ 0 and 1.  This figure shows the effect of the
case of a very low C/O ratio originating from a large
$M {\rm (ZAMS)}$ and a large $^{12}$C$(\alpha, \gamma)^{16}$O rate.
In the following sections, we will discuss the evolutions of
the 25 and 35 M$_{\odot}$ stars to clarify how these differences
in $\xi_{2.5}$ appear.

\subsubsection{Chemical Evolution of $25$ M$_\odot$ Star}

The following figures show the chemical evolution of the 25 M$_{\odot}$
stars for $\sigma_{C12\alpha}=$0, $-$1 and $-$2.  Figure
\ref{fig:25M_core} shows Kippenhahn diagrams.  The time

$\tau = t_{\rm f} -t$ is measured from the final stage of $t_{\rm f}$
(denoted as $t_{\rm final}$ in the axes of Figures) where the central
temperature reaches log $T_{\rm c}$ (K) $=$ 10.0 as in the 28
M$_\odot$ star.

{\bf Reaction rate dependence}:

Figure~\ref{fig:25M_central_Xcore} shows the evolution of the mass
fractions of several isotopes at the center (left) and $M_r=$ 2.0
M$_\odot$ (right) from He burning through Fe core-collapse.  It is
seen that the chemical evolution at the center (a, c, e) does not much
depend on $\sigma_{C12\alpha}$, while the evolution at $M_r=$2.0
M$_\odot$ (b, d, f) is affected by $\sigma_{C12\alpha}$ owing to the
strength of C shell burning ($\tau=$ 10$^{0.6}$ - 10$^{-0.5}$ yr).
and O shell burning ($\tau=$ 10$^{-2}$ - 10$^{-3}$ yr).

When He is exhausted in the core ($\tau \sim$ 10$^{4.5}$),
$X$($^{12}$C) at $M_r=$ 2.0 M$_\odot$ is larger for smaller
$\sigma_{C12\alpha}$.  For $\sigma_{C12\alpha}=-2$, $X$($^{12}$C) is
large enough for the C burning convective shell to reach the overlying C-rich
layer.  Resulting convective mixing enhances $X$($^{12}$C) and make
C-shell burning active.  For $\sigma_{C12\alpha}=0$ and $-1$, such
C-mixing does not occur and $^{12}$C is exhausted at $M_r=$ 2.0
M$_\odot$. Then, C-shell burning becomes inactive.  Such inactivity of
C-shell burning for $\sigma_{C12\alpha}=-1$ in the 25 M$_{\odot}$ star
is different from the 28 M$_{\odot}$ star in
which C-shell burning is active until core-collapse for
$\sigma_{C12\alpha}=-1$.

\begin{figure*}[tbp]
\centering
\begin{minipage}[c]{0.8\textwidth}
\centerline{$M({\rm ZAMS})=25 $ M$_\odot$}
\includegraphics [width=140mm]{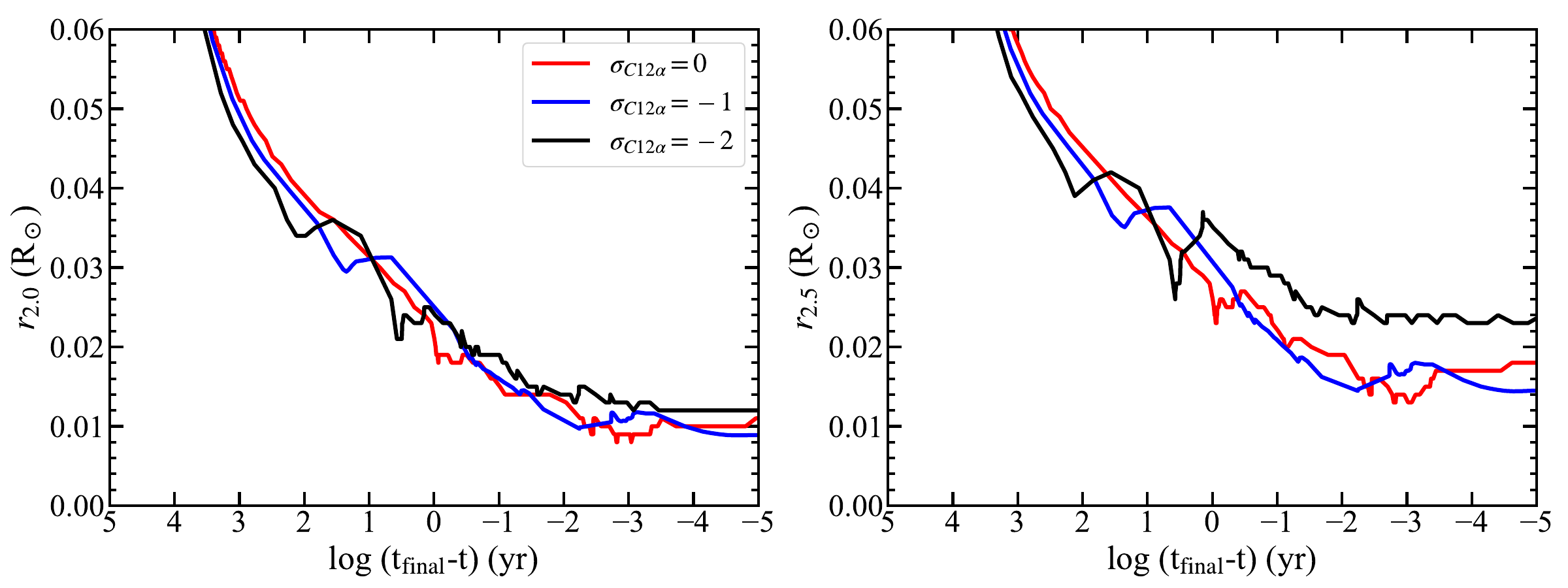}
\end{minipage}%
\caption{The time evolution of radius at $M_r=$ 2.0 M$_\odot$
  ($\rho_{2.0}$) and $M_r=$ 2.5 M$_\odot$ ($\rho_{2.5}$) of stars with
  $M {\rm (ZAMS)}$ = 25 M$_{\odot}$ for $\sigma_{C12\alpha}=$ 0, $-1$, and
  $-2$.
\label{fig:25M_radius}}
\end{figure*}

\begin{figure*}[htbp]
\centering
\begin{minipage}[c]{0.72\textwidth}
\centerline{$M({\rm ZAMS})=25 $ M$_\odot$}
\includegraphics [width=130mm]{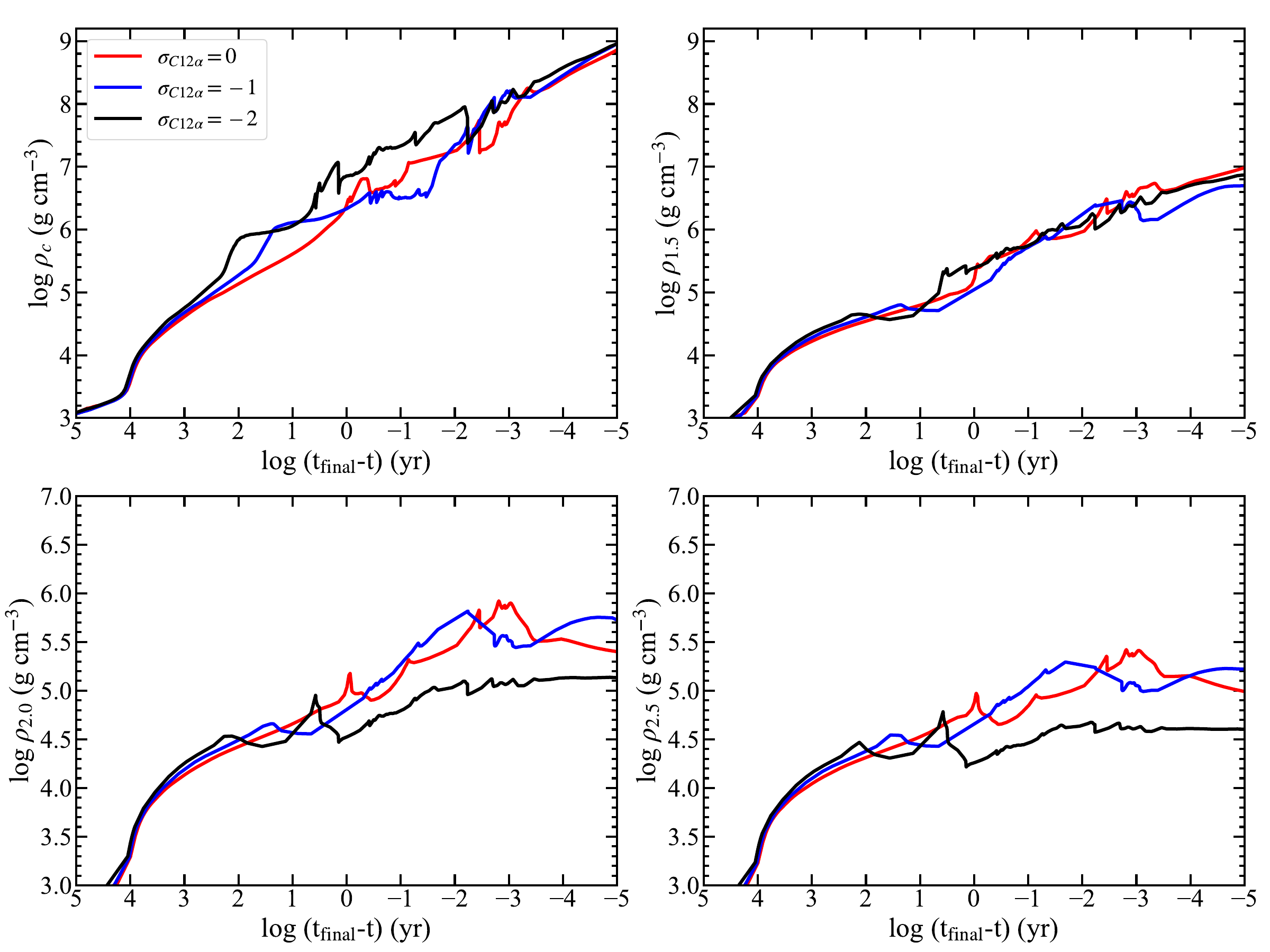}
\end{minipage}%
\caption{The time evolution of density at the center ($\rho_{\rm c}$),
$M_r=$ 1.5 M$_\odot$ ($\rho_{1.5}$), $M_r=$ 2.0 M$_\odot$
($\rho_{2.0}$) and $M_r=$ 2.5 M$_\odot$ ($\rho_{2.5}$) of stars
with $M {\rm (ZAMS)}$ = 25 M$_{\odot}$ for $\sigma_{C12\alpha}=$ 0
$-$1 and $-$2.
\label{fig:25M_rhos}}
\end{figure*}

The main difference in C-shell burning with $\sigma_{C12\alpha}=-2$ from
$\sigma_{C12\alpha}=-1$ and 0 is that, because of larger $X$($^{12}$C) after
He exhaustion in $\sigma_{C12\alpha}=-2$, C shell burning at $M_r \sim$
2.0 M$_\odot$ is active and convective during the evolution from central O burning to
Fe core-collapse (Figure \ref{fig:25M_core}).  Resultant heating of
shell burning causes the expansion of the outer layer as seen from
$r_{2.5}$ and $r_{2.0}$ (Figure \ref{fig:25M_radius}) and $\rho_{2.5}$
and $\rho_{2.0}$ (Figure \ref{fig:25M_rhos}).

Figures~\ref{fig:28M_compact} and \ref{fig:25M_compact}
show the mass-dependence of $\xi_{2.5}$ for $M {\rm (ZAMS)}$ = 25 and 28
M$_\odot$, respectively.  There appears to be a difference at
$\tau <$ 10$^{-3.5}$ yr for $\sigma_{C12\alpha}=0$.

This difference stems from {\sl O-shell burning}.  For the 25
M$_\odot$ star, convective mixing of fresh O from the outer layer
occurs at the shell near $M_r$ = 2.0 M$_\odot$ as seen from the almost
constant $X$(O) in Figure~\ref{fig:25M_central_Xcore}(b).  Then
O-shell burning continues to be active.  This is different from the 28
M$_\odot$ star with $\sigma_{C12\alpha}=0$ where O shell burning is
weak because $X$(O) $<$ 0.1 at $M_r=$ 2.0 M$_\odot$ around $\tau <
10^{-3}$ yr (Figure \ref{fig:28M_central_Xcore}(e)).  Such a difference
in O-shell burning leads to the difference in $\xi_{2.5}$ between the
25 and 28 M$_\odot$ stars even for the same $\sigma_{C12\alpha}=0$.

\begin{figure*}[htbp]
\centering
\centerline{$M({\rm ZAMS})=25$ M$_\odot$}
\begin{minipage}[c]{0.40\textwidth}
\includegraphics [width=72mm]{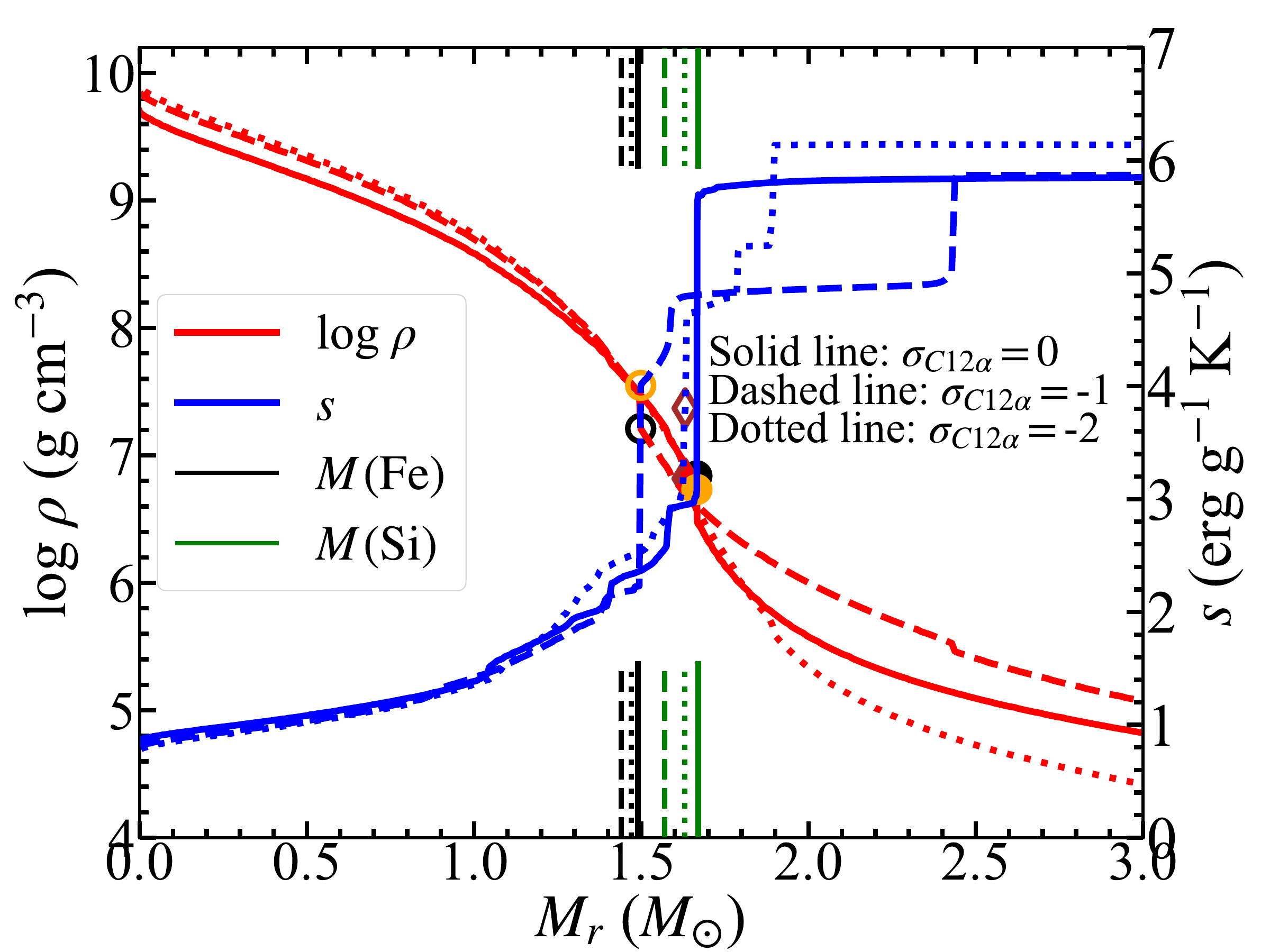}
\end{minipage}%
\begin{minipage}[c]{0.4\textwidth}
\includegraphics [width=72mm]{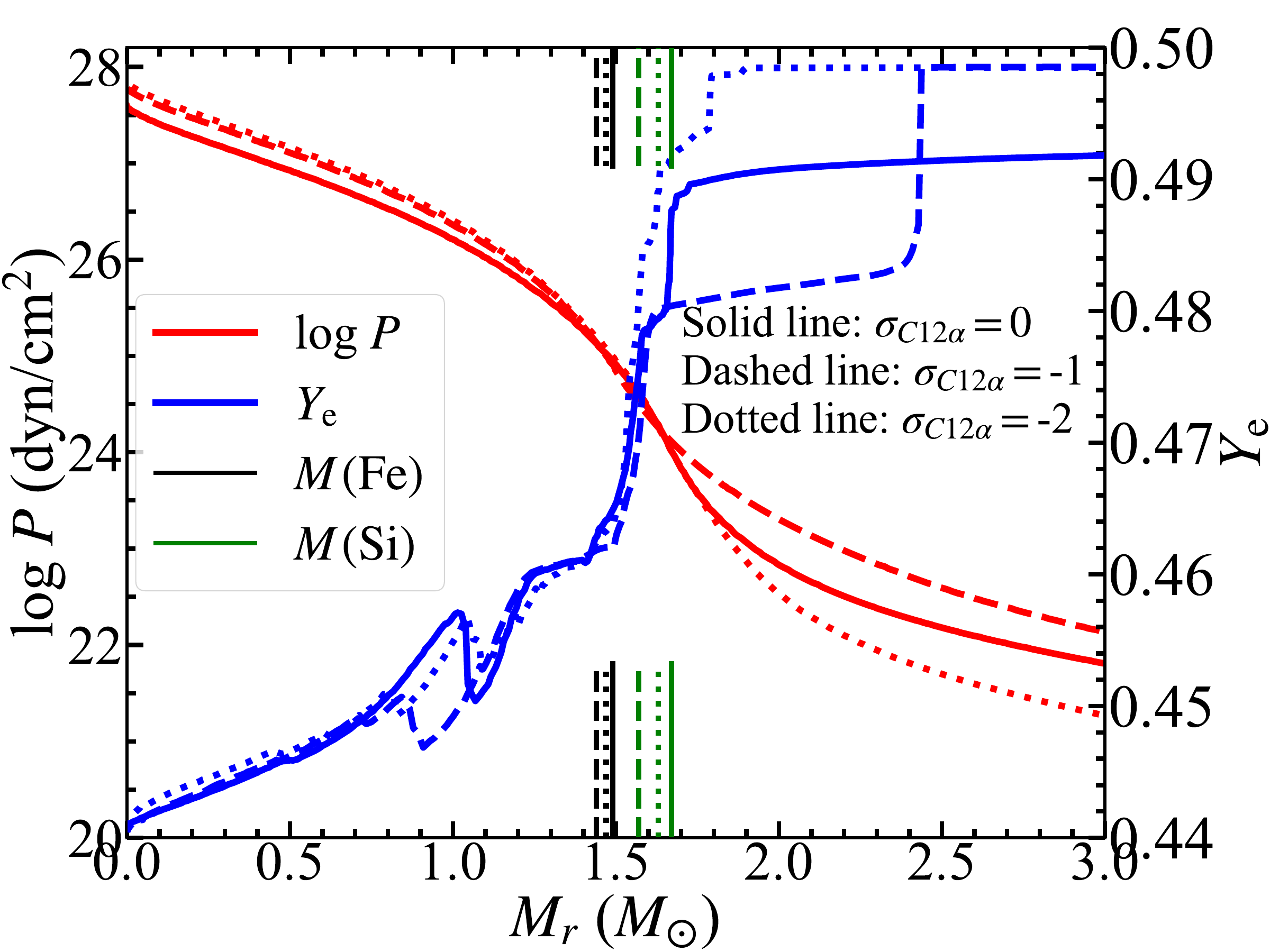}
\end{minipage}%
\caption{Same as Figure \ref{fig:28M_logp_rho}, but for $M ({\rm ZAMS})$ = 25 M$_{\odot}$
and $\sigma_{C12\alpha}=$ 0, $-$1 and $-$2.
\label{fig:25M_logp_rho}}
\end{figure*}

\begin{figure*}[htbp]
\centering
\begin{minipage}[c]{0.75\textwidth}
\centerline{$M({\rm ZAMS})=25$ M$_\odot$}
\includegraphics [width=132mm]{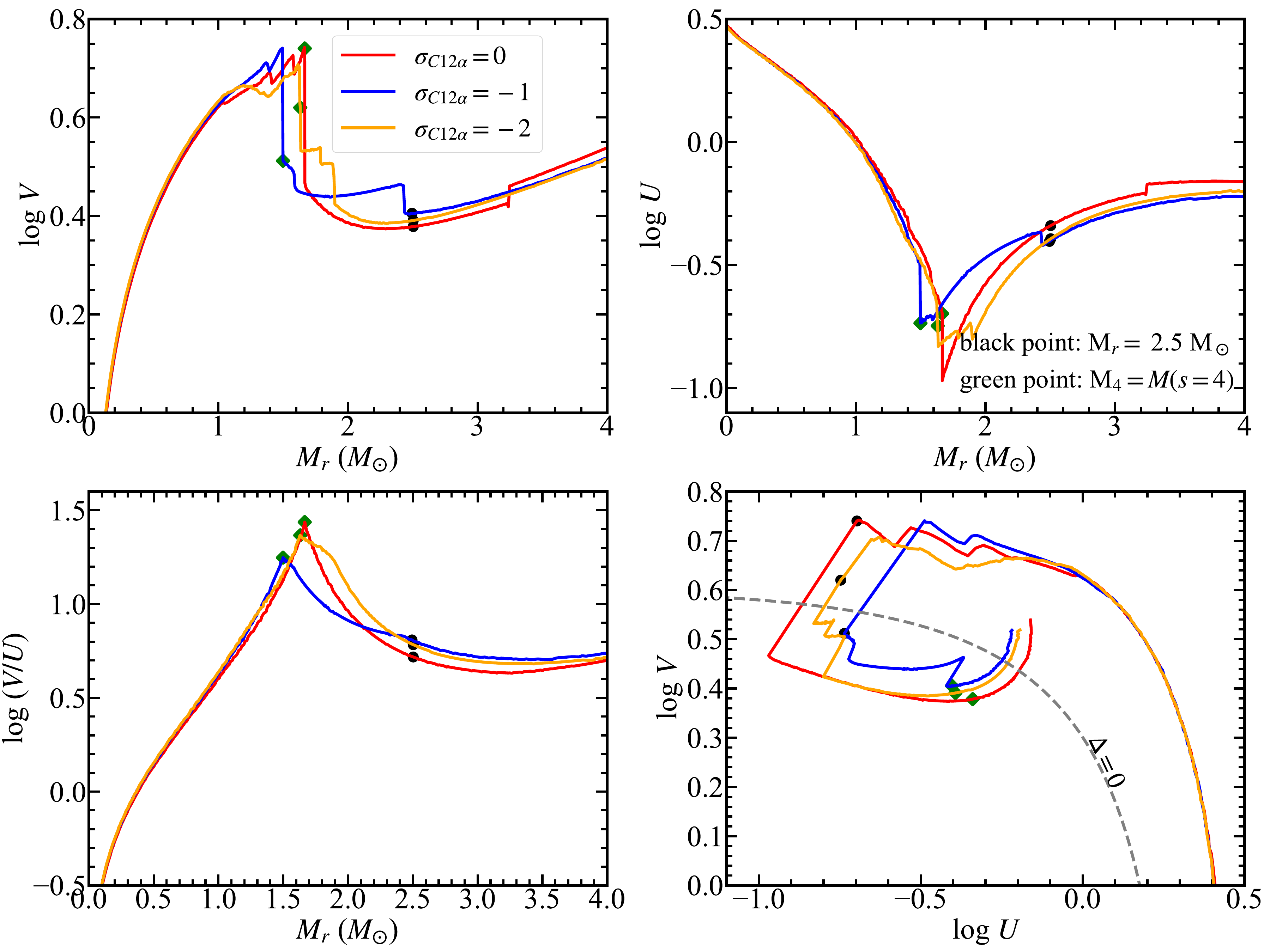}
\end{minipage}%
\caption{The $U-V$ curves of 25 M$_{\odot}$ star at $\tau=t_{\rm f}$
The orange points represent the place where $M_r = 2.5$ M$_{\odot}$,
while the green diamonds show the place where $M_r = M(s=4)$.
\label{fig:25M_uv_final}}
\end{figure*}

\begin{figure*}[htp]
\centering
\begin{minipage}[c]{0.75\textwidth}
\centerline{$M({\rm ZAMS})=25 $ M$_\odot$}
\includegraphics [width=132mm]{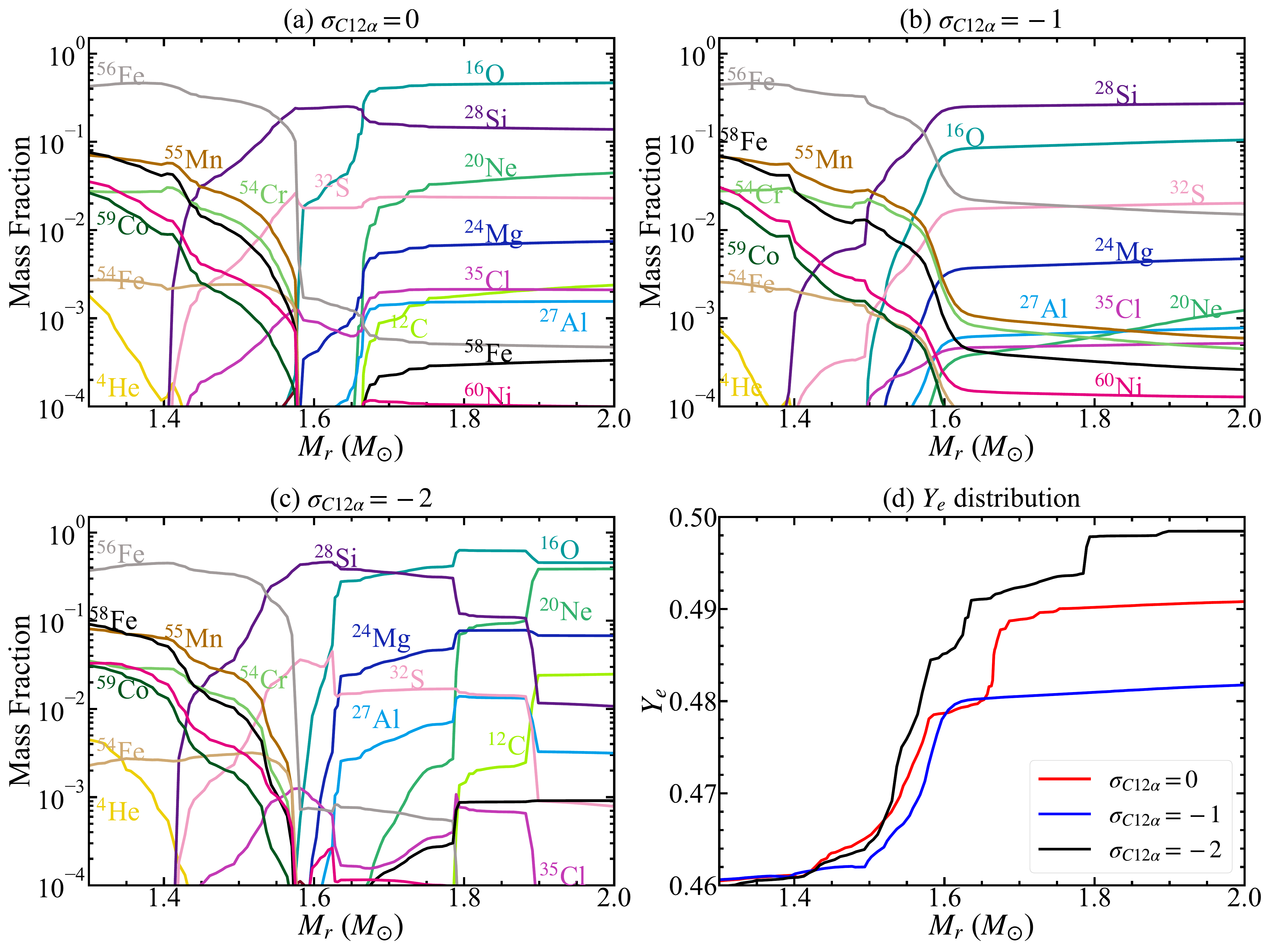}
\end{minipage}%
\caption{The abundance distributions in stars with $M({\rm ZAMS}) =$
25 M$_{\odot}$ for $\sigma_{C12\alpha}=$ 0 (a), $-$1 (b) and
$-$2 (c) at log $T_{\rm c}$ (K) $=10$ ($t=t_{\rm f}$).
Also, $Y_{\rm e}$ distribution is shown (d).
\label{fig:25M_abund}}
\end{figure*}

\subsubsection{Thermal and Dynamical Evolution of 25 M$_\odot$ Star} \label{sec:thermal}

{\bf Entropy}:

Figure~\ref{fig:25M_entr} shows the evolution of
specific entropy at the center ($s_{\rm c}$), $M_r=$ 1.5 M$_\odot$
($s_{1.5}$), $M_r=$ 2.0 M$_\odot$ ($s_{2.0}$), and $M_r=$ 2.5 M$_\odot$
($s_{2.5}$).

We should note that $s_{2.0}$ is enhanced around $\tau \sim 10^{0.6}$ yr
for $\sigma_{C12\alpha}=-2$ (black) owing to the mixing of C.
This is similar to the increase in $s_{2.0}$ in the 28 M$_\odot$ star
with $\sigma_{C12\alpha}=-1$.

Note also that the decrease in $\xi_{2.5}$ during $\tau \sim$ $10^{-2.3}$ -
$10^{-3}$ yr for $\sigma_{C12\alpha}=-1$ (blue) and after $\tau \sim$ $10^{-3}$
for $\sigma_{C12\alpha}=0$ (red) are originated from the increase in
$s_{2.0}$ owing to O-mixing (see Figure \ref{fig:25M_central_Xcore} (b) and (d)).
Such an increase is also seen in the 28 M$_\odot$ star in $s_{2.0}$ after
$\tau \sim 10^{-2.8}$ with $\sigma_{C12\alpha}=0$ and $s_{1.5}$ after
$\tau \sim 10^{-3}$ with $\sigma_{C12\alpha}=-1$.

{\bf Temperature}:

Figure~\ref{fig:25M_temp} shows the evolution of
temperature at the center ($T_{\rm c}$), $M_r=$ 1.5 M$_\odot$ ($T_{1.5}$),
$M_r=$ 2.0 M$_\odot$ ($T_{2.0}$) and $M_r=$ 2.5 M$_\odot$ ($T_{2.5}$).

Note that $T_{2.0}$ reaches the O-burning temperature for
$\sigma_{C12\alpha}=0$ (red) and $-1$ (blue).  For
$\sigma_{C12\alpha}=-2$, $T_{2.0}$ and $T_{2.5}$ increase slowly
staying at the C-burning temperature.

{\bf Radius}:

Figure~\ref{fig:25M_radius} shows the evolution of radius at $M_r=$
2.0 M$_\odot$ ($r_{2.0}$) and $M_r=$ 2.5 M$_\odot$ ($r_{2.5}$).  These
radii generally decrease during the core contraction.  However,
$r_{2.5}$ increases for $\sigma_{C12\alpha}= -2$ at $\tau \sim
10^{0.4} - 10^{0}$ yr because of the heating effect of C shell
burning.  This is the reason that the compactness parameter of
$\sigma_{C12\alpha}= -2$ decreases during that period
(Figure~\ref{fig:28M_compact} (right)).

{\bf Density}:

Figure~\ref{fig:25M_rhos} shows the evolution of
densities at the center ($\rho_{\rm c}$), $M_r=$ 1.5 M$_\odot$
($\rho_{1.5}$), $M_r=$ 2.0 M$_\odot$ ($\rho_{2.0}$), and $M_r=$ 2.5
M$_\odot$ ($\rho_{2.5}$).  These figures show that the evolution of
the density at each layer of the 25 M$_\odot$ stars is not so different
between $\sigma_{C12\alpha}=$ 0 and $-$1, in contrast to the large
difference for the 28 M$_{\odot}$ star.  As a result, the density and
entropy distributions at $t=t_{\rm f}$ are not so different between
$\sigma_{C12\alpha}= 0$ and $-1$ (Figure \ref{fig:25M_logp_rho}).

For $\sigma_{C12\alpha}= -2$, on the other hand, the density evolution
and the final distributions of the density and entropy are different,
being similar to the 28 M$_\odot$ stars with $\sigma_{C12\alpha}= -1$
as seen in Figures \ref{fig:25M_logp_rho}.

\subsubsection {Presupernova Structure and U-V Curves of 25 M$_{\odot}$ Stars}

Behaviors of shell burning at $\tau=t_{\rm f}$ are shown in the
$U-V$ curves in Figure \ref{fig:25M_uv_final}. For $\sigma_{C12\alpha}=0$ 
and -2, maximum of $V/U$-curves appears near $M_r$ = 1.7
M$_\odot$, which is the bottom of the O burning shell.  On the other hand, 
maximum for $\sigma_{C12\alpha}=-1$ appears between the O-burning shell
and the Si-burning shell ($M_r$ = 1.5 M$_\odot$).  Similar to the 28
M$_\odot$ models, these peaks in $V/U$-curves are located 
at $M_4$ and correspond to the steepest gradient of log $P$
with respect to $M_r$ (see Fig \ref{fig:25M_logp_rho}). The maximum 
values of log $V/U$ in 25 M$_\odot$ models are 1.44, 1.25, and 1.37
for $\sigma_{C12\alpha}=$0, $-1$, and $-2$, respectively.
As will be discussed later, these values of the steepness are marginal
for the criterion of the explosion vs. collapse.

\begin{figure*}[hbp]
\centering
\begin{minipage}[c]{0.42\textwidth}
\centerline{$M({\rm ZAMS})=35$ M$_\odot$}
\includegraphics [width=75mm]{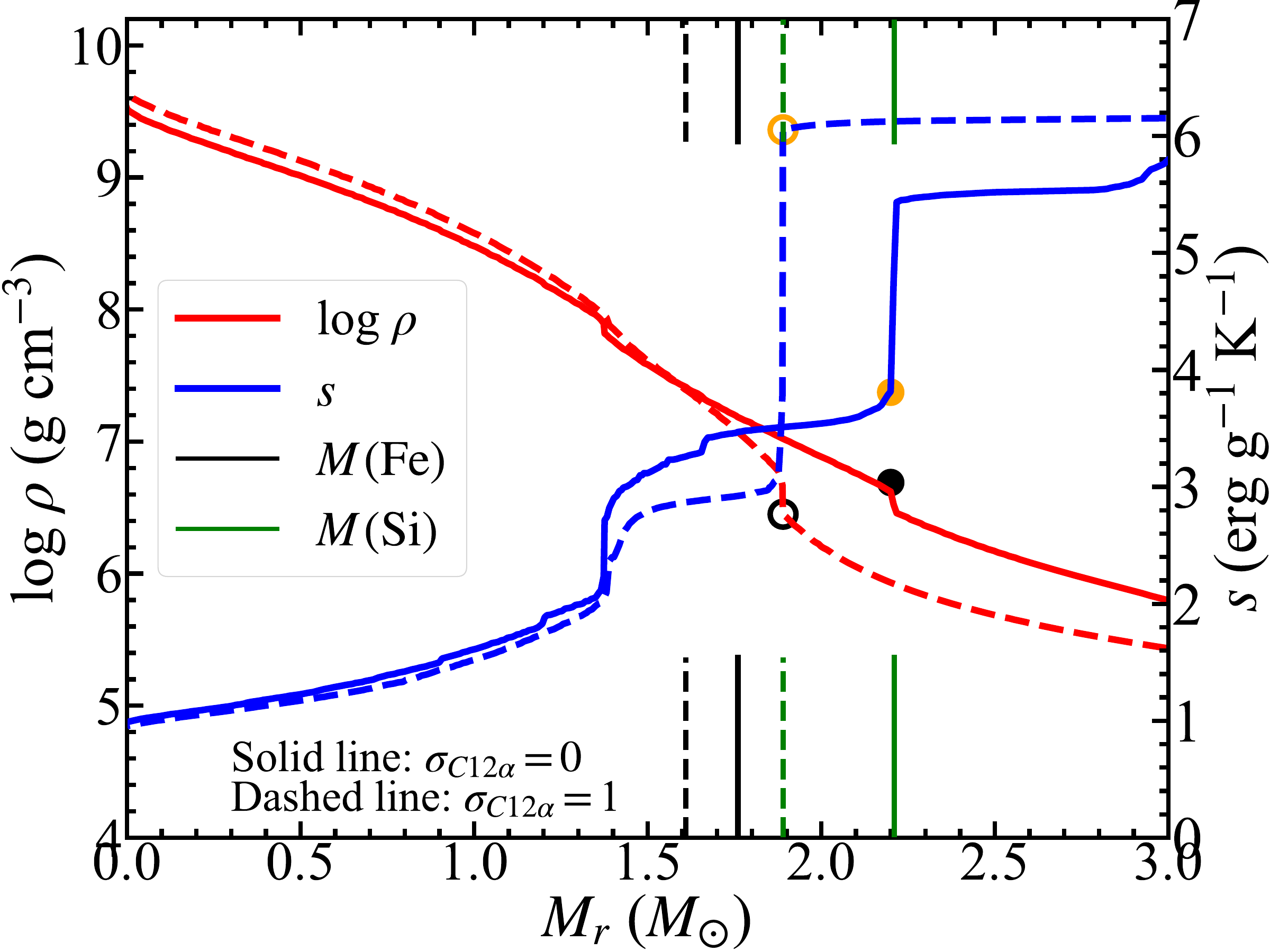}
\end{minipage}%
\caption{Same as Figure \ref{fig:28M_logp_rho} (left), but for stars with
$M {\rm (ZAMS)}$ = 35 M$_{\odot}$ and $\sigma_{C12\alpha}=$ 1 and 0.
\label{fig:rho_s_mr35}}
\end{figure*}

\subsubsection{Evolution of 35 M$_\odot$ Stars}

\begin{figure*}[htbp]
\centering
\begin{minipage}[c]{0.42\textwidth}
\centerline{$M({\rm ZAMS})=35$ M$_\odot$, $\sigma_{C12\alpha}=0$}
\includegraphics [width=75mm]{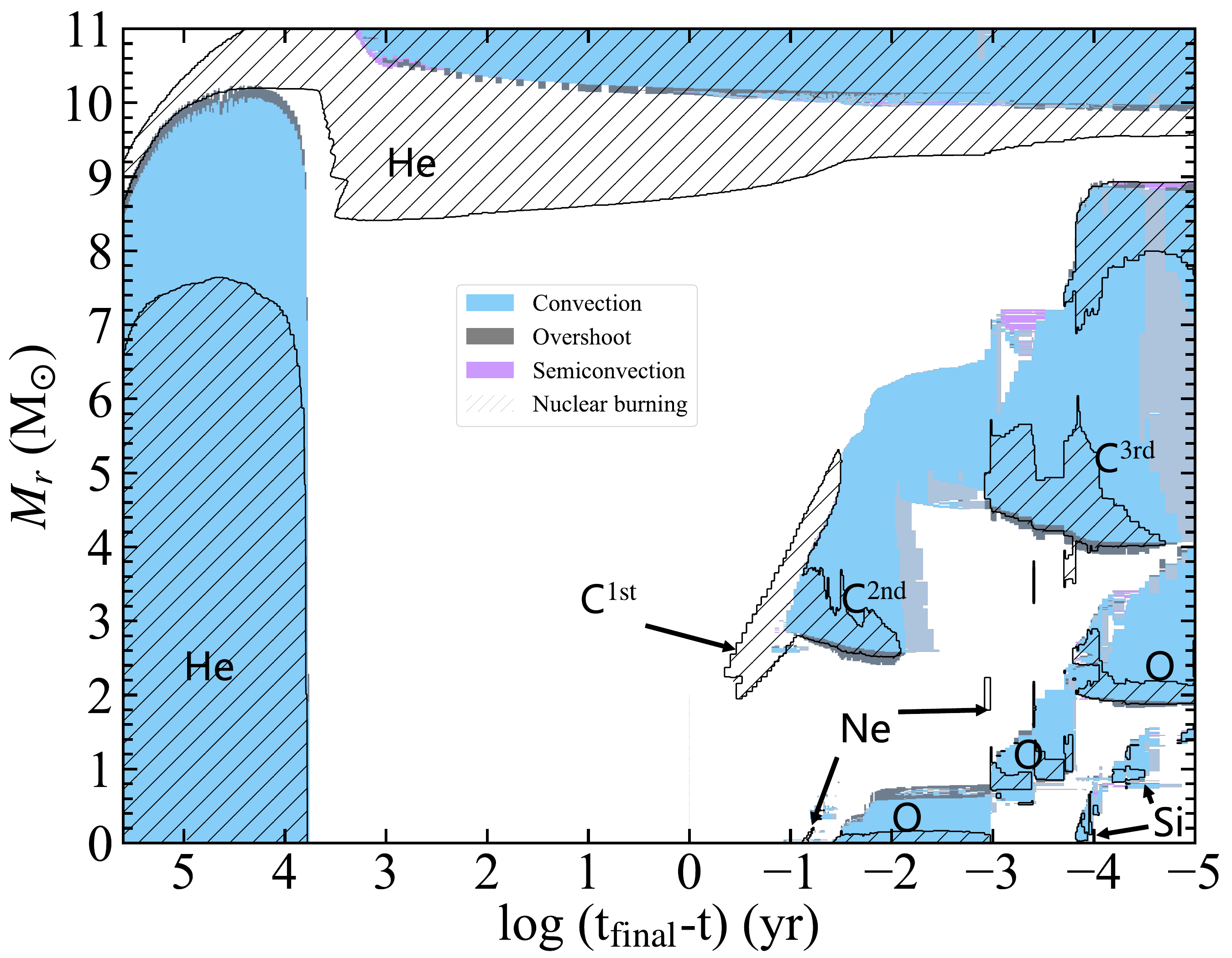}
\end{minipage}%
\begin{minipage}[c]{0.42\textwidth}
\centerline{$M({\rm ZAMS})=35$ M$_\odot$, $\sigma_{C12\alpha}=1$}
\includegraphics [width=75mm]{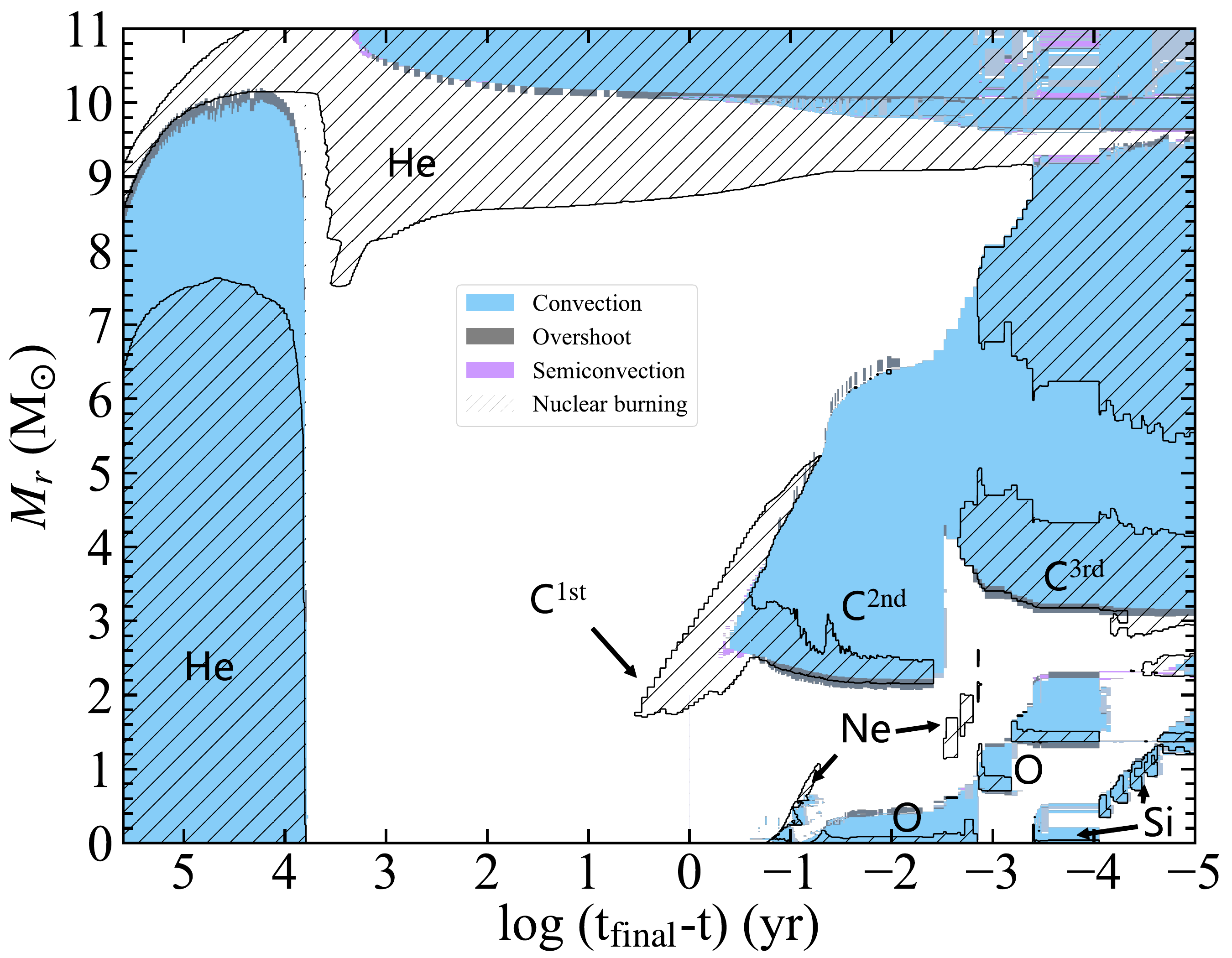}
\end{minipage}%
\caption{Same as Figure \ref{fig:28M_core} but for stars with
$M {\rm (ZAMS)}$ = 35 M$_{\odot}$ and $\sigma_{C12\alpha}=$ 1 (left)
and 0 (right).
\label{fig:35M_core}}
\end{figure*}

\begin{figure*}[htbp]
\centering
\begin{minipage}[c]{0.88\textwidth}
\centerline{$M({\rm ZAMS})=35 $ M$_\odot$}
\includegraphics [width=154mm]{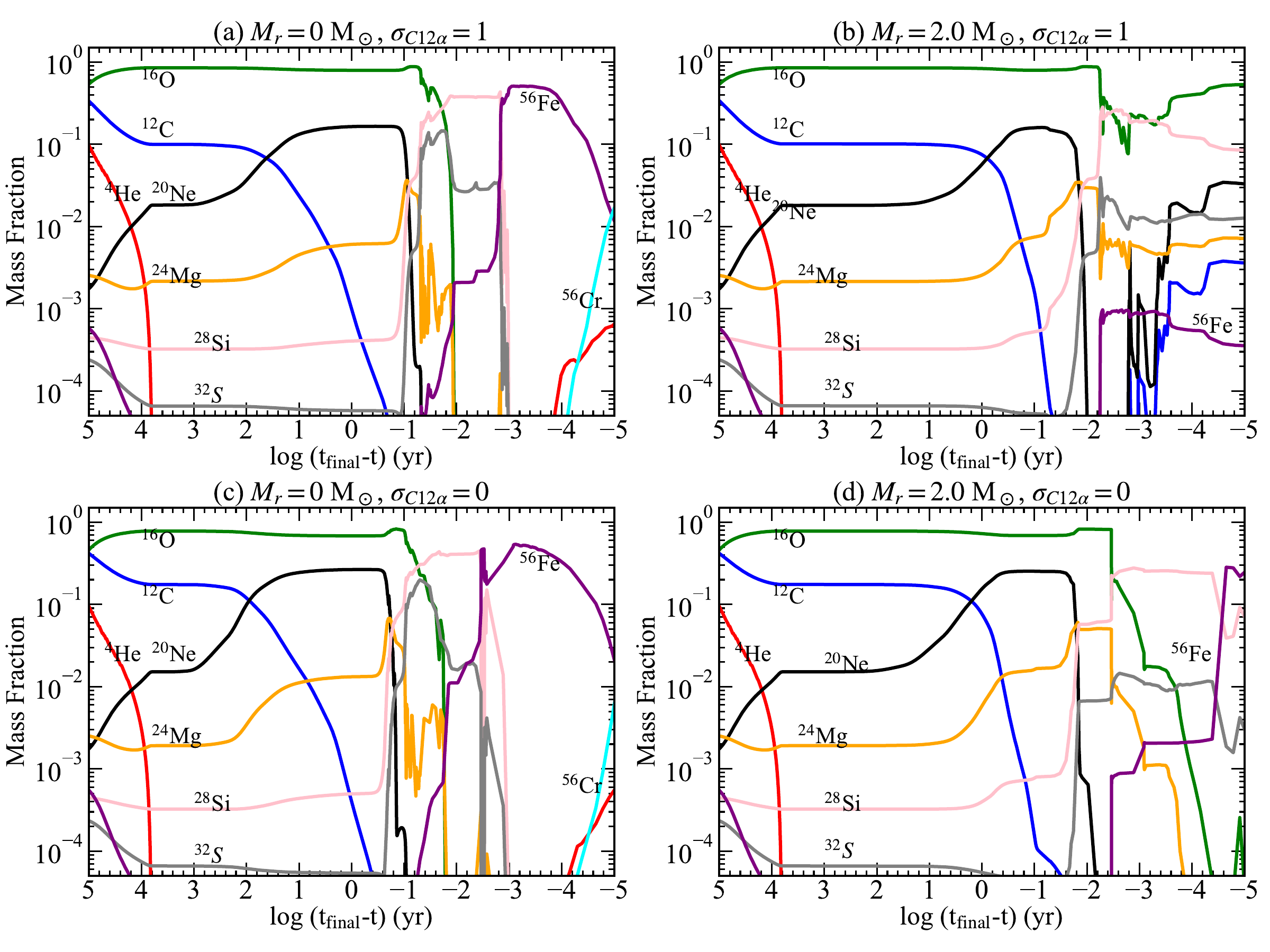}
\end{minipage}%
\caption{The time evolution of the mass fractions of some isotopes
at the center and $M_r=$2.0 M$_\odot$ for a star with 
$M ({\rm ZAMS})$ = 35 M$_{\odot}$ for $\sigma_{C12\alpha}=$ 1 (top) and 0 (bottom).
\label{fig:35M_Xcore_p10}}
\end{figure*}

\begin{figure*}[htbp]
\centering
\begin{minipage}[c]{0.75\textwidth}
\centerline{$M({\rm ZAMS})=35 $ M$_\odot$}
\includegraphics [width=132mm]{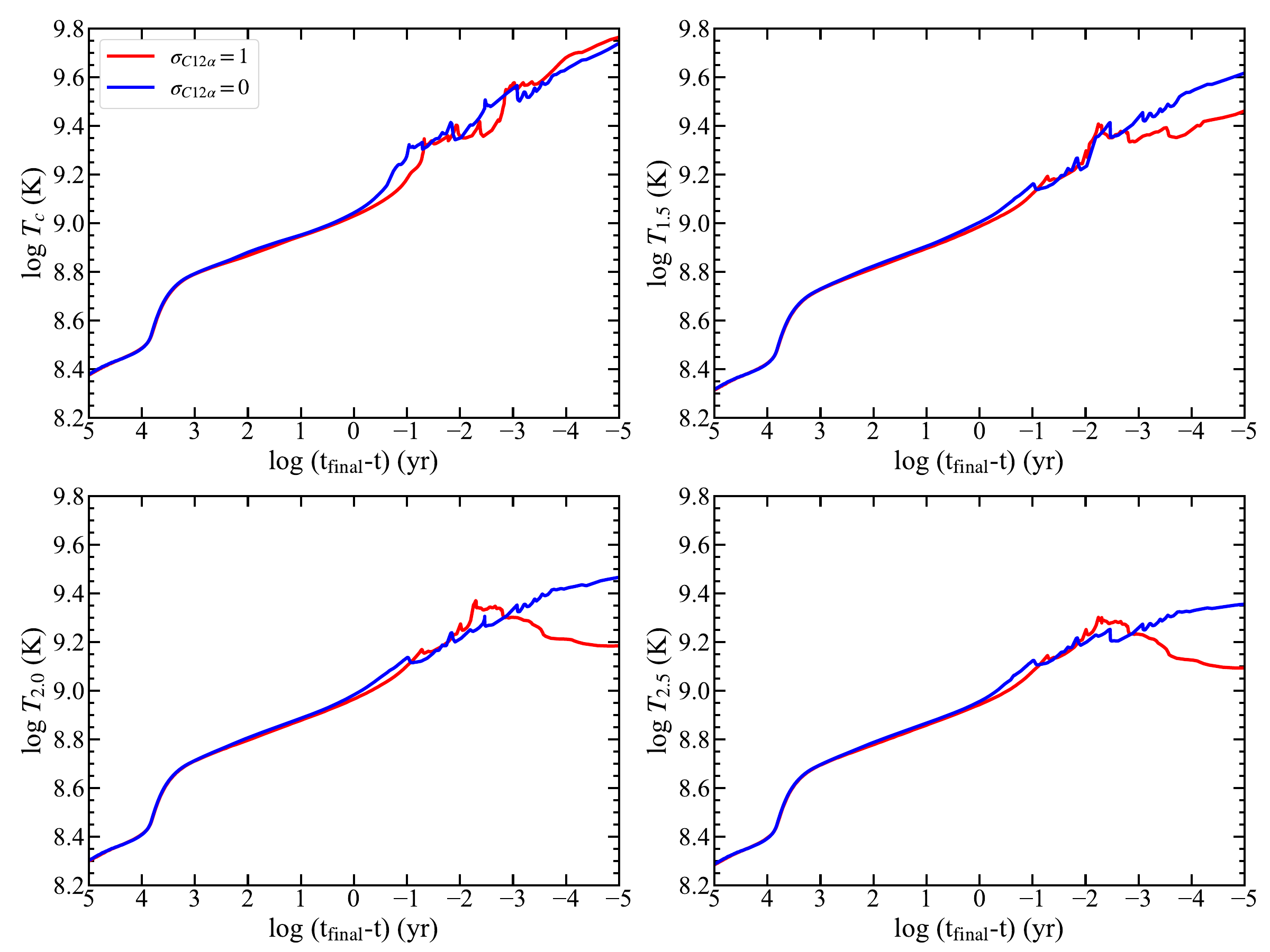}
\end{minipage}%
\caption{The time evolution of temperatures at the center ($T_{\rm c}$),
  $M_r=$ 1.5 M$_\odot$ ($T_{1.5}$), $M_r=$ 2.0 M$_\odot$ ($T_{2.0}$)
  and $M_r=$ 2.5 M$_\odot$ ($T_{2.5}$) of stars with $M {\rm (ZAMS)}$ = 35
  M$_{\odot}$ for $\sigma_{C12\alpha}=$ 1 and 0.
\label{fig:35M_temp_p10}}
\end{figure*}

\begin{figure*}[htbp]
\centering
\begin{minipage}[c]{0.75\textwidth}
\centerline{$M({\rm ZAMS})=35 $ M$_\odot$}
\includegraphics [width=132mm]{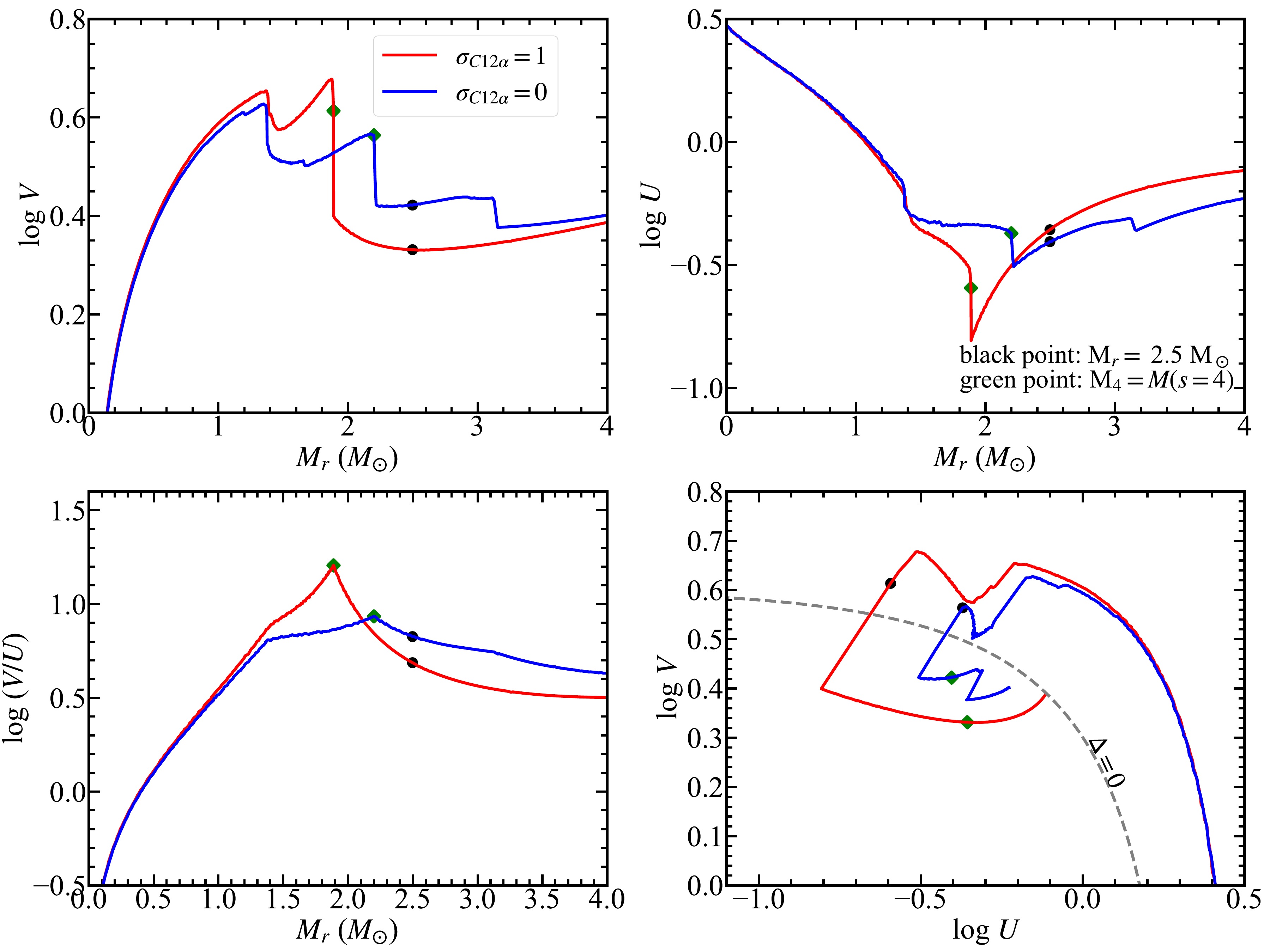}
\end{minipage}%
\caption{The $U-V$ curves of 35 M$_{\odot}$ star at $\tau=t_{\rm f}$
The orange points represent the place where $M_r = 2.5$ M$_{\odot}$,
while the green diamonds show the place where $M_r = M(s=4)$.
\label{fig:35M_uv_final}}
\end{figure*}

For 35 M$_\odot$ stars, we investigate the case of
$\sigma_{C12\alpha}=1$, i.e., the effect of the low C/O ratio.
Figure~\ref{fig:35M_compact_p10} shows that $\xi_{2.5}$ increases up
to 0.4 for $\sigma_{C12\alpha}=0$ but decreases down to 0.28 for
$\sigma_{C12\alpha}=1$ from $\tau =10^{-3}$ to 10$^{-5}$ yr.

As shown in Figure \ref{fig:35M_core} of Kippenhahn diagrams, 
the behavior of C shell burning is similar between these cases.
Both models ignite C shell burning at $M_r \sim$ 2 M$_\odot$
and the convection extends to $M_r \sim$ 6.5 M$_\odot$.
After $\tau \sim 10^{-2}$ yr, the C burning shell moves to $M_r \sim$
4.7 M$_\odot$ (4.0 M$_\odot$) and forms an oxygen core of $M$(O) = 4.0
M$_\odot$ (3.2 M$_\odot$) for $\sigma_{C12\alpha}=1$ 
($\sigma_{C12\alpha}=0$).

Owing to the low $X$($^{12}$C) in both models, the heating effect of C
shell burning is weak.  In Figure \ref{fig:35M_compact_p10},
evolutions of $\xi_{2.5}$ do not show a large difference between the
two models up to $\tau \sim 10^{-3}$ yr.

However, $\xi_{2.5}$ decrease after $\tau = 10^{-3}$ yr for
$\sigma_{C12\alpha}=1$.  This is due to the heating effect of O shell
burning, which extends from $M_r$= 2.0 to 4.0 M$_\odot$.
Figure~\ref{fig:35M_Xcore_p10} (b) shows that $X$($^{16}$O) at $M_r=$
2 M$_\odot$ decreases from $\tau=10^{-2.3}$ yr to $10^{-2.8}$ yr due
to O shell burning and increases after $\tau=10^{-2.8}$ yr due to
mixing of oxygen from outer layers.  After O shell burning is ignited,
the layers above $M_r=$ 2.0 M$_\odot$ begin to expand as seen from the
time evolution of log $T$ at $M_r=$ 2.0 M$_\odot$ and $M_r=$ 2.5
M$_\odot$ in Figure~\ref{fig:35M_temp_p10}.

The core structure (log $\rho$ and $s$) at the final stage for these
two models are compared in Figure~\ref{fig:rho_s_mr35}. The steepest
gradient of density and specific entropy occurs around $s =$ 3-4 erg
g$^{-1}$ K$^{-1}$ just below the O burning shell.
We conclude that the heating effect of O shell burning is important
when and where $X$($^{12}$C) is low enough.

\section{Dependence of Explodability on $M{\rm (ZAMS)}$
and $^{12}$C$(\alpha, \gamma)^{16}$O Rate} \label{sec:xi_rate}

We calculate the evolution of massive stars for various combinations
of $M({\rm ZAMS})$ and $\sigma_{C12\alpha}$ and obtain the
compactness parameter $\xi_{2.5}$ (Equation \ref{equ:compact}) at the final
stage of evolution ($t=t_{\rm f}$) when log $T_{\rm c}$ (K) = 10.0
(see Tables \ref{sec:exp_info}).  $\xi_{2.5}$ has been suggested to be
useful to evaluate the explodability of massive stars
\citep{2012ApJ...757...69U, 2014Sukhbold, 2016MNRAS.460..742M, 2016ApJ...818..124E, 
2018ApJ...860...93S, 2020ApJ...890...51E}.

\begin{figure}[htb]
\centering
\begin{minipage}[c]{0.45\textwidth}
\includegraphics [width=80mm]{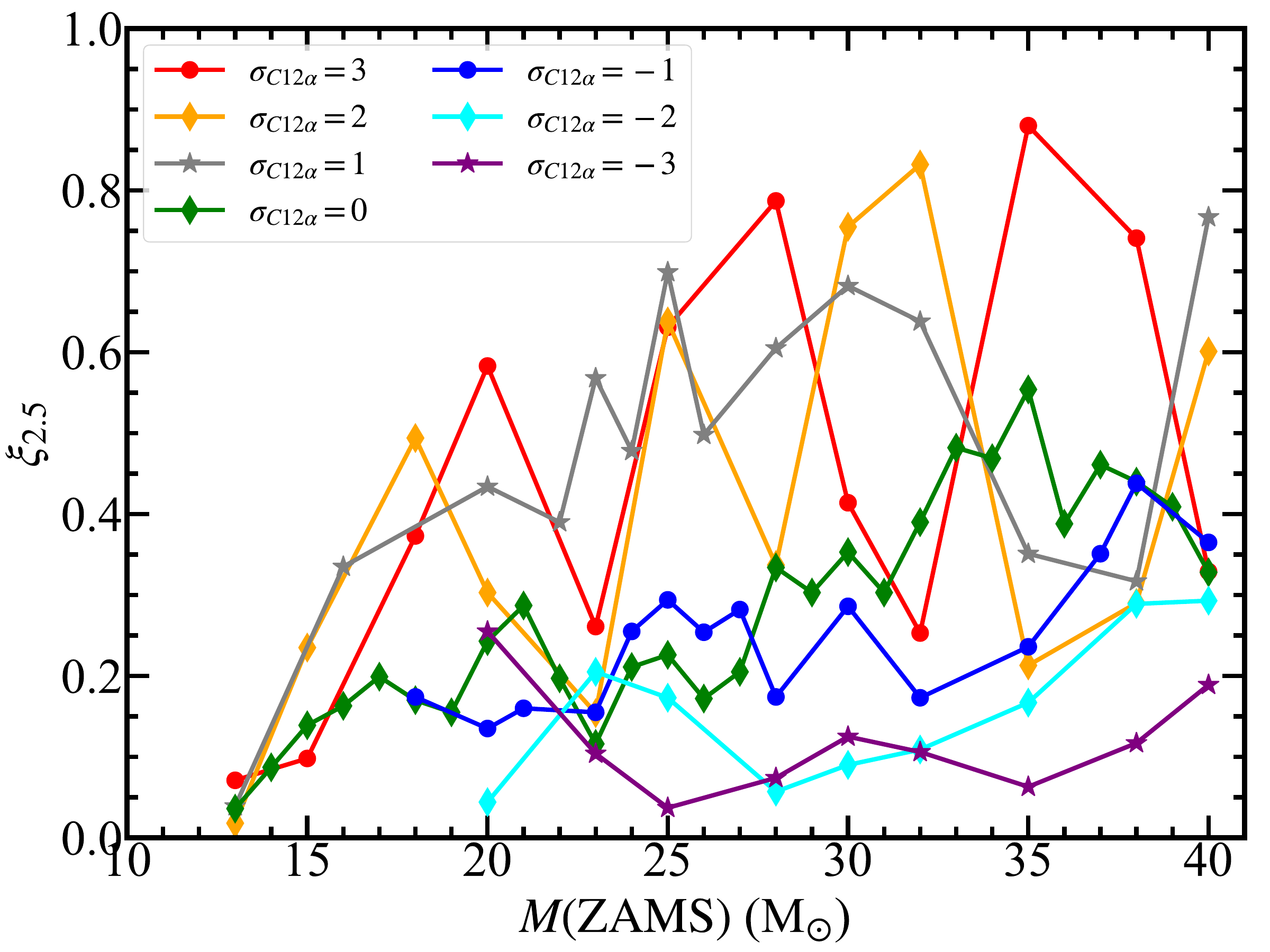}
\end{minipage}%
\caption{The compactness parameter $\xi_{2.5}$ as a function
$({\rm ZAMS})$ for $\sigma_{C12\alpha}$ = -3, -2, -1, 0, 1, 2, and 3.
\label{fig:compact_sigma}}
\end{figure}

Figure~\ref{fig:compact_sigma} shows how $\xi_{2.5}$ depends on
$\sigma_{C12\alpha}$ from -3 to +3 and $M{\rm (ZAMS)}$ for 13 - 40
M$_\odot$.  The dependence on the $^{12}$C$(\alpha, \gamma)^{16}$O
rate has also been studied by \citet{2014Sukhbold}.  Our results cover
a wider range of the reaction rate, i.e., $\sigma_{C12\alpha} = -3$ to
$+3$.  Figure~\ref{fig:compact_sigma} shows complicated dependences of
$\xi_{2.5}$ on $M{\rm (ZAMS)}$ and $\sigma_{C12\alpha}$, but the
details of the dependencies are shown in the following figures with
discussion.

\subsection{Compactness Parameter and Shell Burning} \label{sec:xi_shell}

As a source of such complicated dependences, the effect of C shell
burning on $\xi_{2.5}$ is important \citep{2020ApJ...890...43C}.  
We have investigated in sections \ref{sec:c_shell} and
\ref{sec:evo_r_rho} how the mixing of C and resultant C shell burning
cause the difference in $\xi_{2.5}$ between $\sigma_{C12\alpha}$ = 0
and $-1$ as summarized below.

The evolution of $\xi_{2.5}$ for $\sigma_{C12\alpha}$ = -1 and 0
(Figure \ref{fig:28M_compact}) is simply the evolution of 1/$r_{2.5}$
as shown in Figure \ref{fig:28M_radius}.  During the evolution until
$\tau \sim 10^0$ yr, the shell at $M_r =$ 2.5 M$_\odot$ simply
contracts, i.e., $r_{2.5}$ decreases, so that $\xi_{2.5}$ increases
for both $\sigma_{C12\alpha}$.

Around $\tau \sim 10^0$ yr, the contraction of the shell stops for
$\sigma_{C12\alpha} = -1$ because $X$(C) at $M_r =$ 2.0 M$_\odot$ is
enhanced (Figure \ref{fig:28M_central_Xcore} (f)) by convective mixing and
C shell burning heats up to cause slight expansion of the overlying layers
(see Figure \ref{fig:28M_entr} for enhancement of $s_{2.0}$ and $s_{2.5}$).
Thus 1/$r_{2.5}$ and $\xi_{2.5}$ slightly decrease.

For $\sigma_{C12\alpha}=0$, such enhancement of $X$(C) and $s$ does
not occur and C is depleted in the convective shell
(Figure \ref{fig:28M_central_Xcore} (e)) to cease C shell burning. Thus
1/$r_{2.5}$ and $\xi_{2.5}$ continue to increase
(Figs. \ref{fig:28M_radius} and \ref{fig:28M_compact}).  The temporal
decreases in 1/$r_{2.5}$ and $\xi_{2.5}$ around $\tau \sim 10^{-3}$ yr
are caused by the activation of O shell burning.

Such a difference in the evolution of $\xi_{2.5}$ and $X$(C) are also
found for $M{\rm (ZAMS)}=$ 25 M$_\odot$ and $\sigma_{C12\alpha}=-2$ to
0 as seen in Figure \ref{fig:25M_compact}.

\subsection{Compactness Parameter and $M{\rm (ZAMS)}$} \label{sec:xi_zams}

The effect of C shell burning on $\xi_{2.5}$ appears not only in the
$\sigma_{C12\alpha}$ dependence as described in
subsection \ref{sec:xi_shell} above, but also on the $M({\rm ZAMS})$
dependence. As seen in Figure~\ref{fig:compact_sigma}, 
$\xi_{2.5}$ changes non-monotonically with the
$M({\rm ZAMS})$ as has been investigated in several works
\citep[e.g.,][]{2014Sukhbold, 2018ApJ...860...93S, 2020ApJ...890...43C}.

\begin{figure}[htb]
\centering
\begin{minipage}[c]{0.45\textwidth}
\includegraphics [width=80mm]{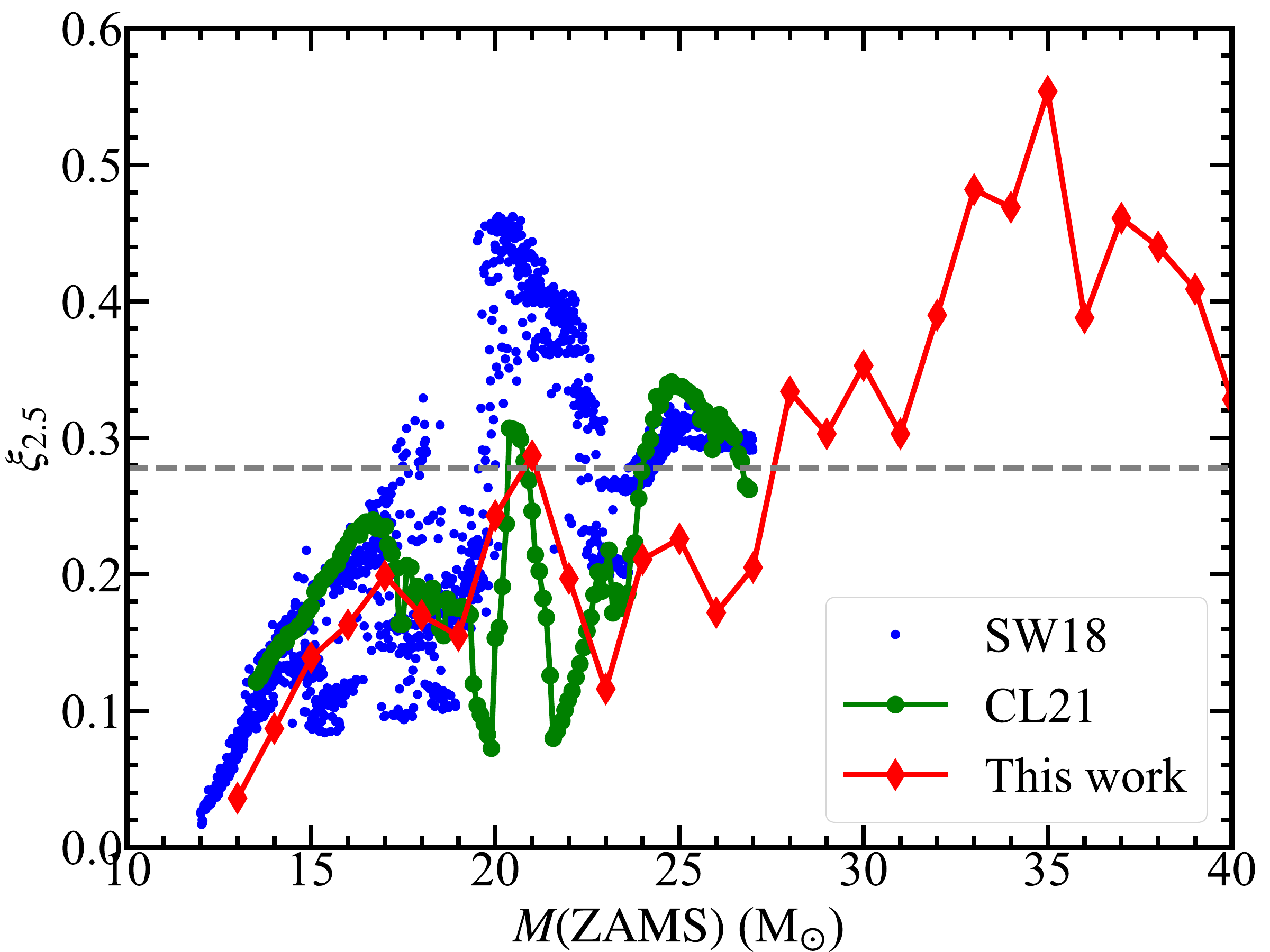}
\end{minipage}%
\caption{The compactness parameter as a function of $M ({\rm ZAMS})$
from different works.
Green points show the results from \citet{2021ApJ...916...79C}
and blue points represent the results from \citet{2018ApJ...860...93S}, respectively. 
The results of this work are marked by red diamonds.
The grey dashed line represents the critical value of $\xi_{2.5}$ = 0.278
from \citet{2016MNRAS.460..742M}.
\label{fig:compact_works}}
\end{figure}

Figure \ref{fig:compact_works} compares our $\xi_{2.5}$ as a function
of $M({\rm ZAMS})$ for $\sigma_{C12\alpha}$ = 0 (red) with those from
\citet{2018ApJ...860...93S} (blue) and \citet{2021ApJ...916...79C}
(green). Dependencies on $M{\rm (ZAMS)}$ are basically similar.

$\xi_{2.5}$ generally increases with $M{\rm (ZAMS)}$ as seen in Figure
\ref{fig:compact_works}.  For larger $M{\rm (ZAMS)}$, the mass
fraction $q = M_r/M{\rm (ZAMS)}$ of the shell of
$M_r = 2.5 M_\odot$ is smaller. 
For nearly homologous models, the density structure as a
function of $q$ is similar at the same stage.  The shell of smaller
$q$ is deeper and $r_{2.5}$ is smaller.  Thus $\xi_{2.5}$ tends to be
larger for larger $M{\rm (ZAMS)}$.

However, the dependence of $\xi_{2.5}$ on $M{\rm (ZAMS)}$ is not
monotonic.  For example, the green \citep{2021ApJ...916...79C}
points in Figure \ref{fig:compact_works} show the decrease
of $\xi_{2.5}$ with $M{\rm (ZAMS)}$ for several mass ranges.
\citet{2021ApJ...916...79C} discussed such a change as follows.  At
the mass ranges of $M{\rm (ZAMS)} =$ 17.4 - 19.4 and 20.3 - 21
$M_\odot$, the heating effect of convective C shell burning is strong
enough to decrease $\xi_{2.5}$.  Thus, the three peaks are formed at 
$M{\rm (ZAMS)} =$ 19.4, 21.0, and 25 $M_\odot$.  Our results shown by
the red-line in Figure~\ref{fig:compact_works} are basically similar,
although the peak value of $\xi_{2.5}$ around $M{\rm (ZAMS)} =$ 25
$M_\odot$ is smaller.

\begin{figure}[hbp]
\centering
\begin{minipage}[c]{0.42\textwidth}
\includegraphics [width=75mm]{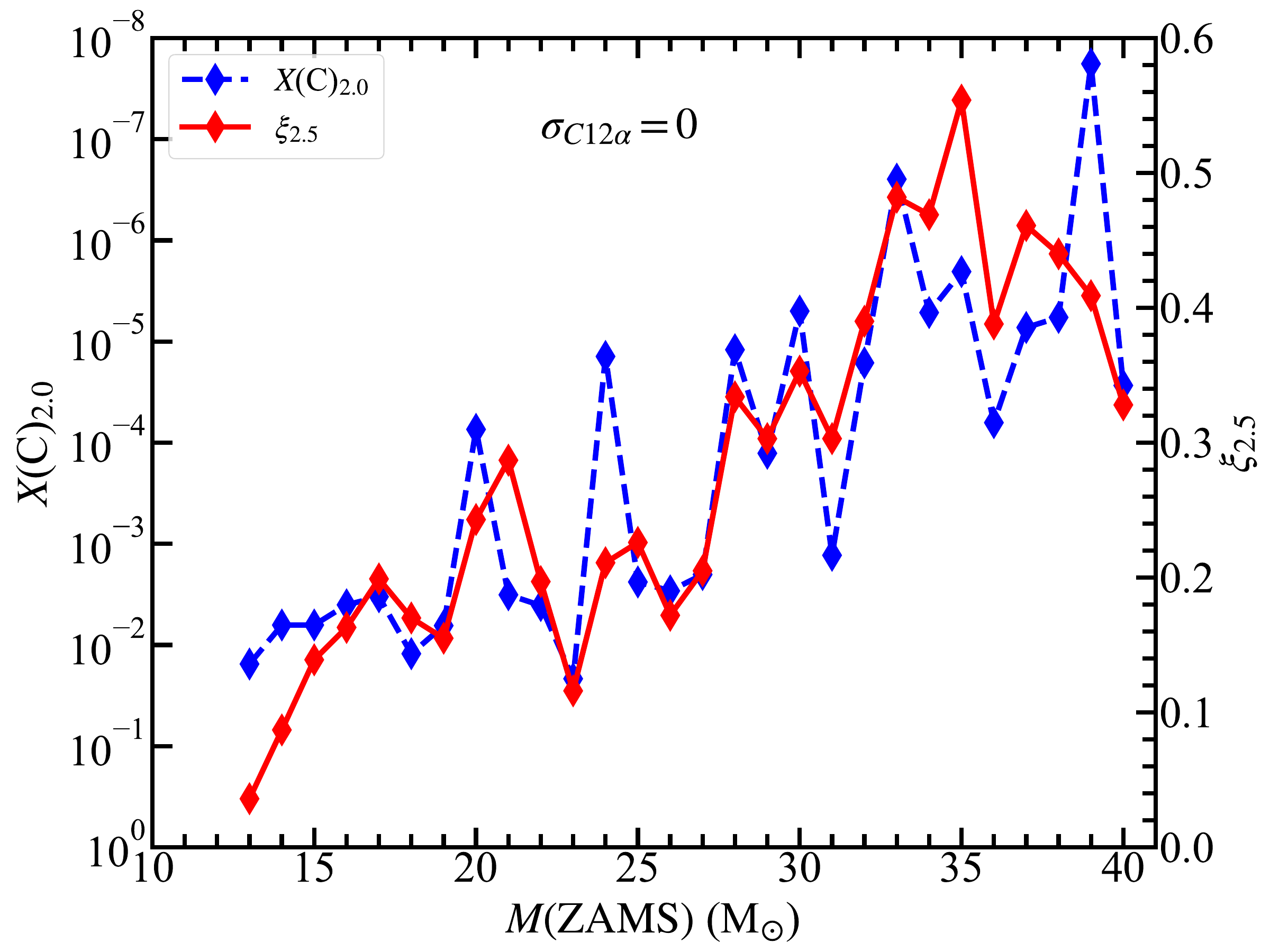}
\end{minipage}%
\begin{minipage}[c]{0.42\textwidth}
\includegraphics [width=75mm]{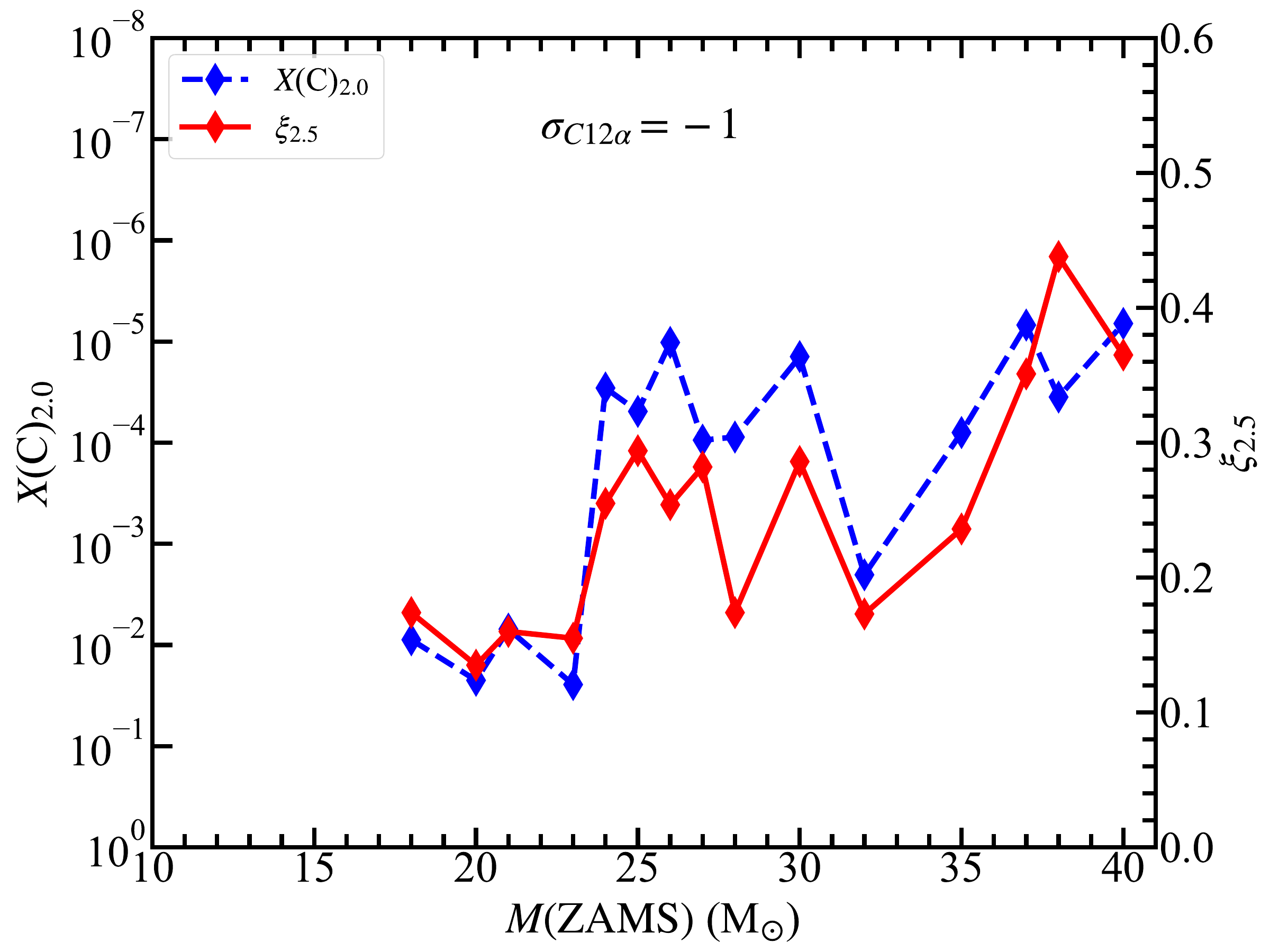}
\end{minipage}%
\caption{$X$(C)$_{2.0}$ compared with $\xi_{2.5}$ and 
as a function of $M ({\rm ZAMS})$ for $\sigma_{C12\alpha}=0$ and $-$1. 
Here $X$(C)$_{2.0}$ denotes $X$($^{12}$C) and $X$($^{16}$O) at
$M_r=$ 2.0 M$_\odot$.
\label{fig:xco20_xi25}}
\end{figure}

For the peak of $\xi_{2.5}$ around $M{\rm (ZAMS)} =$ 21 $M_\odot$ in
Figure~\ref{fig:compact_works}, \citet{2020ApJ...890...43C} discussed
the relation with the critical mass $M_{\rm crit}$.
As mentioned in \S~\ref{sec:c_ign}, for stars with
increasing $M ({\rm ZAMS})$, core C burning changes from convective to
radiative around $M_{\rm crit}$ because of the decreasing C/O ratio.
This transition may cause a sharp change in $\xi_{2.5}$ with
$M ({\rm ZAMS})$ as discussed by \cite{2020ApJ...890...43C} for
$\sigma_{C12\alpha} = 0$ (Figure~\ref{fig:compact_works}).

The critical mass $M_{\rm crit}$ is larger for smaller $\sigma_{C12\alpha}$ 
(Table \ref{tab:mcrit}), so that the peak of $\xi_{2.5}$ appears at
larger $M{\rm (ZAMS)}$ as compared in Figure~\ref{fig:xco20_xi25}.

To examine whether such C mixing affects the compactness
parameter for other stars, we compare $X$(C)$_{2.0}$ and
$\xi_{2.5}$ as a function of $M{\rm (ZAMS)}$ in Figure
\ref{fig:xco20_xi25} for $\sigma_{C12\alpha}=0$ and $-1$.
Here $X$(C)$_{2.0}$ denotes $X$($^{12}$C) at $M_r=$ 2.0 M$_\odot$.
(Note, in the vertical axis of Figure \ref{fig:xco20_xi25},
$X$(C)$_{2.0}$ decreases upward, while $\xi_{2.5}$ increases upward.)

We find a good correlation between $\xi_{2.5}$ and $X$(C)$_{2.0}$,
thus concluding that the non-monotonicity of $\xi_{2.5}$ is mainly
determined by the behavior of the C burning shell, which is related to
$X$($^{12}$C) near the bottom of the C burning shell (around $M_r
\sim$ 2.0 M$_\odot$).  For higher $M{\rm(ZAMS)}$, the effect of O
shell burning is significant.

\begin{figure*}[htbp]
\centering
\begin{minipage}[c]{0.7\textwidth}
\includegraphics [width=130mm]{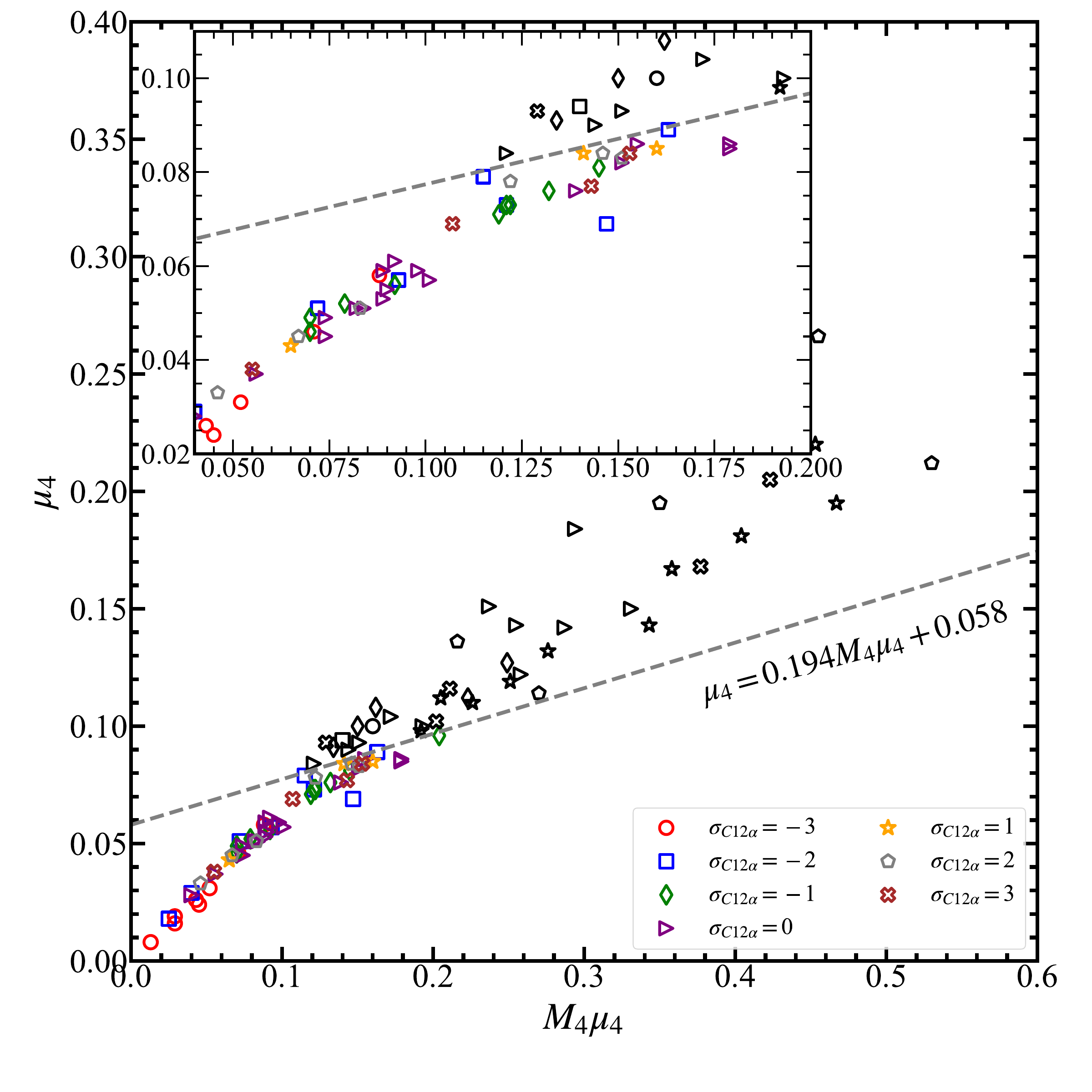}
\end{minipage}%
\caption{
Distribution of the presupernova models in the $M_4\mu_4-\mu_4$ plane.
The explosion and collapse models are distinguished with
$\mu_4=0.194\mu_4M_4+0.058$ from \citet{2016ApJ...818..124E}. 
The collapse models are in black, and the explosion models are in other colors.
\label{fig:mu4m4_ertl}}
\end{figure*}

\begin{figure}[htbp]
\centering
\begin{minipage}[c]{0.42\textwidth}
\includegraphics [width=75mm]{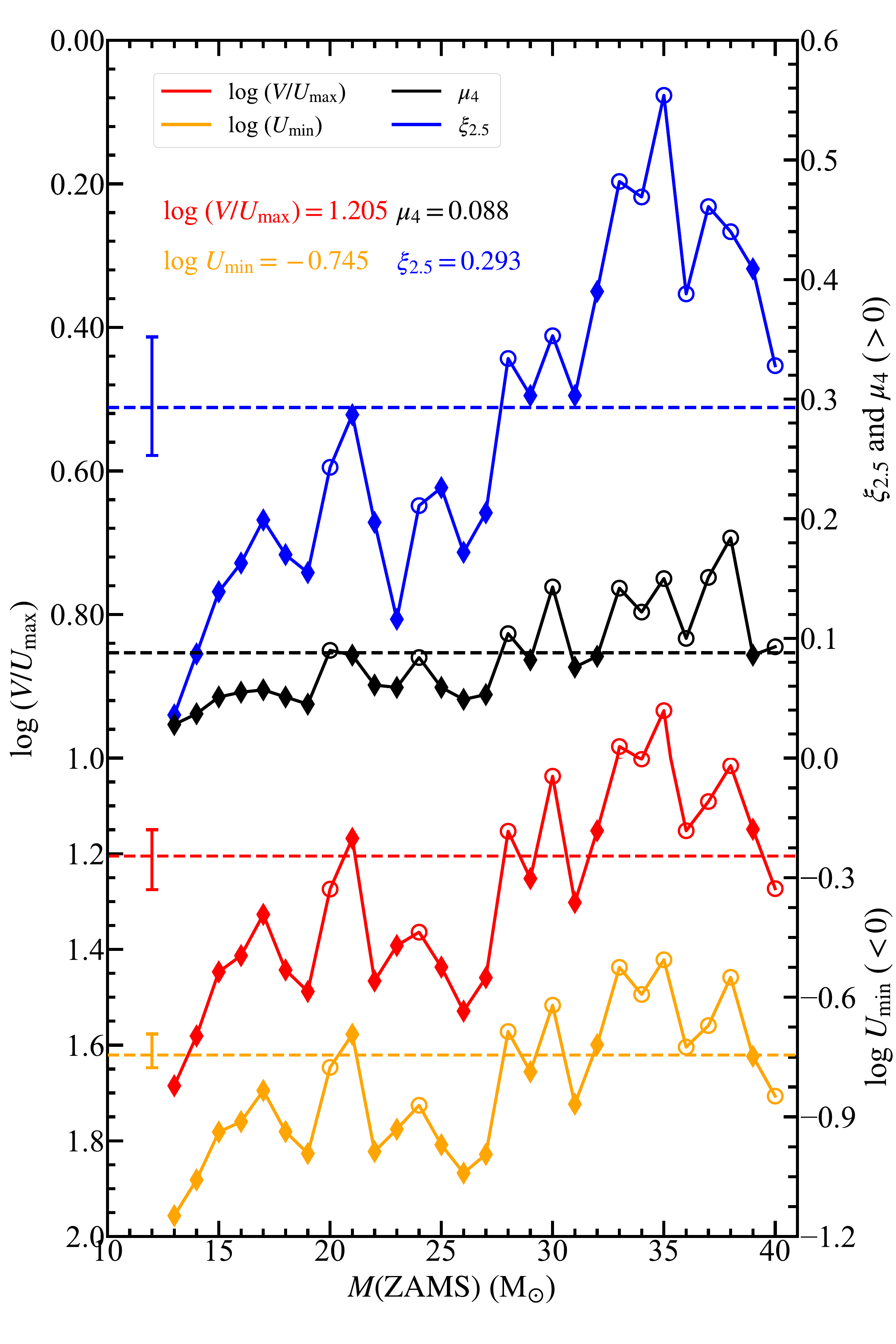}
\end{minipage}%
\caption{The log ($V/U_{\rm max}$), log $U_{\rm min}$, $\xi_{2.5}$
and $\mu_4$ as a function of $M ({\rm ZAMS})$ for $\sigma_{C12\alpha}=0$.
The left Y-axis is shown upside down. On the right Y-axis, $\xi_{2.5}$
and $\mu_4$ are shown with the upper part (positive)
while log $U_{\rm min}$ is shown in the bottom part (negative).
The filled diamonds represent explosion models,
while open circles represent collapse models.
The explosion and collapse are predicted by the results of 
Figure~\ref{fig:mu4m4_ertl}.
The critical value of each parameter is shown as dashed 
lines and an uncertainty of $\pm6$\%.
We assume that the models beyond $\pm6$\% of critical
value should collapse and explode,
while those within this range are uncertain
because of the uncertainties of progenitor evolution
and explosion mechanism.
\label{fig:xi_vu_p00}}
\end{figure}

\begin{figure*}[htbp]
\centering
\begin{minipage}[c]{0.96\textwidth}
\includegraphics [width=170mm]{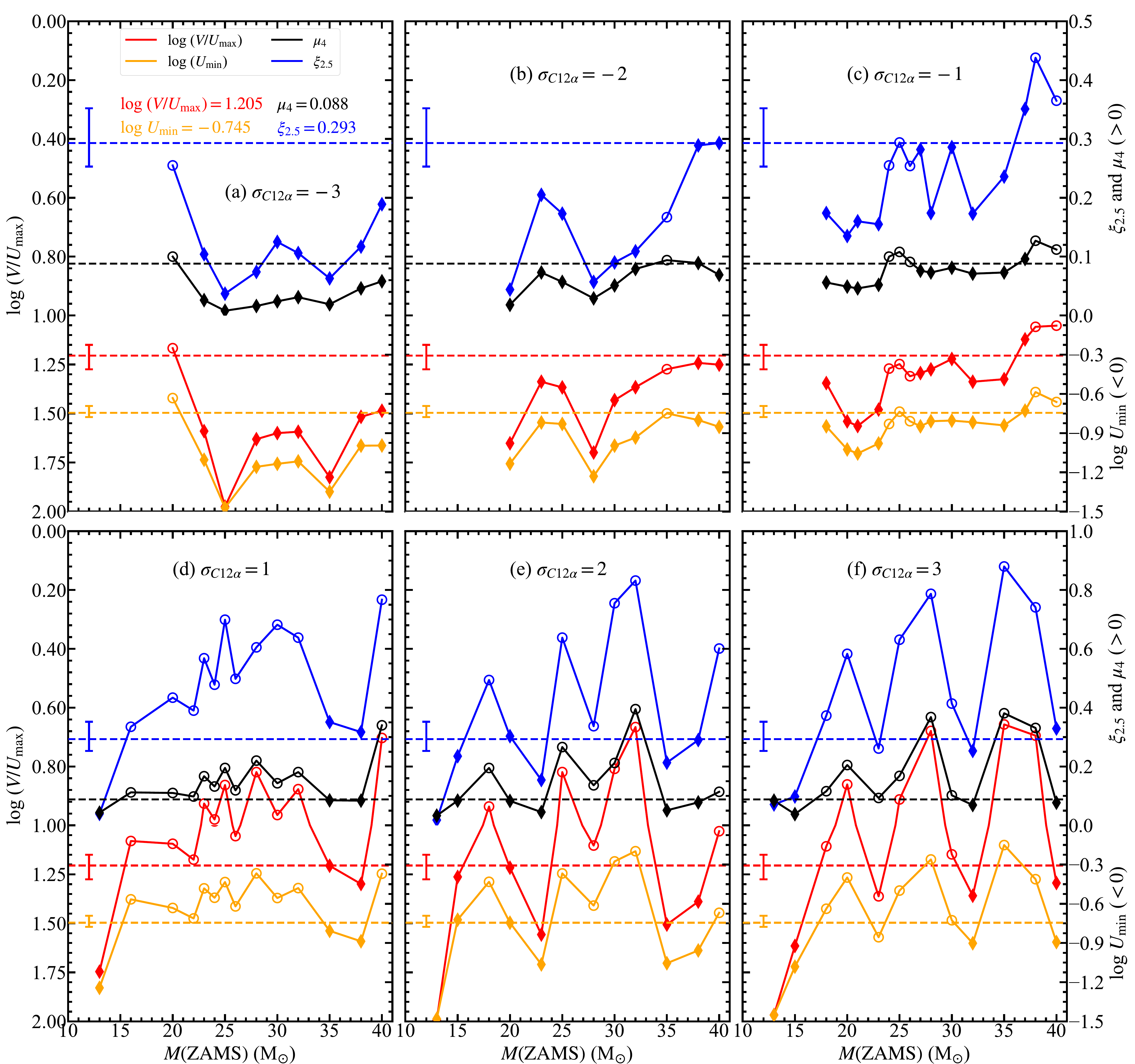}
\end{minipage}%
\caption{The log ($V/U_{\rm max}$), log $U_{\rm min}$, $\xi_{2.5}$ and $\mu_4$
as a function of $M ({\rm ZAMS})$ for $\sigma_{C12\alpha}= \pm1, \pm2$ and $\pm3$.
\label{fig:xi_vu}}
\end{figure*}

\subsection{Explodability} \label{sec:m4mu4_core}

\subsubsection{Criteria for Explodability}

The compactness parameter has been suggested to be a criterion of the
explodability of presupernova models
\citep{2012ApJ...757...69U, 2016MNRAS.460..742M}. Based on the 1D
hydrodynamical simulations of neutrino-driven supernova models, a
two-parameter criterion in the $M_4 \mu_4$ - $\mu_4$ plane for the
explodability has been proposed by \citet{2016ApJ...818..124E} and
\citet{2020ApJ...890...51E}.  Here $\mu_4$ is defined in Equation
\ref{equ:m4mu4}.

We plot all of our 99 models with the combination of $M({\rm ZAMS})$
and $\sigma_{C12\alpha}$ in the $M_4 \mu_4$ - $\mu_4$ plane of Figure
\ref{fig:mu4m4_ertl}, where $M_4$ and $\mu_4$ are given in
Table \ref{tab:structure}.
We adopt the line of $\mu_4 = 0.194 \mu_4 M_4 + 0.058$ \footnote{We do not
use the separation line of $\mu_4 = 0.33 \mu_4 M_4 + 0.09$
proposed by \citet{2016MNRAS.460..742M}, because among our 99 models
there is only one ``collapse'' model, which seems too small.}
from \citet{2016ApJ...818..124E} and draw in this plane.
According to the hydrodynamical simulations by \citet{2016ApJ...818..124E},
models above this line are likely to form BHs without much mass
ejection and models below this line are likely to explode to form
NSs.

Note the presupernova models adopted by \citet{2016ApJ...818..124E} are different
from ours and we need hydrodynamical simulations of core-collapse for our
own models. Nevertheless, we assume that
the line of $\mu_4 = 0.194\mu_4M_4+0.058$ divides the BH formation and
the NS formation for our presupernova models with different
$\sigma_{C12\alpha}$.  For simplicity in the present paper, we call
the models above this line ``collapse'' models and indicate them
with black markers.  We call the models below this line as
``explosion'' models and indicate them with color markers.
Different shapes and colors are used for different $\sigma_{C12\alpha}$s.

In Tables \ref{tab:structure} and \ref{tab:core_mass}, we indicate
``explosion'' or ``collapse'' for each model.  To show clearly how the
separation between ``explosion'' or ``collapse'' depend on model
parameters ($\sigma_{C12\alpha}$ and $M({\rm ZAMS})$) and model
properties ($\xi_{2.5}$, $\mu_4$, $V/U_{\rm max}$
\footnote{$V/U_{\rm max}$ denotes $(V/U)_{\rm max}$.}, and
$U_{\rm min}$), we plot in Figures \ref{fig:xi_vu_p00} and
\ref{fig:xi_vu} these model properties as a function of
$M({\rm ZAMS})$ for $\sigma_{C12\alpha}=-3$ to +3.

In these figures, $V/U_{\rm max}$ of the left Y-axis decreases upward.
On the right Y-axis, the scale of the positive value part for
$\xi_{2.5}$ and $\mu_4$ is different from the negative value part for
log $U_{\rm min}$.  The filled diamonds indicate ``explosion'' models,
while open circles indicate ``collapse'' models.  These explosion and
collapse models are predicted by the results of
Figure~\ref{fig:mu4m4_ertl}.

We find that ``explosion'' and ``collapse'' can be separated by the
critical values of $\mu_4 =$ 0.088 $\pm0$, $\xi_{2.5} = 0.293^{+0.059}
_{-0.040}$, log $V/U_{\rm max} = 1.205^{+0.070}_{-0.055}$,
and log $U_{\rm min} = -0.745^{+0.053}_{-0.032}$.
In Figures \ref{fig:xi_vu_p00} and \ref{fig:xi_vu}, these critical
values are shown by the dashed lines with the uncertainty bars.

For models outside the uncertainty bars, the numbers of correct
predictions of ``explosion'' and ``collapse'', and false predictions
(``false'') for $\mu_4$, $\xi_{2.5}$, $V/U_{\rm max}$, and $U_{\rm min}$
are given in Table \ref{tab:predict} for $\sigma_{C12\alpha}$ from -3 to +3.
The uncertainty bars are chosen to make the fraction of
``false'' prediction less than 5 \%.  (Remember that the total
number of models is 99).  Inside the uncertainty bars, the final fate of
the model is regarded as ``uncertain'', where the numbers of correct and
false predictions are also given in Table \ref{tab:predict}.  These
uncertainties may be related to the uncertainties of presupernova
models and hydrodynamical simulations.  To make the uncertainties
smaller, we need core-collapse calculations using our own presupernova
models calculated in this paper, which will be done in the forthcoming
study.

\begin{table}[htb]  
\small
\centering
\caption{The number of ``explosion", ``collapse" and ``uncertain" models for different
criteria.}
\label{tab:predict}
\begin{tabular}{cccccc}
\toprule
\multirow{2}{*}{prediction} & \multirow{2}{*}{``explosion"} & \multirow{2}{*}{``collapse"} & \multirow{2}{*}{``false"} & \multicolumn{2}{l}{``uncertain"} \\
                            &                            &                           &                        & correct        & false        \\
\midrule
$\mu_4$                     & 54                         & 42                        & 3                      & 0              & 0            \\
$\xi_{2.5}$                 & 41                         & 31                        & 5                      & 11             & 11           \\
$V/U_{\rm max}$             & 45                         & 31                        & 5                      & 11             & 7            \\
$U_{\rm min}$               & 50                         & 32                        & 5                      & 7              & 5            \\
\bottomrule
\end{tabular}
\end{table}

We note from Figures \ref{fig:xi_vu_p00} and \ref{fig:xi_vu} and Table
\ref{tab:predict} several properties of presupernova structures.

$\xi_{2.5}$, $\mu_4$, $V/U_{\rm max}$, and $1/U_{\rm min}$ show
similar dependencies on $M$(ZAMS).  As discussed in Section \ref{sec:uv_curve},
$V/U_{\rm max}$ and $1/U_{\rm min}$ are the
steepest gradients of log $P$ and log $r$ with respect to $M_r$ in the
CO core.  $\mu_4$ is also the average gradient.  Thus, the compactness
parameter $\xi_{2.5}$ is closely related to the gradients of pressure
and density in the presupernova core.  This is the reason why larger
$\mu_4$ and $\xi_{2.5}$, smaller $V/U_{\rm max}$ and $1/U_{\rm min}$
tend to collapse.  As shown in Figure \ref{fig:28M_logp_rho}, the
core-collapse models would more easily explode with the steeper
gradients of pressure and density.

During the core-collapse and bounce, the mass accretion onto the
collapse object produces lamb pressure against the bouncing shock and
the accretion rate would be lower if the density structure is steeper.
This would make the explosion easier to occur.  Also, the shock wave
generated at the bounce and propagates outward would be more
strengthened at the steeper gradient, which would also make the
explosion easier.

Although $\xi_{2.5}$ would be a good measure of ``collapse''
vs. ``explosion'', we propose that $V/U_{\rm max}$ and $1/U_{\rm min}$
are physically reasonable measures of the explodability.

\subsubsection{Explodability and $^{12}$C$(\alpha, \gamma)^{16}$O Rate}

Figures \ref{fig:xi_vu_p00} and \ref{fig:xi_vu} and Table
\ref{tab:predict} show the dependence of the explodability on
$\sigma_{C12\alpha}$.  For smaller $\sigma_{C12\alpha}$, generally,
$V/U_{\rm max}$ and $1/U_{\rm min}$ tend to be larger, while
$\xi_{2.5}$ and $\mu_4$ tend to smaller.  Thus, a larger number of
models tend to undergo explosions for smaller $\sigma_{C12\alpha}$,

The dependence of ``collapse vs. explosion'' on $M({\rm ZAMS})$ also
depends on $\sigma_{C12\alpha}$ as seen in these figures.  For
$\sigma_{C12\alpha} > 0$, e.g., the $M({\rm ZAMS})$ dependence is
different because the effect of O shell burning is dominant over C
shell burning.  How $\sigma_{C12\alpha}$ affects the presupernova
structures and $V/U_{\rm max}$ will be shown for $M({\rm ZAMS})$ in
the next section.

\begin{figure*}[htbp]
\centering
\begin{minipage}[c]{0.4\textwidth}
\includegraphics [width=70mm]{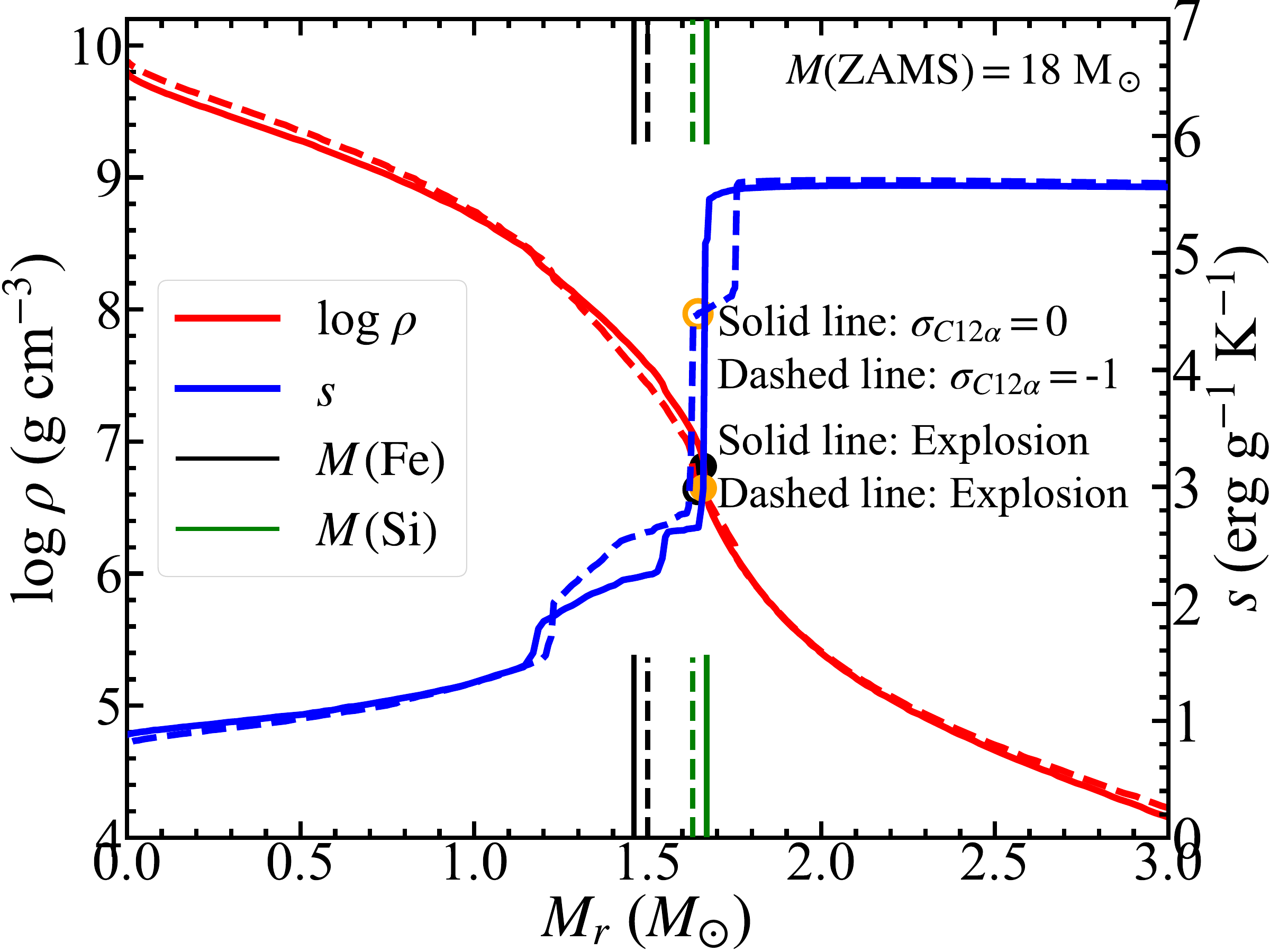}
\end{minipage}%
\begin{minipage}[c]{0.4\textwidth}
\includegraphics [width=70mm]{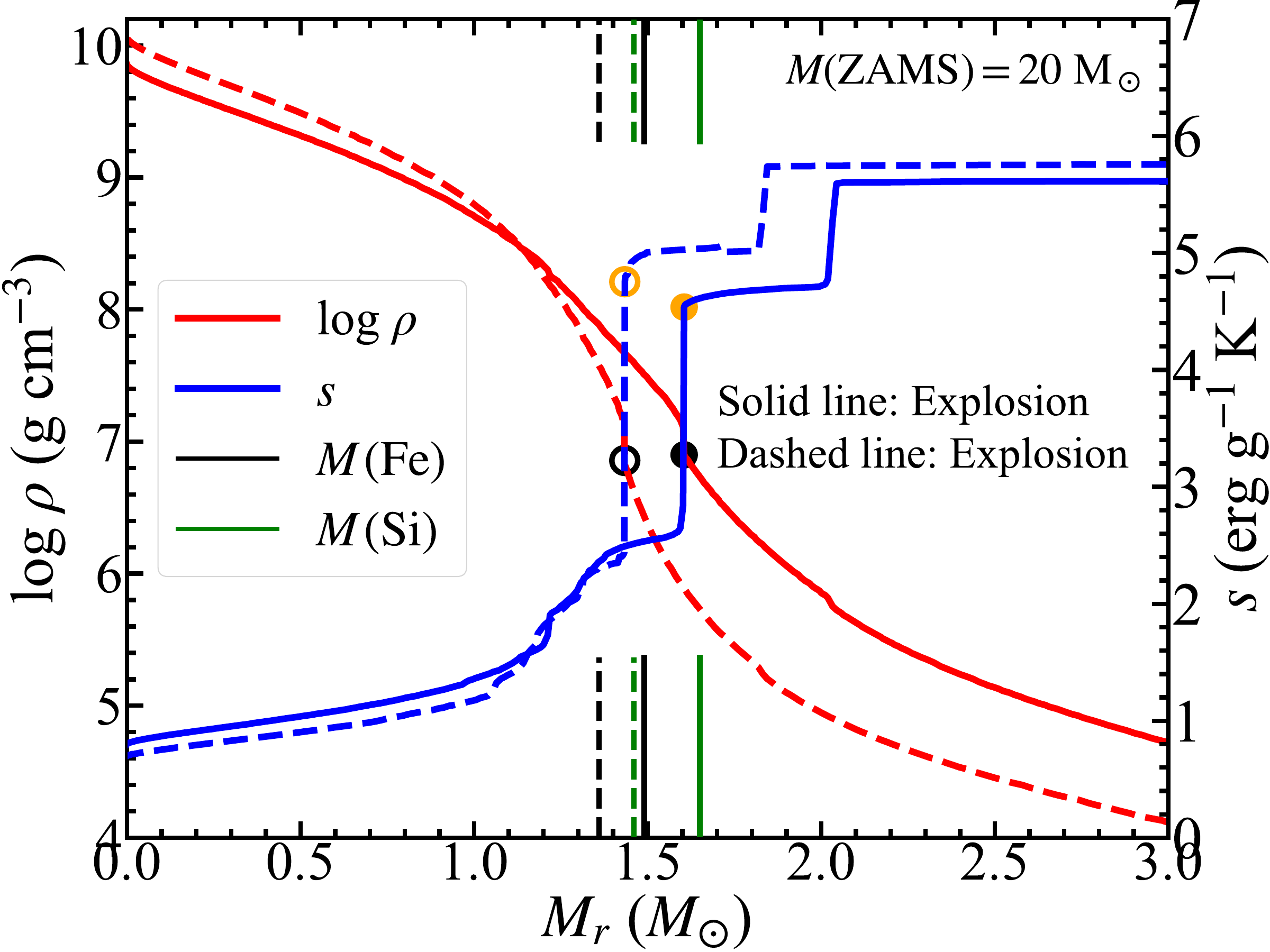}
\end{minipage}%
\begin{minipage}[c]{0.4\textwidth}
\includegraphics [width=70mm]{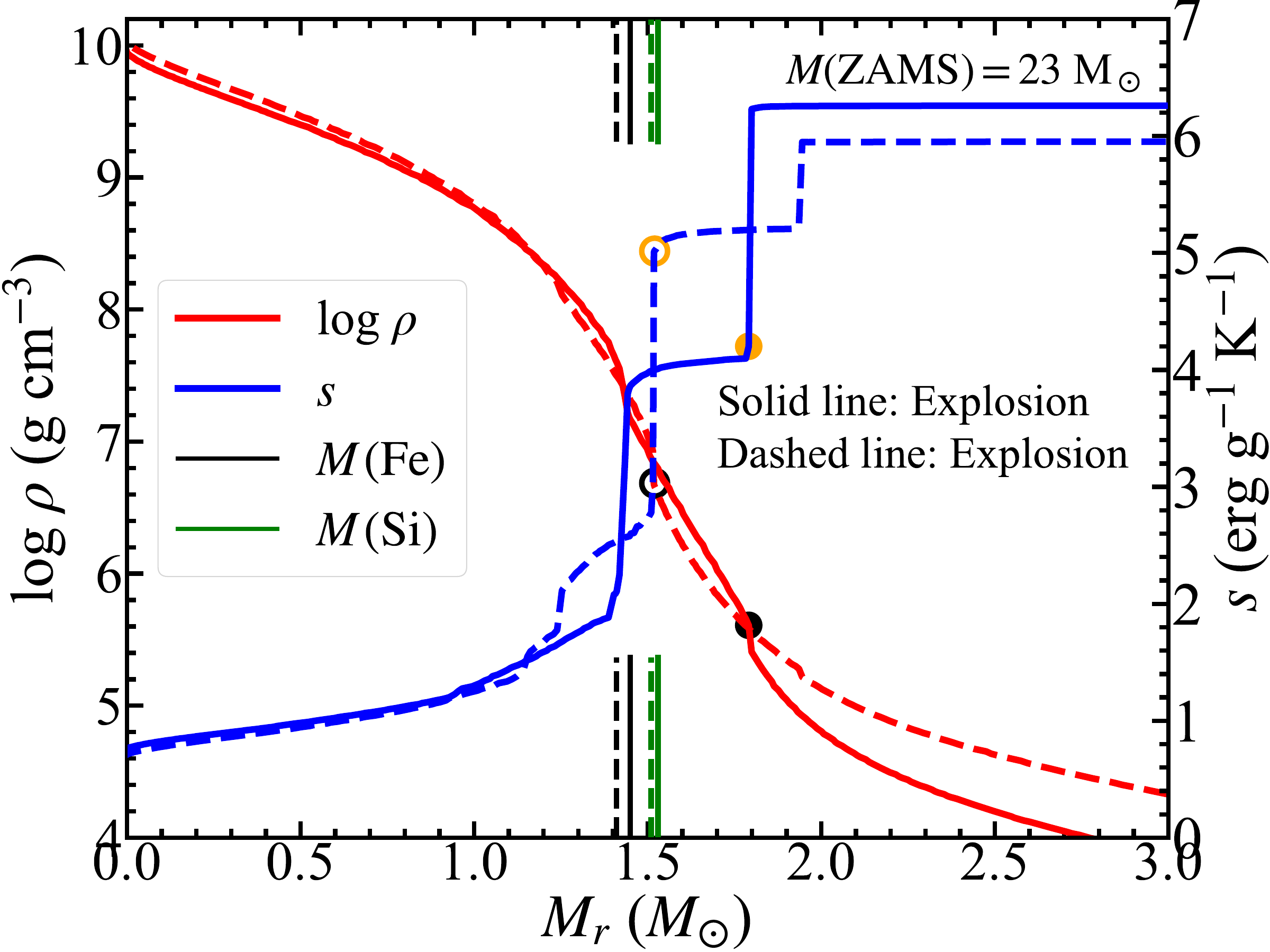}
\end{minipage}%
\begin{minipage}[c]{0.4\textwidth}
\includegraphics [width=70mm]{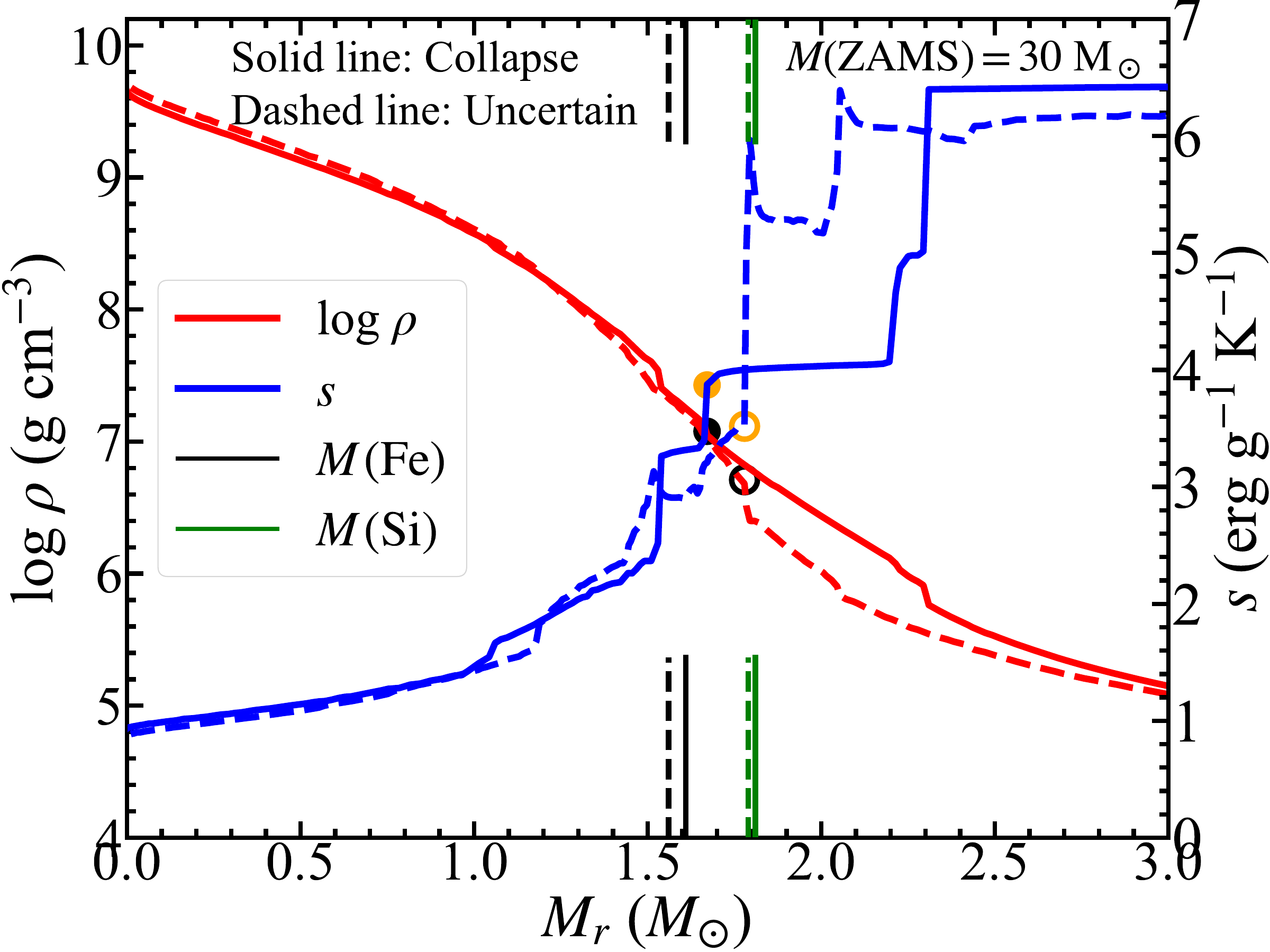}
\end{minipage}%
\begin{minipage}[c]{0.4\textwidth}
\includegraphics [width=70mm]{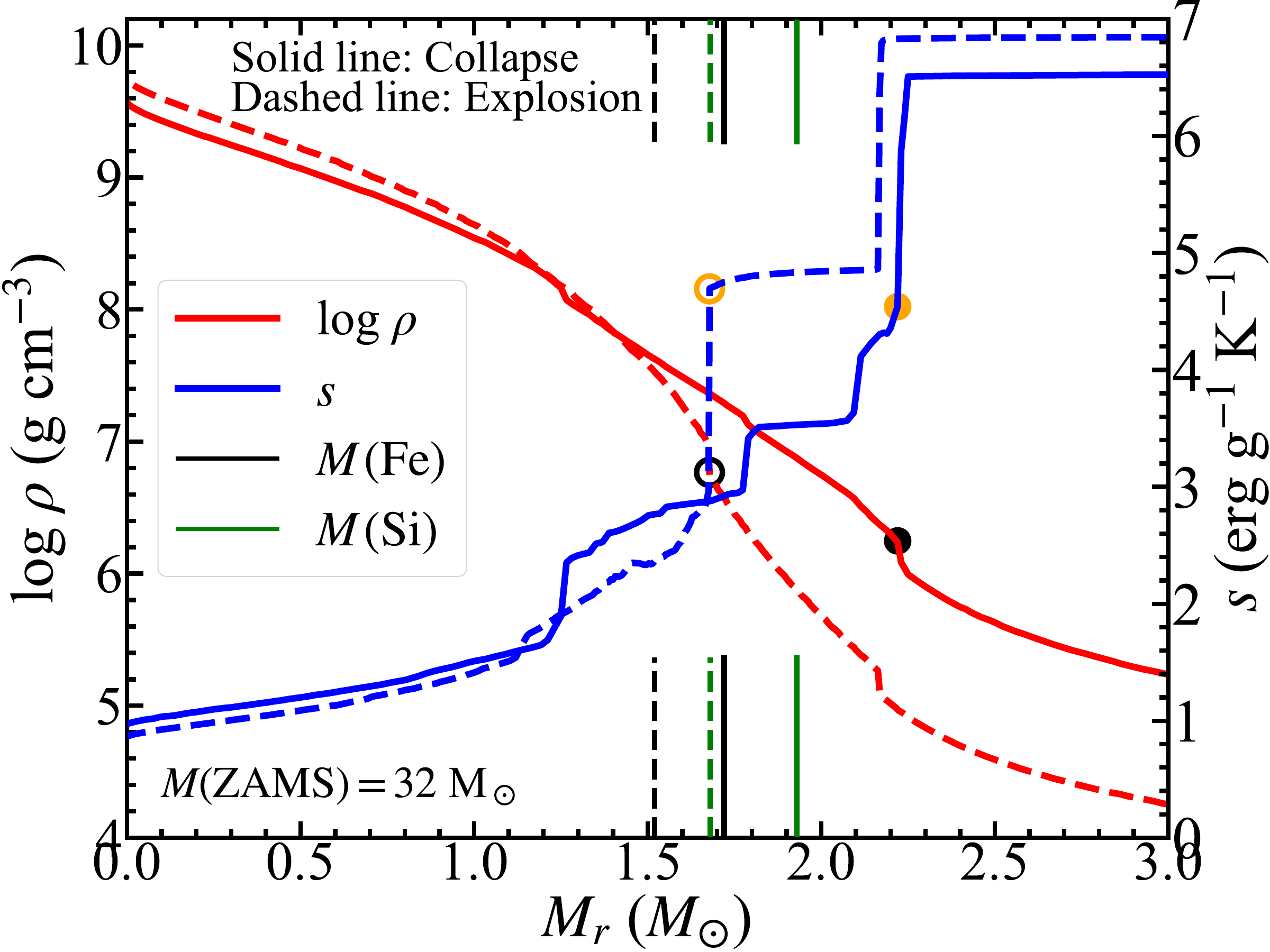}
\end{minipage}%
\begin{minipage}[c]{0.4\textwidth}
\includegraphics [width=70mm]{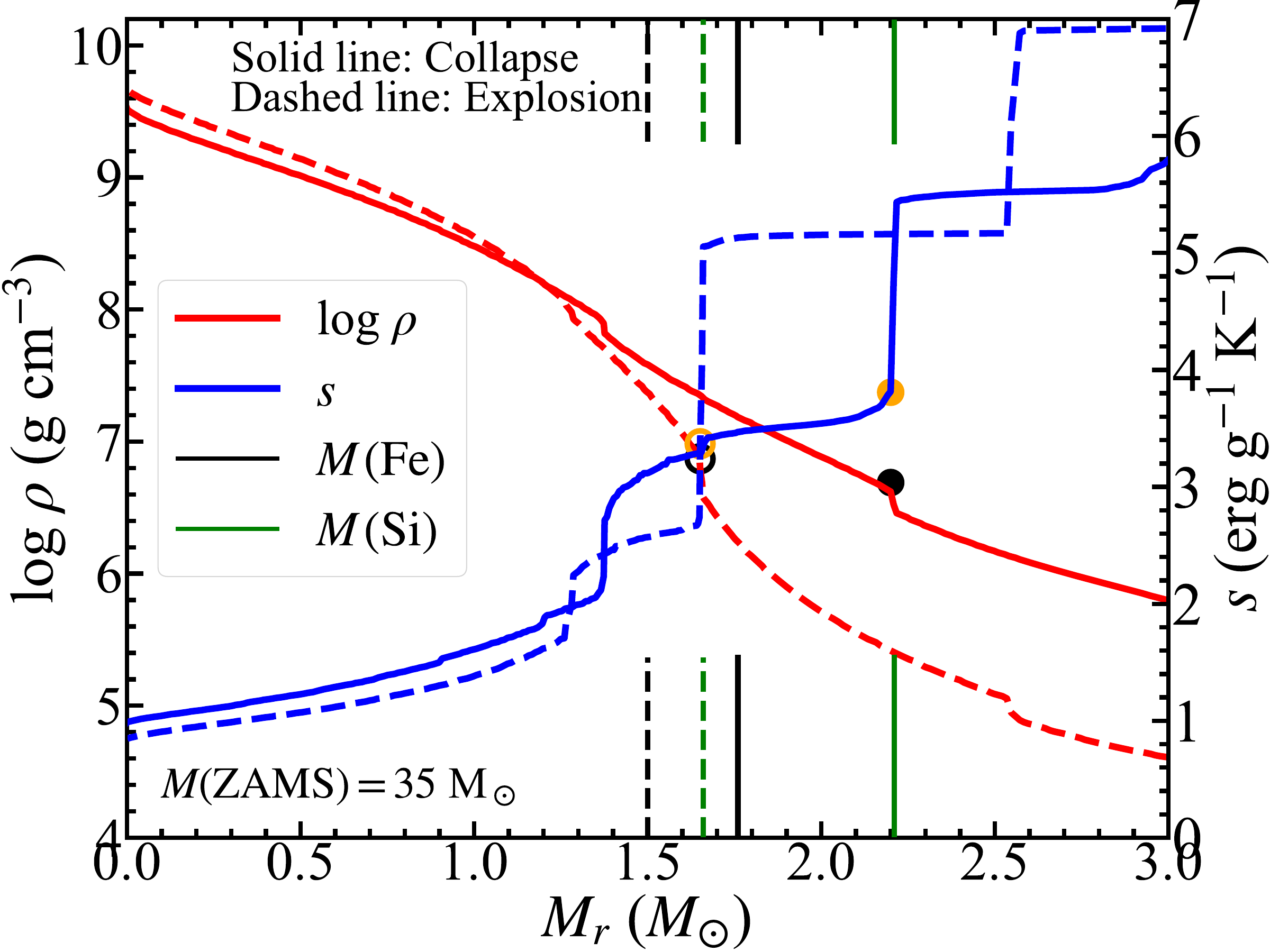}
\end{minipage}%
\begin{minipage}[c]{0.4\textwidth}
\includegraphics [width=70mm]{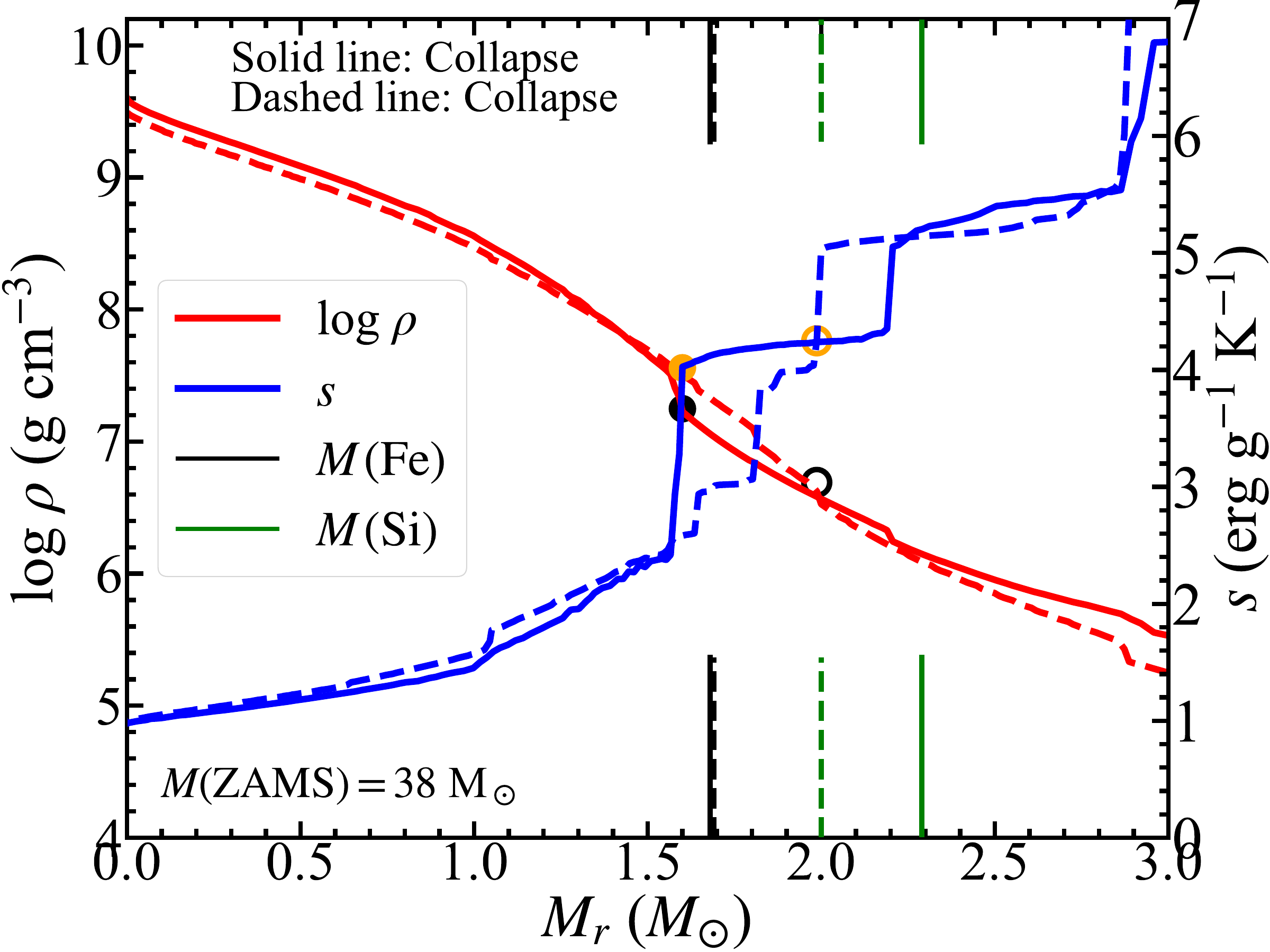}
\end{minipage}%
\begin{minipage}[c]{0.4\textwidth}
\includegraphics [width=70mm]{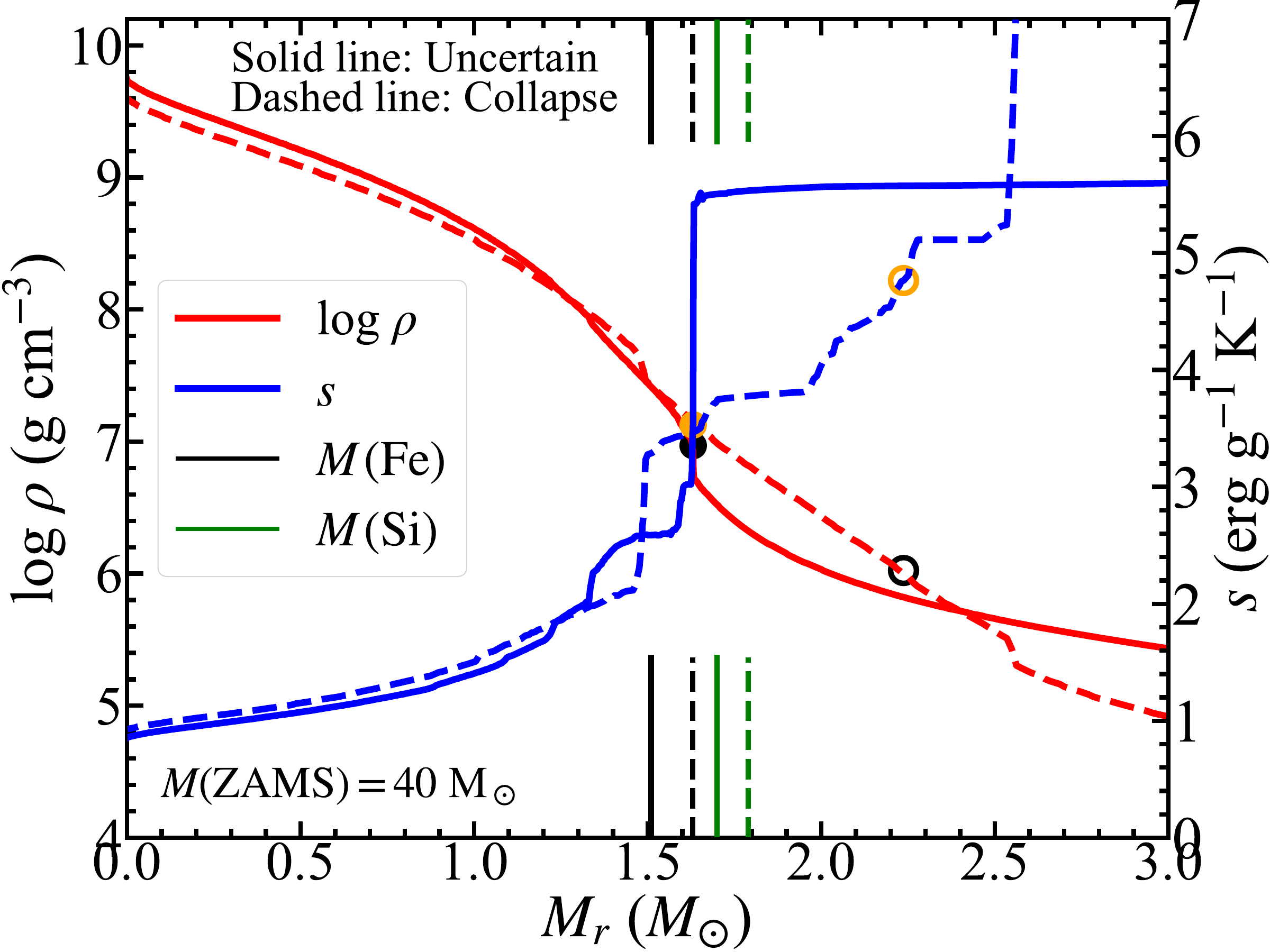}
\end{minipage}%
\caption{The presupernova density and entropy distributions of stars
  of several $M ({\rm ZAMS})$ for $\sigma_{C12\alpha}=$ 0 (solid line)
  and -1 (dashed line).  It is indicated whether these models are
  ``collapse'' or ``explosion'' (or ``uncertain'').  The lines and
  marks are the same as used in Figure \ref{fig:28M_logp_rho} for
  $M ({\rm ZAMS})$ = 28 M$_{\odot}$.
\label{fig:logrho_s}}
\end{figure*}

\begin{figure*}[htbp]
\centering
\begin{minipage}[c]{0.8\textwidth}
\includegraphics [width=150mm]{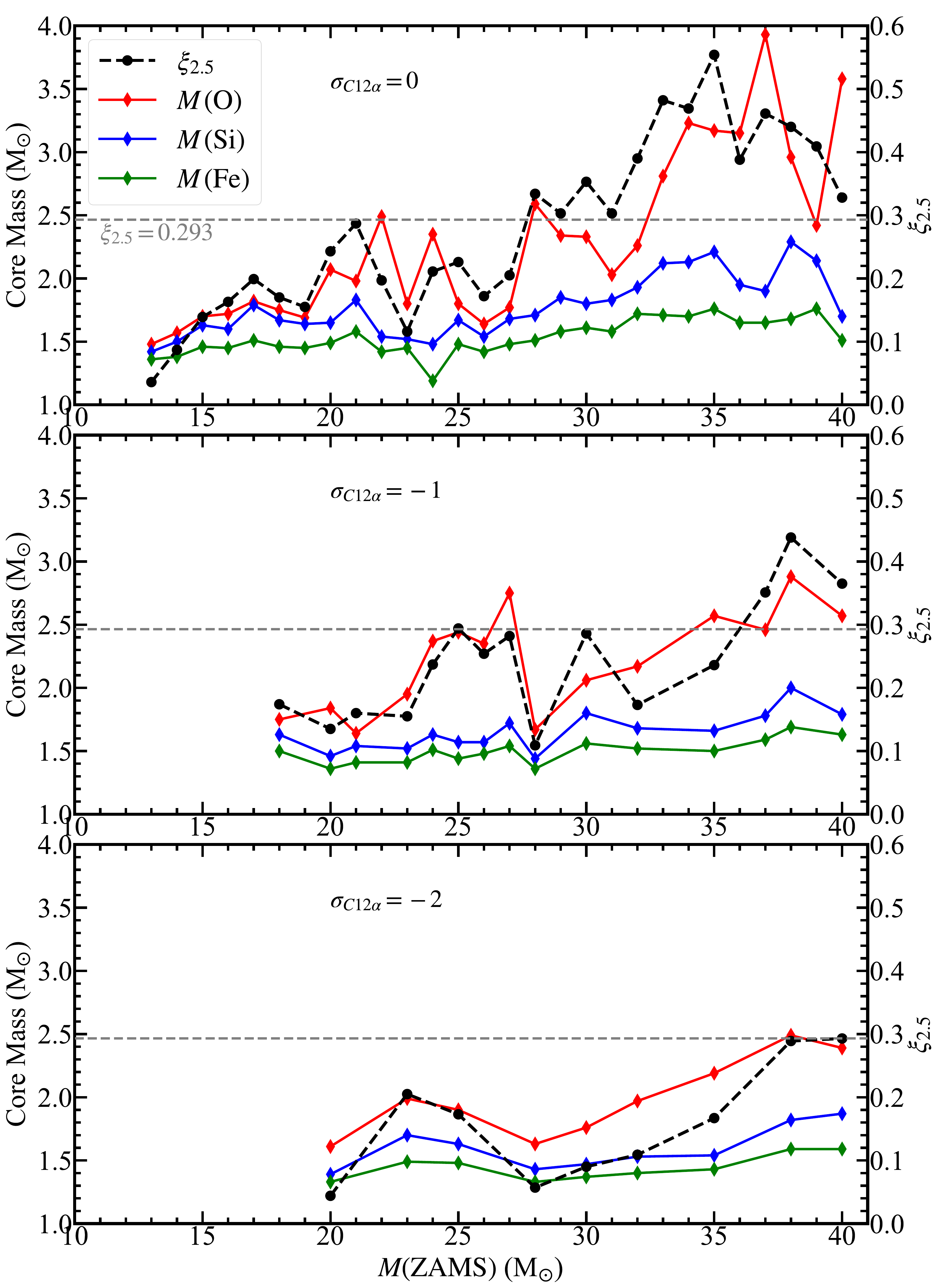}
\end{minipage}%
\caption{The core masses, $M({\rm O})$, $M({\rm Si})$, $M({\rm Fe})$
and $\xi_{2.5}$ as a function of $M({\rm ZAMS})$ for
$\sigma_{C12\alpha}=0$ (top), -1 (median) and -2 (bottom).
The gray dashed line shows the critical value of $\xi_{2.5}=0.293$.
\label{fig:core_xi}}
\end{figure*}

\subsection{Presupernova Structure and $^{12}$C$(\alpha, \gamma)^{16}$O Rate} \label{sec:vu-criterion}

\subsubsection{Oxygen Shell Burning, $V/U_{\rm max}$, and $U_{\rm min}$}

\begin{figure*}[htb]
\centering
\begin{minipage}[c]{0.96\textwidth}
\includegraphics [width=170mm]{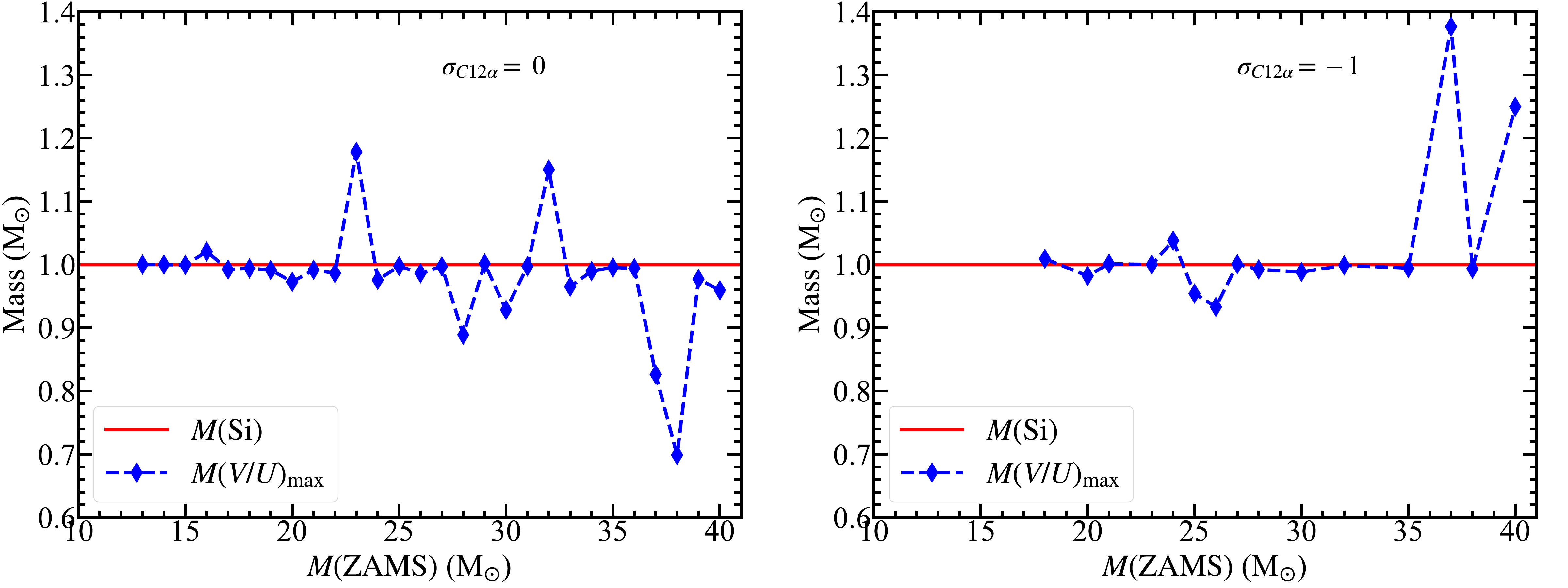}
\end{minipage}%
\caption{$M(V/U_{\rm max})$ normalized by $M({\rm Si})$
as a function of $M ({\rm ZAMS})$ for $\sigma_{C12\alpha}$ = 0 and $-$1.
\label{fig:core_eff}}
\end{figure*}

Figures~\ref{fig:logrho_s} show how $\sigma_{C12\alpha}$ affects the
presupernova structures of the density and entropy and the locations
of $V/U_{\rm max}$ for $M ({\rm ZAMS})$ = 18 - 40 M$_{\odot}$ with
$\sigma_{C12\alpha} = 0$ (solid lines) and $-1$ (dashed lines).  It is
indicated whether these models are ``collapse'' or ``explosion'' (or
``uncertain'').  The lines and marks are the same as used in Figure
\ref{fig:28M_logp_rho} for $M ({\rm ZAMS})$ = 28 M$_{\odot}$.

These figures show that $V/U_{\rm max}$ is located in most cases at
the sharp increase in $s$, which is produced by O shell burning (i.e.,
at the outer edge of the Si core $M_r=M$(Si)).  When shell burning is
active, the released heat prevents the contraction of the outer layers
and the density there remains almost constant and even decreases as seen
in Figure \ref{fig:28M_rho}.  On the other hand, the inner core continues
to increase the density of the inner core.  This creates a very steep
density gradient, almost a density jump, and a decrease in $U$,
which is the ratio between $\rho$ and the mean density of the inner core.

For example, two density jumps are produced by C shell burning and O
shell burning at $M_r=$ 1.95 M$_\odot$ and 1.44 M$_\odot$,
respectively, in the 28 M$_\odot$ star for $\sigma_{C12\alpha}=-1$
(Figure \ref{fig:28M_logp_rho}).  These two density jumps correspond to
the sharp jumps of $V$, two valley points of $U$, and the highest peak
and knee points of $V/U$ in Figure \ref{fig:28M_uv_tfinal}.

Thus, the steepest gradient of density is created by the most active
shell burning, mostly oxygen burning.  This implies
$M(V/U_{\rm max})$ $\simeq$ $M$(Si).  For $\sigma_{C12\alpha} = -1$,
$M$(Si) tends to be smaller than those for $\sigma_{C12\alpha} = 0$,
which would affect the explodability as has been discussed for $M
({\rm ZAMS})$ = 28 M$_{\odot}$.

\subsubsection{Core Masses and $\sigma_{C12\alpha}$}

Given the importance of these core masses for the explodability,
Figure \ref{fig:core_xi} shows the masses of Fe core $M({\rm Fe})$, Si
core $M({\rm Si})$, and O core $M({\rm O})$, as well as $\xi_{2.5}$ as
a function of $M({\rm ZAMS})$ at the final stages ($\tau=t_{\rm f}$)
for $\sigma_{C12\alpha}=$ 0, $-$1 and $-2$.  The data for all models
are also listed in Table \ref{tab:structure} and \ref{tab:core_mass}.
As mentioned earlier, the boundaries of $M({\rm Fe})$, $M({\rm Si})$,
and $M({\rm O})$ are defined at the location where the energy
generation rates of the Si-burning shell, O-burning shell, and
C-burning shell are the highest.

In these figures, $M({\rm O})$ and $\xi_{2.5}$ change with 
$M({\rm ZAMS})$ non-monotonously.  Their changes are similar, which
implies that $\xi_{2.5}$ is mainly affected by C shell burning,
as discussed earlier.  For smaller $\sigma_{C12\alpha}$,
$M({\rm O})$ tends to be smaller because $X({\rm C})$ is larger in the
deeper region, as has been shown in Figure \ref{fig:xco20_xi25} for
$M({\rm ZAMS})=$ 28 M$_\odot$.
Then the smaller $M({\rm O})$ leads to the formation of
smaller $M({\rm Si})$ and $M({\rm Fe})$.

\begin{figure}[htbp]
\centering
\begin{minipage}[c]{0.42\textwidth}
\includegraphics [width=75mm]{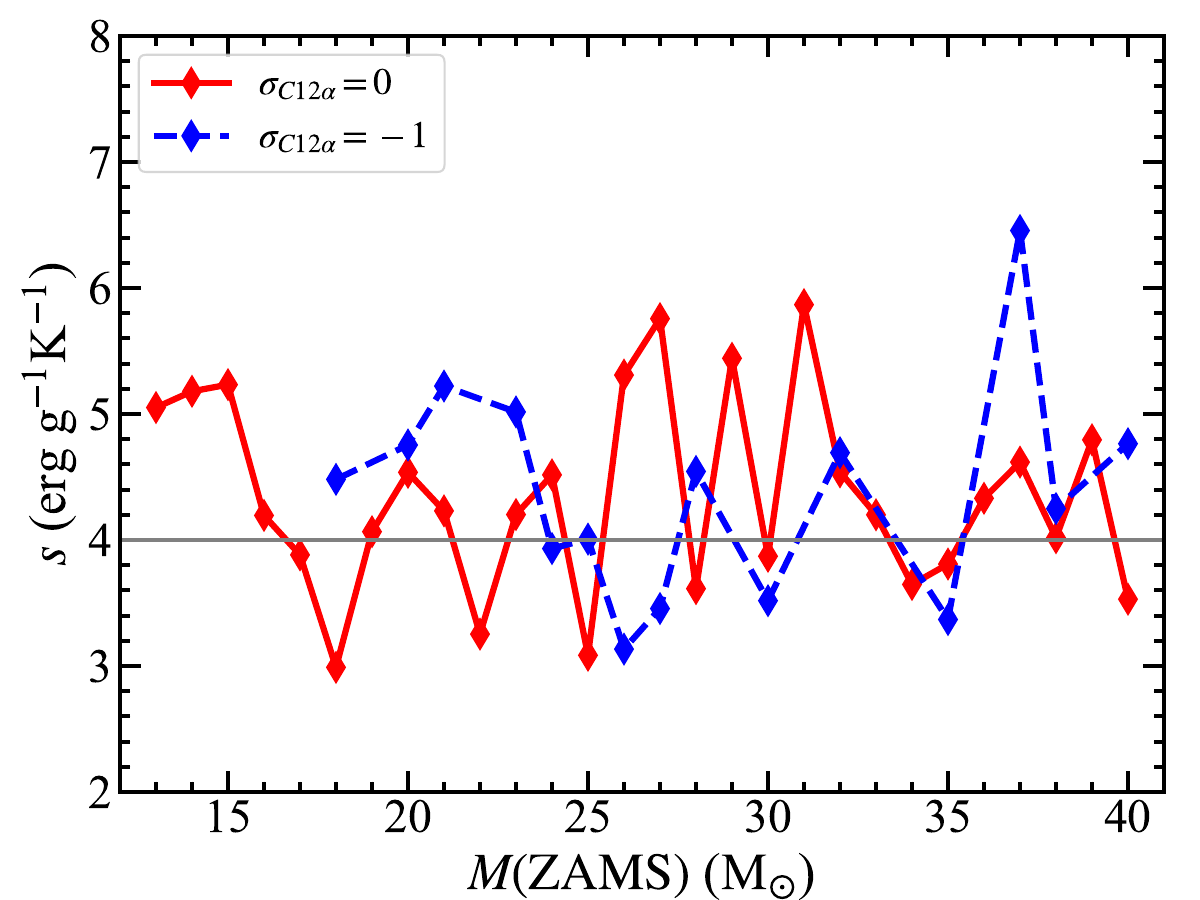}
\end{minipage}%
\caption{The entropy at $M(V/U_{\rm max})$ as a function of
$M({\rm ZAMS})$ for $\sigma_{C12\alpha}=0$ and -1.
The grey line shows where $s=4$ erg g$^{-1}$K$^{-1}$.
\label{fig:s_loc} }
\end{figure}

Figure~\ref{fig:core_eff} shows $M(V/U_{\rm max})$ normalized by
$M({\rm Si})$, which shows that $M(V/U_{\rm max})$ is almost identical
to $M({\rm Si})$ except for a few $M({\rm ZAMS})$.  This is because
O-shell burning is the strongest shell burning at the onset of
collapse and forms a very steep increase in entropy (Figure \ref{fig:28M_logp_rho})
and thus the steepest gradients of pressure and density as discussed
above for Figure~\ref{fig:logrho_s}.

Figure~\ref{fig:s_loc} shows the entropy at $(V/U_{\rm max})$, which
ranges from $s$(erg g$^{-1}$ K$^{-1}$) $=$ 3 to 6 including 4.  Thus
$M_4 = M(V/U_{\rm max})$ except for a few $M({\rm ZAMS})$ 
(see Table \ref{tab:structure}).
This is also due to the entropy jump at the O burning shell.

As discussed in Section \ref{sec:uv_curve}, $M(V/U_{\rm max}) = M_{\rm eff}$.
Therefore, smaller $M(V/U_{\rm max})$ for smaller $\sigma_{C12\alpha}$
leads to smaller $s_c$ and thus higher $\rho_c$ for the same $T_c$
(Equation \ref{equ:trho} and \ref{equ:s2}).            

\begin{figure*}[htbp]
\centering
\begin{minipage}[c]{0.48\textwidth}
\includegraphics [width=85mm]{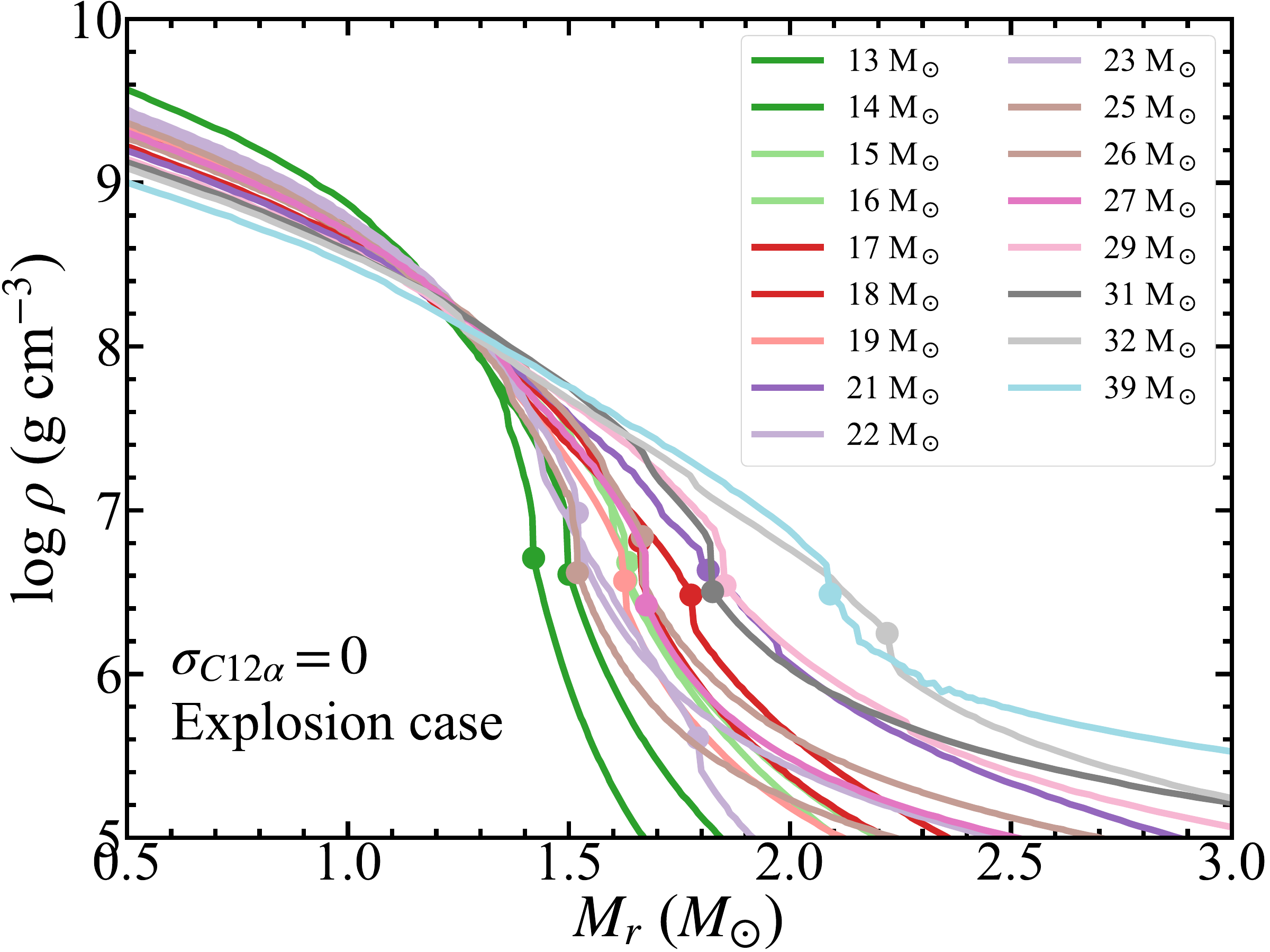}
\end{minipage}%
\begin{minipage}[c]{0.48\textwidth}
\includegraphics [width=85mm]{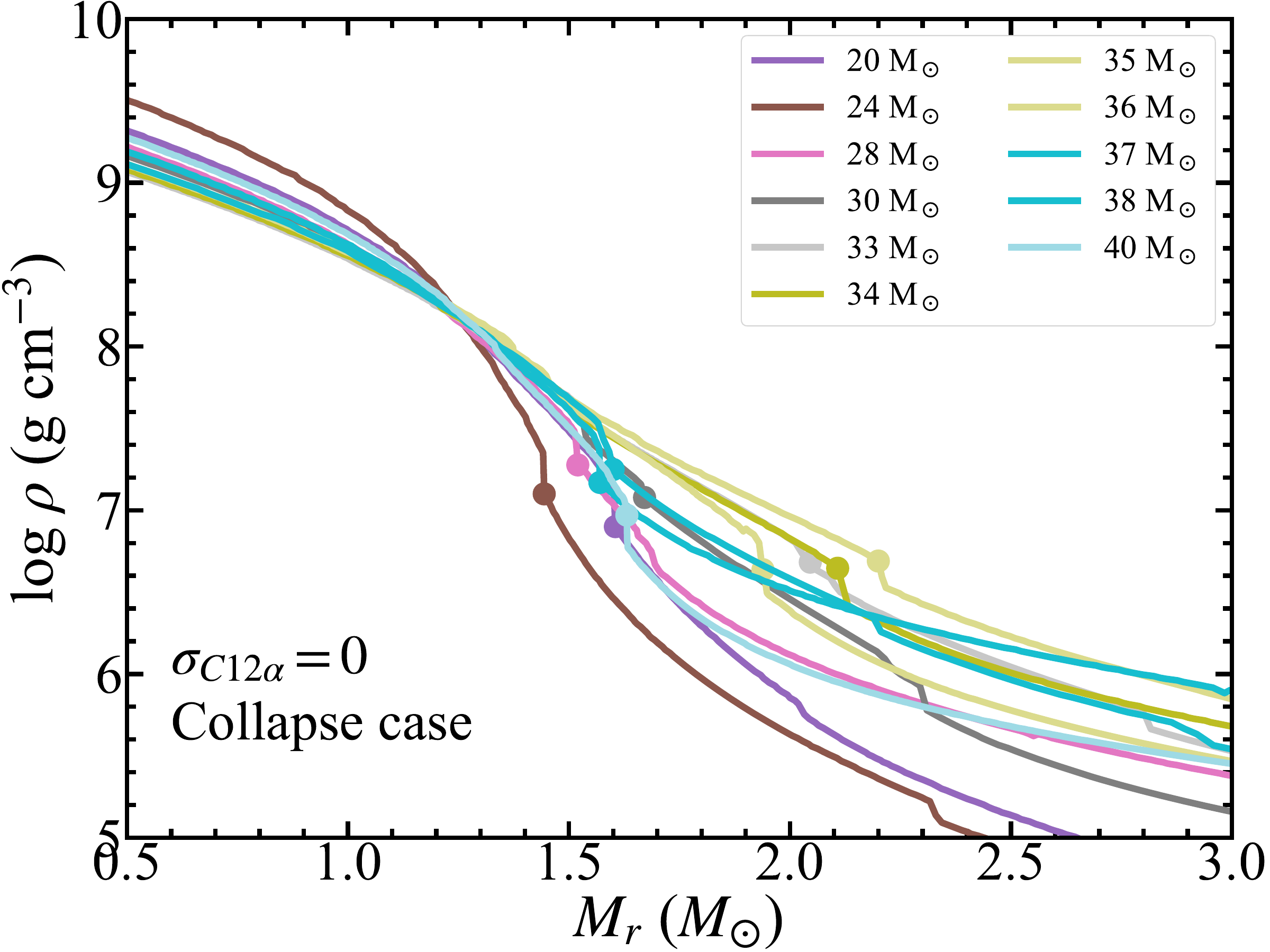}
\end{minipage}%

\begin{minipage}[c]{0.48\textwidth}
\includegraphics [width=85mm]{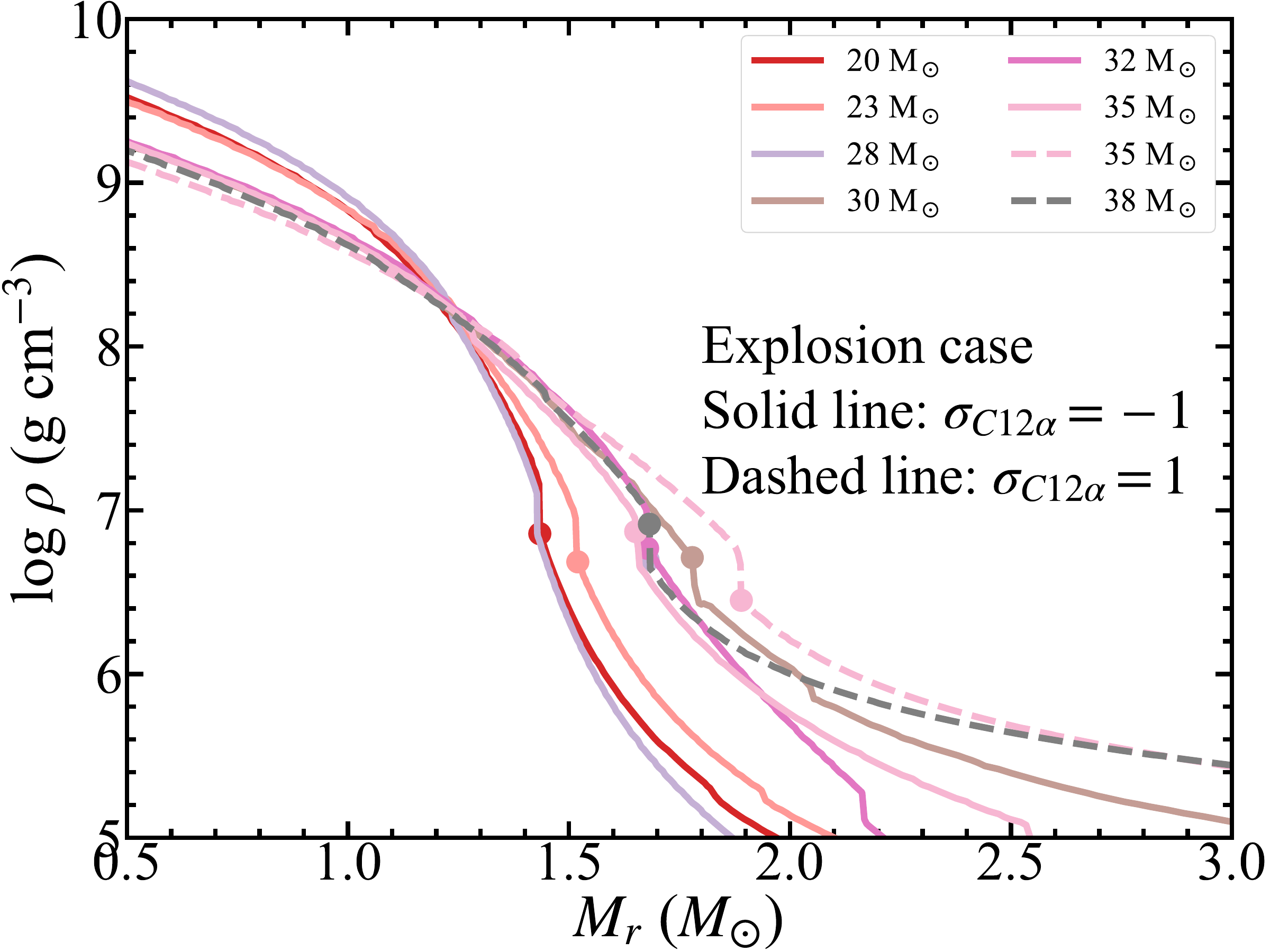}
\end{minipage}%
\begin{minipage}[c]{0.48\textwidth}
\includegraphics [width=85mm]{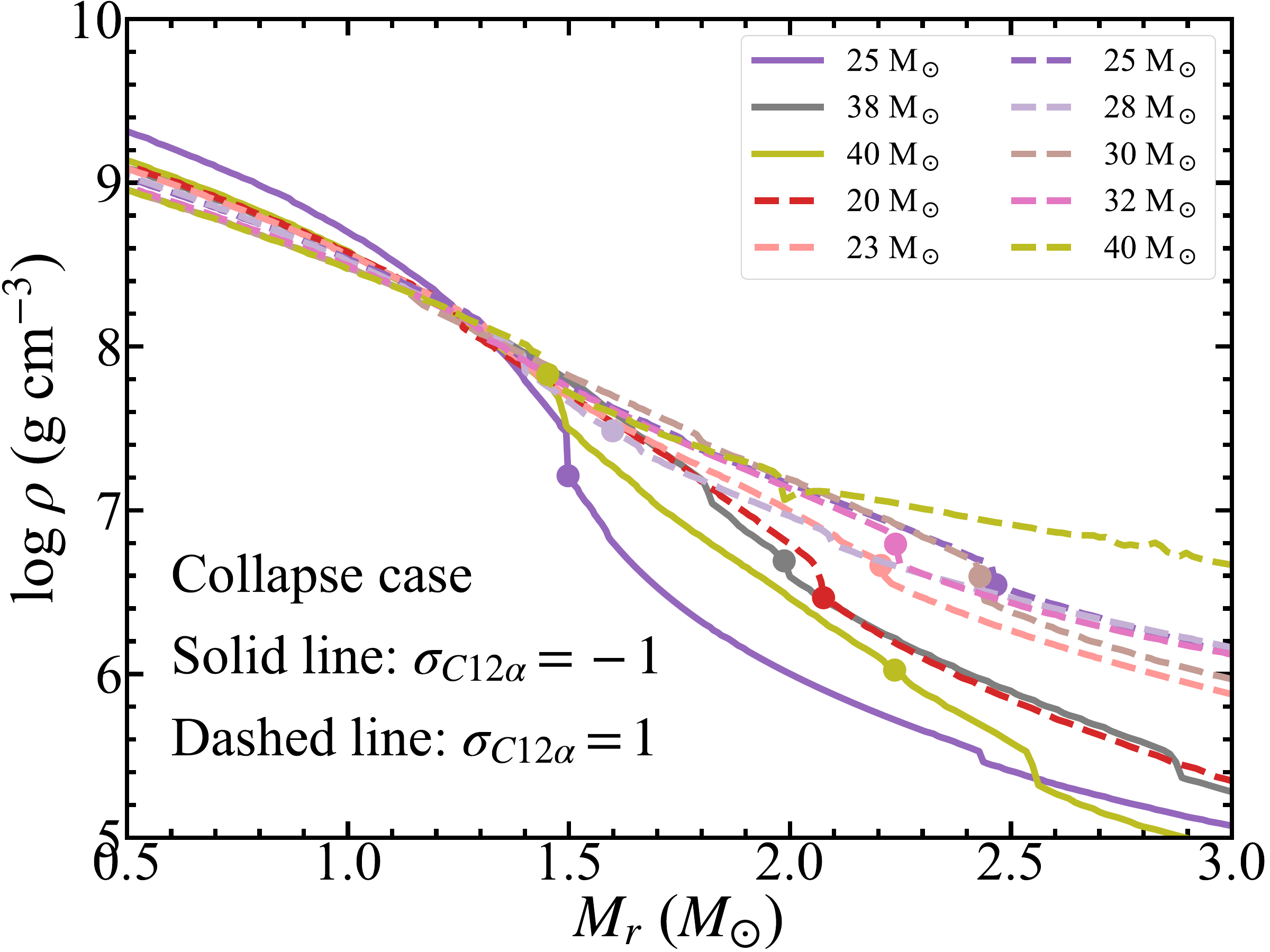}
\end{minipage}%
\caption{The density profile ranging from $0.7 <M_r<2.75$ M$_\odot$ for
different $M ({\rm ZAMS})$ and $\sigma_{C12\alpha}$.
(Top) The explosion (left) and collapse (right) models for $\sigma_{C12\alpha}=$ 0
are shown.
(Bottom) The explosion (left) and collapse (right) models for $\sigma_{C12\alpha}=$
-1 and 1 are shown by the solid lines and dashed lines, respectively.
The points on the lines show the location of $M(V/U_{\rm max})$ and its local density.
\label{fig:rho_jump}}
\end{figure*}

\subsubsection{Density Profiles and $\sigma_{C12\alpha}$}

Figures \ref{fig:rho_jump} show the density profiles at the final
pre-collapse models (i.e., at log $T_c$(K) $=$ 10.0) for
$\sigma_{C12\alpha}=$ 0, -1 and 1, On top figures, models for
``explosion'' (left) and ``collapse'' (right) are shown for
$\sigma_{C12\alpha}=0$.  At the bottom, models for ``explosion''
(left) and ``collapse'' (right) are shown for $\sigma_{C12\alpha}=$ -1
(solid lines) and 1 (dashed lines). The points on the lines show the
location of $V/U_{\rm max}$.

By comparing the models with the same $M({\rm ZAMS})$,
$M(V/U_{\rm max})$ ($= M_{\rm eff}$) is smaller and thus $\rho_c$ is
lower for smaller $\sigma_{C12\alpha}$.  This means that the density
distribution concerning $M_r$ tends to be steeper for smaller
$\sigma_{C12\alpha}$.  Also, the ``explosion'' models (left) tend to
have a steeper density structure than the ``collapse'' models (right).
This suggests that models with smaller $\sigma_{C12\alpha}$ explode
more easily, although the exact explodability depends on the behavior
of C shell burning.

To judge whether the modes are ``explosion'' or ``collapse'',
\citet{2021Natur.589...29B}, \citet{2022MNRAS.517..543W},
\citet{2023ApJ...949...17B} and \citet{2025MNRAS.537.1182B} have
suggested that the models with stronger density jumps at the Si/O
interface are more likely to explode and the location of this density
jump may be the initial boundary of the proto-NS based on their 2D and
3D simulations of the core collapse.  This is consistent with our
findings of the importance of $V/U_{\rm max}$, but our results can
provide the {\sl quantitative} criterion, i.e., the critical steepness
of the density for the ``explosion vs. collapse''.

\section{Summary and Discussion} \label{sec:summary}

Among the uncertainties of stellar evolution theory, we investigate
how the $^{12}$C($\alpha, \gamma$)$^{16}$O reaction rate affects the
evolution of massive stars through the beginning of the Fe core
collapse.  We adopt stars with the initial masses of $M ({\rm ZAMS})=$
13 - 40 M$_\odot$ and the solar metallicity.  For the reaction rate,
we adopt $\sigma_{C12\alpha}= -3, -2, -1, 0, 1, 2$, and 3, i.e., by a
factor of 0.37, 0.52, 0.78, 1, 1.28, 1.93, and 2.69 of the standard
rate.

Comparison between $\sigma_{C12\alpha}= -1$ and 0 is made particularly
in detail for the evolution of the $M ({\rm ZAMS})=$ 28 M$_\odot$
stars.  For $M ({\rm ZAMS})=$ 25 M$_\odot$, models with
$\sigma_{C12\alpha}= -2$, $-1$, and 0 are compared, and for the
$M ({\rm ZAMS})=$ 35 M$_\odot$ models, $\sigma_{C12\alpha}= 0$ and $+1$
are examined.  For the explodability, models with the whole range of
$M ({\rm ZAMS})$ and $\sigma_{C12\alpha}$ are calculated.

We find the following differences:

(1) Smaller $\sigma_{C12\alpha}$ leads to the synthesis of a larger mass
fraction of C, $X$(C), which makes C shell burning stronger.
Then, the convection zone extends to mix C from the overlying layer,
further strengthening C shell burning.  The extra heating by C shell
burning tends to prevent the contraction of outer layers and decrease
the {\sl compactness parameter} at $M_r$ = 2.5 M$_\odot$.

(2) This effect leads to the formation of smaller mass cores of Si and
Fe and the steeper density and pressure gradients at the O burning
shell in the presupernova models.

(3) This difference affects the {\sl explodability} of the models,
i.e., whether a NS or a BH is formed,
The steeper pressure gradient results in more likely the explosion forming
a NS rather than a BH.

(4) We describe the pressure gradient against $M_r$ with $V/U$ and the
density drop with $1/U$, where $U$ and $V$ are non-dimensional
variables to describe the stellar structure (Equation.~\ref{equ:u} and~\ref{equ:v}).

(5) By applying the Ertl-criterion for explodability, we estimate
the critical values of $V/U$ and $1/U$ at the O-burning shell above
which the model is more likely to explode.

(6) We conclude that the smaller $^{12}$C($\alpha, \gamma$)$^{16}$O
reaction rate makes the mass range of $M ({\rm ZAMS})$ that forms a
NS larger.

(7) The dependencies of the compactness parameter on $M ({\rm ZAMS})$ and
the reaction rate is consistent with the earlier results by
\citet{2014Sukhbold} and \citet{2020ApJ...890...43C}
but our study covers a much wider range of input parameters.

(8) In view of the importance of the overshooting of C shell burning
to enhance $X$(C) by mixing fresh C, multi-D simulations of
convective overshooting \citep{2019MNRAS.484.4645C, 2022MNRAS.515.4013R}
are critically important.

In the forthcoming papers, we will present the explodability by
hydrodynamical calculations of collapse for our progenitor models with
different $\sigma_{C12\alpha}$ rather than using the Ertl-criterion.
We will also present how the detailed nucleosynthesis yields depend on
$\sigma_{C12\alpha}$.

\section{Acknowledgements} \label{sec:ack}

This work was supported by the National Natural Science Foundation of China under
Grant Nos. 11988101 and the National Key R\&D Program of China No.
2024YFA1611900.
W. Y. X. is supported by the Cultivation Project for LAMOST Scientific Payoff
and Research Achievement. W. Y. X. thanks Xianfei Zhang and Shaolan Bi for 
their financial support (the grants 12073006, 12090040 and 12090042 from the
National Natural Science Foundation of China) and also their help in using MESA.
K. N. is supported by the World Premier International Research Center Initiative
(WPI), MEXT, Japan, and the Japan Society for the Promotion of Science
JSPS KAKENHI Grant Numbers JP20K04024, JP21H044pp, and JP23K03452.
We would like to thank Marco Limongi and Raphael Hirschi for useful discussion.

\appendix

\section{Property of Explosion} \label{sec:exp_info}

\begin{longtable*}{ccccccccccc}
\caption{Presupernova properties for varying $\sigma_{C12\alpha}$ and $M ({\rm ZAMS})$
at the final stage ($T_c=10^{10}$ K).}

\label{tab:structure}\\
\toprule[1pt]
$\sigma_{C12\alpha}$  & $M ({\rm ZAMS})$& log $(V/U_{\rm max})$ & log $V_{\rm max}$ & log $U_{\rm min}$  & $M(V/U_{\rm max})$ &$M_4$& $\mu_4$  & $M_4 \mu_4$ & $\xi_{2.5}$ & fate\\
&(M$_\odot$)&    &   &    & (M$_\odot$) & (M$_\odot$) &      &      &    &\\
\midrule[1pt]
\endfirsthead

\multicolumn{7}{l}{\autoref{tab:structure} $-$ continued}\\
\toprule[1pt]
$\sigma_{C12\alpha}$  & $M ({\rm ZAMS})$& log $(V/U_{\rm max})$ & log $V_{\rm max}$ & log $U_{\rm min}$  & $M(V/U_{\rm max})$ &$M_4$& $\mu_4$  & $M_4 \mu_4$ & $\xi_{2.5}$ & fate\\
&(M$_\odot$)&    &   &    & (M$_\odot$) & (M$_\odot$) &      &      &    &\\
\midrule[1pt]
\endhead

\bottomrule[1pt]
\endlastfoot

\multirow{9}{*}{-3$\sigma$}  & 20  & 1.167 & 0.696 & -0.632 & 1.531    & 1.597 & 0.100 & 0.160 & 0.255      & uncertain \\
                     & 23    & 1.590 & 0.723 & -1.104 & 1.680      & 1.677 & 0.026 & 0.043 & 0.104      & explosion\\
                     & 25    & 1.975 & 0.803 & -1.471 & 1.545      & 1.544 & 0.008 & 0.013 & 0.037      & explosion\\
                     & 28    & 1.631 & 0.779 & -1.158 & 1.793      & 1.762 & 0.016 & 0.029 & 0.074      & explosion\\
                     & 30    & 1.600 & 0.745 & -1.135 & 1.877      & 1.868 & 0.024 & 0.045 & 0.125      & explosion\\
                     & 32    & 1.593 & 0.764 & -1.116 & 1.863      & 1.687 & 0.031 & 0.052 & 0.106      & explosion\\
                     & 35    & 1.825 & 0.809 & -1.348 & 1.654      & 1.504 & 0.019 & 0.029 & 0.063      & explosion\\
                     & 38    & 1.517 & 0.795 & -0.996 & 1.523      & 1.529 & 0.046 & 0.071 & 0.117      & explosion\\
                     & 40    & 1.486 & 0.799 & -0.995 & 1.512      & 1.511 & 0.058 & 0.088 & 0.189      & explosion\\
\midrule[1pt]
\multirow{4}{*}{-2$\sigma$}    & 20  & 1.652 & 0.756 & -1.134 & 1.395     & 1.394 & 0.018 & 0.025 & 0.044     & explosion \\
                     & 23    & 1.338 & 0.754 & -0.819 & 1.667      & 1.662 & 0.073 & 0.121 & 0.205      & explosion\\
                     & 25    & 1.367 & 0.707 & -0.830 & 1.631      & 1.631 & 0.057 & 0.093 & 0.173      & explosion \\
                     & 28    & 1.699 & 0.821 & -1.230 & 1.633      & 1.386 & 0.029 & 0.040 & 0.057      & explosion \\
\multirow{5}{*}{-2$\sigma$}    & 30    & 1.432 & 0.763 & -0.996 & 1.429      & 1.429 & 0.051 & 0.072 & 0.090      & explosion \\
                     & 32    & 1.366 & 0.749 & -0.934 & 1.956      & 1.463 & 0.079 & 0.115 & 0.109      & explosion \\
                     & 35    & 1.274 & 0.754 & -0.749 & 1.354      & 1.489 & 0.094 & 0.140 & 0.167      & explosion \\
                     & 38    & 1.242 & 0.705 & -0.800 & 1.830      & 1.826 & 0.089 & 0.163 & 0.289      & uncertain \\
                     & 40    & 1.251 & 0.683 & -0.852 & 2.377      & 2.112 & 0.069 & 0.147 & 0.293      & uncertain \\
\midrule[1pt]
\multirow{15}{*}{-1$\sigma$} & 18    & 1.345 & 0.715 & -0.848 & 1.645    & 1.631 & 0.056 & 0.092 & 0.174    & explosion \\
                     & 20    & 1.540 & 0.795 & -1.025 & 1.434    & 1.434 & 0.049 & 0.070 & 0.135     & explosion \\
                     & 21    & 1.564 & 0.807 & -1.056 & 1.542    & 1.541 & 0.046 & 0.070 & 0.160     & explosion \\
                     & 23    & 1.480 & 0.735 & -0.979 & 1.520    & 1.515 & 0.052 & 0.079 & 0.155     & explosion \\
                     & 24    & 1.244 & 0.713 & -0.770 & 1.692    & 1.692 & 0.073 & 0.124 & 0.237     & uncertain \\
                     & 25    & 1.248 & 0.741 & -0.736 & 1.498    & 1.498 & 0.108 & 0.162 & 0.294     & uncertain \\
                     & 26    & 1.310 & 0.740 & -0.809 & 1.465    & 1.471 & 0.091 & 0.134 & 0.254     & uncertain \\
                     & 27    & 1.294 & 0.693 & -0.848 & 1.721    & 1.721 & 0.076 & 0.132 & 0.282     & uncertain \\
                     & 28    & 1.566 & 0.769 & -1.035 & 1.429    & 1.428 & 0.044 & 0.063 & 0.109     & explosion \\
                     & 30    & 1.222 & 0.657 & -0.804 & 1.779    & 1.779 & 0.081 & 0.145 & 0.286     & uncertain \\
                     & 32    & 1.338 & 0.724 & -0.817 & 1.678    & 1.678 & 0.071 & 0.119 & 0.173     & explosion \\
                     & 35    & 1.325 & 0.729 & -0.841 & 1.651    & 1.657 & 0.073 & 0.121 & 0.236     & explosion \\
                     & 37    & 1.122 & 0.688 & -0.728 & 2.450    & 2.119 & 0.096 & 0.204 & 0.351     & uncertain \\
                     & 38    & 1.058 & 0.634 & -0.586 & 1.987    & 1.958 & 0.127 & 0.249 & 0.438     & collapse \\
                     & 40    & 1.052 & 0.684 & -0.662 & 2.237    & 1.988 & 0.112 & 0.223 & 0.365     & collapse \\
\midrule[1pt]
\multirow{26}{*}{0$\sigma$}  & 13    & 1.685 & 0.814 & -1.148 & 1.420     & 1.417 & 0.028 & 0.040 & 0.036   & explosion \\
                     & 14    & 1.581 & 0.820 & -1.058 & 1.500      & 1.497 & 0.037 & 0.056 & 0.087    & explosion \\
                     & 15    & 1.447 & 0.751 & -0.938 & 1.629      & 1.625 & 0.051 & 0.082 & 0.139    & explosion \\
                     & 16    & 1.413 & 0.748 & -0.912 & 1.633      & 1.627 & 0.055 & 0.090 & 0.163    & explosion \\
                     & 17    & 1.327 & 0.672 & -0.834 & 1.776      & 1.776 & 0.057 & 0.101 & 0.199    & explosion \\
                     & 18    & 1.443 & 0.779 & -0.937 & 1.660      & 1.660 & 0.051 & 0.084 & 0.170    & explosion \\
                     & 19    & 1.488 & 0.749 & -0.992 & 1.626      & 1.626 & 0.045 & 0.074 & 0.155    & explosion \\
                     & 20    & 1.274 & 0.708 & -0.776 & 1.605      & 1.605 & 0.090 & 0.144 & 0.243    & explosion \\
                     & 21    & 1.168 & 0.644 & -0.693 & 1.815      & 1.802 & 0.086 & 0.155 & 0.287    & uncertain \\
                     & 22    & 1.466 & 0.773 & -0.987 & 1.519      & 1.519 & 0.061 & 0.092 & 0.197    & explosion \\
                     & 23    & 1.392 & 0.777 & -0.931 & 1.791      & 1.514 & 0.059 & 0.089 & 0.116    & explosion \\
                     & 24   & 1.364 & 0.722 & -0.871 & 1.444     & 1.444 & 0.084 & 0.121 & 0.211          & explosion\\
                     & 25     & 1.437 & 0.742 & -0.970 & 1.666      & 1.666 & 0.059 & 0.098 & 0.226    & explosion \\
                     & 26     & 1.529 & 0.798 & -1.041 & 1.519      & 1.519 & 0.049 & 0.074 & 0.172    & explosion \\
                     & 27     & 1.459 & 0.755 & -0.994 & 1.675      & 1.673 & 0.053 & 0.089 & 0.205    & explosion \\
                     & 28     & 1.153 & 0.715 & -0.686 & 1.520      & 1.656 & 0.104 & 0.172 & 0.334    & uncertain \\
                     & 29     & 1.252 & 0.711 & -0.787 & 1.853      & 1.848 & 0.082 & 0.151 & 0.303    & uncertain \\
                     & 30     & 1.038 & 0.655 & -0.620 & 1.671      & 1.786 & 0.143 & 0.255 & 0.353    & collapse \\
                     & 31     & 1.302 & 0.717 & -0.868 & 1.825      & 1.819 & 0.076 & 0.139 & 0.303    & uncertain \\
                     & 32     & 1.152 & 0.637 & -0.719 & 2.220      & 2.114 & 0.085 & 0.179 & 0.390    & collapse\\
                     & 33     & 0.984 & 0.610 & -0.525 & 2.046      & 2.024 & 0.142 & 0.287 & 0.482    & collapse \\
                    & 34     & 1.002 & 0.618 & -0.593 & 2.108      & 2.108 & 0.122 & 0.258 & 0.469    & collapse \\
                     & 35     & 0.934 & 0.628 & -0.506 & 2.200      & 2.200 & 0.150 & 0.331 & 0.554    & collapse \\
                     & 36     & 1.152 & 0.686 & -0.725 & 1.939      & 1.936 & 0.100 & 0.193 & 0.388    & collapse \\
                     & 37     & 1.091 & 0.660 & -0.671 & 1.570      & 1.570 & 0.151 & 0.237 & 0.461    & collapse \\
                     & 38     & 1.016 & 0.664 & -0.550 & 1.600      & 1.600 & 0.184 & 0.294 & 0.440    & collapse \\
\multirow{2}{*}{0$\sigma$}   & 39     & 1.149 & 0.658 & -0.748 & 2.091      & 2.082 & 0.086 & 0.179 & 0.409    & collapse \\
                     & 40     & 1.273 & 0.691 & -0.848 & 1.631      & 1.631 & 0.093 & 0.151 & 0.328    & uncertain \\
\midrule[1pt]
\multirow{14}{*}{+1$\sigma$}  & 13    & 1.746 & 0.775 & -1.243 & 2.363     & 1.498 & 0.043 & 0.065 & 0.039     & explosion \\
                     & 16      & 1.080 & 0.678 & -0.566 & 2.595     & 1.840 & 0.112 & 0.205 & 0.335     & uncertain\\
                     & 20      & 1.094 & 0.625 & -0.632 & 2.076     & 2.058 & 0.110 & 0.226 & 0.434     & collapse \\
                     & 22      & 1.175 & 0.694 & -0.712 & 1.947     & 1.947 & 0.098 & 0.192 & 0.390     & collapse \\
                     & 23      & 0.926 & 0.623 & -0.482 & 2.206     & 2.146 & 0.167 & 0.358 & 0.568     & collapse \\
                     & 24      & 0.980 & 0.636 & -0.554 & 2.075     & 2.093 & 0.132 & 0.276 & 0.478     & collapse \\
                     & 25      & 0.863 & 0.601 & -0.434 & 2.466     & 2.391 & 0.195 & 0.467 & 0.699     & collapse \\
                     & 26      & 1.055 & 0.614 & -0.620 & 2.109     & 2.109 & 0.119 & 0.251 & 0.498     & collapse \\
                     & 28      & 0.819 & 0.631 & -0.367 & 1.599     & 2.057 & 0.220 & 0.453 & 0.605     & collapse \\
                     & 30      & 0.965 & 0.614 & -0.556 & 2.430     & 2.397 & 0.143 & 0.343 & 0.682     & collapse \\
                     & 32      & 0.877 & 0.592 & -0.480 & 2.239     & 2.239 & 0.181 & 0.404 & 0.638     & collapse \\
                     & 35      & 1.206 & 0.678 & -0.807 & 1.890     & 1.888 & 0.085 & 0.160 & 0.351     & uncertain\\
                     & 38      & 1.299 & 0.719 & -0.887 & 1.682     & 1.682 & 0.084 & 0.141 & 0.317     & uncertain\\
                     & 40      & 0.703 & 0.596 & -0.370 & 1.452     & 1.885 & 0.340 & 0.642 & 0.767     & collapse\\
\midrule[1pt]
\multirow{12}{*}{+2$\sigma$}  & 13    & 1.993 & 0.822 & -1.480 & 2.428     & 1.404 & 0.033 & 0.046 & 0.018    & explosion\\
                     & 15      & 1.263 & 0.669 & -0.722 & 2.597     & 1.738 & 0.084 & 0.146 & 0.235    & explosion\\
                     & 18      & 0.936 & 0.674 & -0.432 & 1.405     & 1.796 & 0.195 & 0.350 & 0.494    & collapse\\
                     & 20      & 1.216 & 0.666 & -0.745 & 1.831     & 1.831 & 0.083 & 0.151 & 0.303    & uncertain\\
                     & 23      & 1.557 & 0.813 & -1.064 & 1.482     & 1.480 & 0.045 & 0.067 & 0.154    & explosion \\
                     & 25      & 0.819 & 0.629 & -0.368 & 1.384     & 1.714 & 0.266 & 0.455 & 0.638    & collapse \\
                     & 28      & 1.103 & 0.702 & -0.614 & 1.578     & 1.587 & 0.136 & 0.216 & 0.337    & uncertain \\
                     & 30      & 0.808 & 0.570 & -0.276 & 2.593     & 2.506 & 0.212 & 0.530 & 0.755    & collapse \\
                     & 32      & 0.666 & 0.501 & -0.198 & 2.603     & 2.172 & 0.395 & 0.859 & 0.832    & collapse\\
                     & 35      & 1.506 & 0.794 & -1.055 & 1.630     & 1.630 & 0.051 & 0.083 & 0.213    & explosion \\
                     & 38      & 1.389 & 0.759 & -0.957 & 1.567     & 1.570 & 0.078 & 0.122 & 0.290    & uncertain \\
                     & 40      & 1.030 & 0.645 & -0.669 & 2.358     & 2.358 & 0.114 & 0.270 & 0.601    & collapse \\
\midrule[1pt]
\multirow{12}{*}{+3$\sigma$} & 13    & 1.967 & 0.730 & -1.452 & 2.414   & 1.823 & 0.084 & 0.153 & 0.071   & explosion\\
                     & 15        & 1.615 & 0.819 & -1.081 & 1.458     & 1.458 & 0.038 & 0.055 & 0.098   & explosion\\
                     & 18        & 1.107 & 0.646 & -0.638 & 1.810     & 1.810 & 0.116 & 0.211 & 0.373   & collapse\\
                     & 20    & 0.861 & 0.621 & -0.399 & 1.674   & 2.065 & 0.205 & 0.423 & 0.583   & collapse\\
                     & 23    & 1.362 & 0.765 & -0.856 & 1.386       & 1.385 & 0.093 & 0.129 & 0.261    & uncertain\\
                     & 25    & 0.912 & 0.583 & -0.499 & 2.260       & 2.251 & 0.168 & 0.377 & 0.631    & collapse\\
                     & 28    & 0.679 & 0.546 & -0.261 & 1.789       & 1.867 & 0.368 & 0.686 & 0.787    & collapse\\
                     & 30    & 1.148 & 0.683 & -0.726 & 1.973       & 1.973 & 0.102 & 0.202 & 0.414    & collapse\\
                     & 32    & 1.359 & 0.770 & -0.902 & 1.487       & 1.553 & 0.069 & 0.107 & 0.253    & explosion\\
                     & 35    & 0.657 & 0.506 & -0.151 & 2.596       & 2.696 & 0.381 & 1.028 & 0.880    & collapse\\
                     & 38    & 0.695 & 0.591 & -0.412 & 1.529       & 1.888 & 0.331 & 0.626 & 0.741    & collapse\\
                     & 40    & 1.294 & 0.728 & -0.893 & 1.859       & 1.859 & 0.077 & 0.143 & 0.329    & uncertain
\end{longtable*}

\section{Information of Cores} \label{sec:core_info}
\begin{longtable*}{cccccccc}
\caption{The core masses for varying $\sigma_{C12\alpha}$ and
$M ({\rm ZAMS})$ at the final stage ($T_c=10^{10}$).}\label{tab:core_mass}\\
\toprule[1pt]
$\sigma_{C12\alpha}$ & $M ({\rm ZAMS})$ (M$_\odot$) & $M$(Fe) (M$_\odot$) & $M$(Si) (M$_\odot$) &
$M$(O) (M$_\odot$)  & $M$(CO) (M$_\odot$)  & $M$(He) (M$_\odot$)  \\
\midrule[1pt]
\endfirsthead

\multicolumn{7}{l}{\autoref{tab:core_mass} $-$ continued}\\
\toprule[1pt]
$\sigma_{C12\alpha}$ & $M ({\rm ZAMS})$ (M$_\odot$) & $M$(Fe) (M$_\odot$) & $M$(Si) (M$_\odot$) &
$M$(O) (M$_\odot$)  & $M$(CO) (M$_\odot$)  & $M$(He) (M$_\odot$) \\
\midrule[1pt]
\endhead

\bottomrule[1pt]
\endlastfoot
\multirow{9}{*}{$-$3$\sigma$}& 20 &1.45 & 1.77 & 2.19 & 4.26&  6.43 \\
                              & 23 &1.69 & 1.52 & 1.44 & 5.19&  5.54 \\
                              & 25 &1.34 & 1.38 & 1.54 & 4.76&  8.50 \\
                              & 28 &1.38 & 1.51 & 1.79 & 3.46&  6.36 \\
                              & 30 &1.49 & 1.63 & 1.88 & 7.97&  9.18 \\
                              & 32 &1.45 & 1.58 & 1.75 & 8.23&  9.55 \\
                              & 35 &1.40 & 1.51 & 1.66 & 8.58&  9.89 \\
                              & 38 &1.41 & 1.54 & 1.64 & 10.06&  10.08\\
                              & 40 &1.41 & 1.50 & 1.65 & 11.79&  13.00 \\
\midrule[1pt]
\multirow{10}{*}{$-$2$\sigma$}& 20 &1.33 & 1.39 & 1.61 & 2.41&  6.57 \\
                              & 23 &1.49 & 1.70 & 1.99 & 4.99&  7.22 \\
                              & 25 &1.48 & 1.63 & 1.90 & 5.88&  8.37 \\
                              & 28 &1.33 & 1.43 & 1.63 & 5.93&  7.71 \\
                              & 30 &1.37 & 1.47 & 1.76 & 5.55&  8.22 \\
                              & 32 &1.40 & 1.53 & 1.97 & 3.69&  7.50 \\
                              & 35 &1.43 & 1.54 & 2.19 & 7.46&  9.51 \\
                              & 38 &1.59 & 1.82 & 2.49 & 10.55&  10.60 \\
                              & 40 &1.59 & 1.87 & 2.39 & 11.33&  11.73 \\
\midrule[1pt]
\multirow{15}{*}{$-$1$\sigma$} & 18 &1.50 & 1.63 & 1.75 & 3.73&  5.77 \\
                              & 20 &1.36 & 1.46 & 1.84 & 4.30&  6.42 \\
                              & 21 &1.41 & 1.54 & 1.64 & 4.57&  6.87 \\
                              & 23 &1.41 & 1.52 & 1.95 & 5.33&  6.07  \\
                              & 24 &1.51 & 1.63 & 2.37 & 5.65&  6.51  \\
                              & 25 &1.44 & 1.57 & 2.44 & 5.85&  7.58 \\
                              & 26 &1.48 & 1.57 & 2.35 & 5.86&  8.56 \\
                              & 27 &1.54 & 1.72 & 2.75 & 6.61&  8.21 \\
                              & 28 &1.36 & 1.44 & 1.67 & 6.69&  7.32 \\
                              & 30 &1.56 & 1.80 & 2.06 & 7.78&  8.42 \\
                              & 32 &1.52 & 1.68 & 2.17 & 7.42&  8.71 \\
                              & 35 &1.50 & 1.66 & 2.57 & 7.42&  11.06 \\
                              & 37 &1.59 & 1.78 & 2.46 & 8.36&  12.15 \\
                              & 38 &1.69 & 2.00 & 2.88 & 10.99&  13.07 \\
                              & 40 &1.63 & 1.79 & 2.57 & 10.47&  13.29 \\
\midrule[1pt]
\multirow{12}{*}{0$\sigma$}   & 13 &1.36 & 1.42 & 1.48 & 2.30&  3.83 \\
                             & 14 &1.38 & 1.50 & 1.57 & 2.63&  4.22 \\
                             & 15 &1.46 & 1.63 & 1.70 & 2.92&  4.60 \\
                             & 16 &1.45 & 1.60 & 1.72 & 3.19&  4.93 \\
                             & 17 &1.51 & 1.79 & 1.82 & 3.42&  5.40 \\
                             & 18 &1.46 & 1.67 & 1.75 & 3.54&  5.63 \\
                             & 19 &1.45 & 1.64 & 1.69 & 4.41&  6.34 \\
                             & 20 &1.49 & 1.65 & 2.07 & 4.30&  6.52 \\
                             & 21 &1.58 & 1.83 & 1.98 & 4.53&  6.76 \\
                             & 22 &1.42 & 1.54 & 2.49 & 4.86&  6.04 \\
                             & 23 &1.45 & 1.52 & 1.80 & 4.94&  5.66 \\
                             & 24 &1.19 & 1.48 & 2.35 & 5.74&  7.01 \\
\multirow{16}{*}{0$\sigma$}  & 25 &1.48 & 1.67 & 1.80 & 5.71&  7.13 \\
                             & 26 &1.42 & 1.54 & 1.64 & 5.99&  7.53 \\
                             & 27 &1.48 & 1.68 & 1.77 & 6.66&  8.35 \\
                             & 28 &1.51 & 1.71 & 2.59 & 6.74&  8.79 \\
                             & 29 &1.58 & 1.85 & 2.34 & 7.43&  9.04 \\
                             & 30 &1.61 & 1.80 & 2.33 & 7.97&  9.92 \\
                             & 31 &1.58 & 1.83 & 2.03 & 8.26&  9.77 \\
                             & 32 &1.72 & 1.93 & 2.26 & 9.18&  10.17\\
                             & 33 &1.71 & 2.12 & 2.81 & 8.91&  10.74 \\
                             & 34 &1.70 & 2.13 & 3.23 & 9.31&  12.20 \\
                             & 35 &1.76 & 2.21 & 3.17 & 8.59&  12.62 \\
                             & 36 &1.65 & 1.95 & 3.15 & 9.88&  12.40 \\
                             & 37 &1.65 & 1.90 & 3.93 & 10.41&  13.12 \\
                             & 38 &1.68 & 2.29 & 2.96 & 15.51&  13.69 \\
                             & 39 &1.76 & 2.14 & 2.42 & 12.60&  14.76 \\
                             & 40 &1.51 & 1.70 & 3.58 & 11.56&  14.08 \\
\midrule[1pt]
\multirow{13}{*}{+1$\sigma$}  & 13 &1.35 & 1.50 & 1.57 & 2.31&  3.87 \\
                             & 16 &1.56 & 1.72 & 2.00 & 3.28&  5.17 \\
                             & 20 &1.70 & 2.07 & 2.24 & 4.55&  6.66 \\
                             & 22 &1.67 & 1.95 & 2.67 & 5.30&  7.47 \\
                             & 23 &1.77 & 2.21 & 2.60 & 5.58&  7.88 \\
                             & 24 &1.71 & 2.10 & 2.39 & 5.76&  7.20 \\
                             & 25 &1.88 & 2.47 & 2.93 & 6.07&  8.17 \\
                             & 26 &1.79 & 2.12 & 2.45 & 6.36&  8.34 \\
                             & 28 &1.81 & 2.19 & 3.16 & 7.43&  10.12\\
                             & 30 &1.90 & 2.44 & 2.84 & 7.98&  9.99  \\
                             & 32 &1.98 & 2.45 & 3.57 & 8.71&  10.92 \\
                             & 35 &1.61 & 1.89 & 5.07 & 9.72&  12.02 \\
                             & 38 &1.53 & 1.74 & 4.85 & 10.86& 12.81 \\
                             & 40 &2.11 & 3.33 & 4.72 & 11.88& 14.76 \\
\midrule[1pt]
\multirow{14}{*}{+2$\sigma$} & 13 &1.34 & 1.41 & 1.47 & 2.39&  3.97  \\
                             & 15 &1.54 & 1.74 & 1.87 & 2.82&  4.57  \\
                             & 18 &1.67 & 1.92 & 2.39 & 3.85&  4.91  \\       
                             & 20 &1.57 & 1.84 & 2.00 & 4.48&  5.50  \\
                             & 23 &1.36 & 1.48 & 1.58 & 5.47&  6.55  \\
                             & 25 &1.86 & 2.36 & 3.75 & 6.20&  7.58  \\
                             & 28 &1.58 & 1.74 & 2.49 & 7.25&  8.81  \\
                             & 30 &2.09 & 2.66 & 4.58 & 8.12&  10.07 \\
                             & 32 &2.26 & 3.19 & 5.00 & 8.81&  10.66 \\
                             & 35 &1.46 & 1.63 & 1.76 & 9.48&  11.52 \\
                             & 38 &1.46 & 1.76 & 2.69 & 11.01& 12.89 \\
                             & 40 &1.85 & 2.37 & 3.37 & 11.96& 14.25 \\
\midrule[1pt]
\multirow{4}{*}{+3$\sigma$}  & 13 &1.54 & 1.67 & 2.01 & 2.34&  3.69  \\
                             & 15 &1.36 & 1.46 & 1.53 & 2.79&  3.33  \\
                             & 18 &1.60 & 1.82 & 2.94 & 3.90&  4.61  \\
                             & 20 &1.76 & 2.07 & 3.43 & 4.56&  5.54  \\
\multirow{8}{*}{+3$\sigma$}  & 23 &1.37 & 1.58 & 2.08 & 5.54&  6.81  \\
                             & 25 &1.85 & 2.25 & 4.16 & 6.30&  7.88  \\
                             & 28 &2.09 & 2.99 & 5.26 & 7.42&  9.26  \\
                             & 30 &1.66 & 1.99 & 5.50 & 8.03&  9.83  \\
                             & 32 &1.42 & 1.57 & 4.77 & 8.99&  10.47 \\
                             & 35 &2.52 & 3.10 & 6.54 & 10.00& 12.25 \\
                             & 38 &2.39 & 2.81 & 6.48 & 11.01& 12.88 \\
                             & 40 &1.59 & 1.86 & 6.91 & 11.73& 13.76 \\
\end{longtable*}


\bibliography{sample631}{}
\bibliographystyle{aasjournal}



\end{document}